\documentclass[aip,jcp,reprint,amsmath,amssymb,floatfix,citeautoscript,nofootinbib]{revtex4-2}
\usepackage{amsmath}
\usepackage{graphicx}
\usepackage{amssymb}
\usepackage{bm}
\usepackage{newtxtext,newtxmath}
\allowdisplaybreaks
\usepackage{color}
\usepackage[usenames,dvipsnames]{xcolor}
\definecolor{myblue}{rgb}{0,0,1}
\usepackage[breaklinks=true,colorlinks=true,linkcolor=myblue,urlcolor=myblue,citecolor=myblue]{hyperref}

\makeatletter
\def\@email#1#2{%
 \endgroup
 \patchcmd{\titleblock@produce}
  {\frontmatter@RRAPformat}
  {\frontmatter@RRAPformat{\produce@RRAP{*#1\href{mailto:#2}{#2}}}\frontmatter@RRAPformat}
  {}{}
}%
\makeatother

\usepackage[T1]{fontenc}
\usepackage{makecell}
\usepackage{booktabs}
\usepackage{comment}

\begin{document}

\title{
High-performance parallel implementation of high-order coupled-cluster theories
}

\author{Yu Jin}
\affiliation{Initiative for Computational Catalysis, Flatiron Institute, New York, NY 10010, USA}
\author{Christopher Hillenbrand}
\affiliation{Department of Chemistry, Yale University, New Haven, CT 06520, USA}
\author{Timothy C. Berkelbach}
\email{tberkelbach@flatironinstitute.org}
\affiliation{Initiative for Computational Catalysis, Flatiron Institute, New York, NY 10010, USA}
\affiliation{Department of Chemistry, Columbia University, New York, NY 10027 USA}
\author{Huanchen Zhai}
\email{hzhai@flatironinstitute.org}
\affiliation{Initiative for Computational Catalysis, Flatiron Institute, New York, NY 10010, USA}

\begin{abstract}
High-order coupled-cluster theories with iterative triples (CCSDT), perturbative quadruples [CCSDT(Q)], and iterative quadruples (CCSDTQ) provide benchmark-quality correlation energies, but their steep computational scalings, $O(N^8)$, $O(N^9)$, and $O(N^{10})$, together with the large memory requirements of high-order amplitude tensors, have historically limited their application to small molecules. In this work, we develop efficient open-source implementations of spin-restricted CCSDT (RCCSDT), RCCSDT(Q), RCCSDTQ, and spin-unrestricted CCSDT (UCCSDT) within the \textsc{PySCF} package. The shared-memory implementation combines compact triangular storage of the highest-order amplitude tensors with the multithreaded tensor contraction backend \textsc{pytblis}, enabling efficient use of modern many-core CPU architectures. This design delivers near-ideal thread scaling up to 90 cores and achieves wall times shorter than or comparable to existing single-node implementations for representative benchmark molecules. We further extend RCCSDT, RCCSDT(Q), and RCCSDTQ to distributed-memory architectures using MPI-based algorithms. By distributing compact high-order amplitudes across MPI ranks and overlapping communication with computation through nonblocking data transfers, the distributed implementation achieves near-ideal strong scaling on up to 32 nodes, corresponding to approximately 3,000 CPU cores. These developments substantially extend the practical reach of canonical high-order coupled-cluster theory, enabling CCSDT and CCSDT(Q) calculations on systems with approximately 100 correlated electrons in 450 orbitals and CCSDTQ calculations on systems with approximately 50 correlated electrons in 115 orbitals. Applications to $\pi$-stacked noncovalent dimers, the CO dissociation energy of Cr(CO)$_6$, and the Cope rearrangement of semibullvalene demonstrate that canonical high-order coupled-cluster benchmarks are now feasible for chemically realistic molecular systems.
\end{abstract}

\maketitle

\section{Introduction}

Coupled-cluster (CC) theory provides a systematically improvable hierarchy of approximations for the electronic structure problem~\cite{bartlett2007coupled,shavitt2009many,crawford2007introduction}. CC theory with single, double, and perturbative triple excitations [CCSD(T)] has become the \textit{de facto} ``gold standard'' of quantum chemistry because it balances high accuracy with a tractable $O(N^7)$ computational scaling, where $N$ is the system size. However, the residual errors in CCSD(T) are not always negligible given the stringent accuracies demanded by modern thermochemistry, non-covalent interactions, and condensed-phase correlation benchmarks. In such cases, quantitative predictions require high-order methods that treat triple excitations iteratively (CCSDT)~\cite{noga1987full,noga1988erratum,scuseria1988new}, alongside quadruple corrections introduced either perturbatively [CCSDT(Q)]~\cite{bomble2005coupled} or iteratively (CCSDTQ).~\cite{kucharski1992coupled}

The need for post-CCSD(T) accuracy is especially clear in three settings. First, ``platinum-standard'' thermochemistry and spectroscopy require sub-kilojoule-per-mole precision, for which iterative triples and quadruples corrections are often needed to converge the correlation energy hierarchy~\cite{tajti2004heat,bomble2006high,harding2008high,karton2006w4,kodrycka2019platinum,lesiuk2022gold}. Second, in non-covalent interactions between large $\pi$-conjugated systems, the iterative-triples correction [CCSDT $-$ CCSD(T)] and the perturbative-quadruples correction [CCSDT(Q) $-$ CCSDT] can each be a significant fraction of the interaction energy~\cite{al2021interactions,schafer2025understanding}. Third, in metallic and near-metallic periodic systems, the perturbative triples correction in CCSD(T) exhibits an infrared divergence that is removed only by the full iterative triples treatment in CCSDT~\cite{neufeld2023highly,masios2023averting}. These examples motivate algorithm developments that make high-order CC methods usable beyond the small molecules accessible to conventional implementations.

Several quantum chemistry packages now provide single-node shared-memory parallel implementations of high-order CC methods. \textsc{MRCC}~\cite{mester2025overview} offers a general-order framework based on automated algebraic equation generation and string-based algorithms~\cite{kallay2001higher}, including general-order perturbative corrections such as CCSDT(Q)~\cite{kallay2005approximate,bomble2005coupled,kallay2008approximate}. CFOUR~\cite{matthews2020coupled} provides manually optimized restricted CCSDT, CCSDT(Q), and CCSDTQ implementations that exploit non-orthogonal spin adaptation~\cite{matthews2013revisitation,matthews2015non,matthews2019diagrams}. Related capabilities are also available in \textsc{ORCA}~\cite{lechner2024code}, \textsc{Q-Chem}~\cite{epifanovsky2021software}, \textsc{CCpy}~\cite{ccpy}, $\mathrm{e}^{\mathrm{T}}$~\cite{folkestad20262}, \textsc{Penguin}~\cite{hillers2025penguin}, and other software packages. Nevertheless, the practical reach of single-node implementations remains limited to relatively small systems: for CCSDT(Q), approximately 20 correlated electrons in 160 orbitals without point-group symmetry and 30 correlated electrons in 220 orbitals with point-group symmetry;~\cite{fishman2026development} and for CCSDTQ, approximately 30 correlated electrons in 108 orbitals with point-group symmetry.~\cite{eriksen2020ground} This limitation arises from the steep scaling inherent to such methods. In CCSDT, for instance, the leading amplitude tensors require $O(N^6)$ memory and the dominant computational steps scale as $O(N^8)$.

Reaching substantially larger systems requires distributed parallelism, but efficiently distributing high-order CC methods poses significant challenges due to the multidimensional nature and index-permutation symmetries involved in compact tensor storage and contractions. Parallel CCSD(T) has been implemented in several codes, including \textsc{MPQC/TiledArray}~\cite{peng2016massively}, \textsc{GAMESS}~\cite{datta2021massively}, \textsc{FHI-aims}~\cite{shen2019massive}, \textsc{MRCC}~\cite{ladoczki2025enabling}, and others. Full iterative CCSDT has been parallelized in fewer frameworks: \textsc{CFOUR} through the parallelism of computational tasks~\cite{prochnow2010parallel}; \textsc{NWChem}~\cite{apra2020nwchem} through the Tensor Contraction Engine (TCE)~\cite{hirata2003tensor,baumgartner2005synthesis} using Global Arrays~\cite{nieplocha1996global}; Aquarius with the Cyclops Tensor Framework (CTF)~\cite{solomonik2013cyclops,solomonik2014massively}; and, more recently, DIRAC through the ExaCorr module~\cite{pototschnig2021implementation} using automatically generated equations with CTF~\cite{brandejs2025generating}.

To the best of our knowledge, there are currently no high-performance, distributed-memory implementations of spin-free CCSDT, CCSDT(Q), and CCSDTQ. The existing software landscape often forces a trade-off: highly optimized CC codes are typically written in compiled languages, whose complexity can hinder rapid methodological prototyping, whereas accessible, Python-based open-source implementations generally lack the parallel scalability required for medium-to-large systems. Furthermore, effectively leveraging dense multi-core CPUs ($\sim$96 threads), large node-local memory ($\sim$1~TB per node), and TB-scale inter-node communication to approach peak floating-point throughput in the massive tensor contractions of high-order CC methods remains a major bottleneck. Overcoming these challenges is essential for demonstrating that scalable, high-performance CC frameworks can be achieved without sacrificing methodological accessibility, thereby extending canonical high-order CC calculations to molecular systems beyond the reach of existing implementations.

In this work, we first present Python-based, efficient, shared-memory parallel implementations of spin-restricted CCSDT (RCCSDT), RCCSDT(Q), RCCSDTQ, and spin-unrestricted CCSDT (UCCSDT) within the \textsc{PySCF} quantum chemistry package~\cite{sun2018pyscf,sun2020recent,sun2026python}. The shared-memory implementation reduces memory requirements by up to factors of six for RCCSDT and 24 for RCCSDTQ through compact triangular storage of the highest-order amplitude tensors, while using \textsc{pytblis} for high-performance multithreaded tensor contractions~\cite{huang2018strassen,matthews2018high,tblis,pytblis}. Benchmark calculations demonstrate efficient thread scaling up to 90 cores. We further extend RCCSDT, RCCSDT(Q), and RCCSDTQ to distributed-memory architectures by distributing the compact highest-order amplitudes and residuals across MPI ranks, achieving efficient inter-node scaling up to 32 nodes. These developments bring canonical CCSDT and CCSDT(Q) calculations on systems with approximately 100 correlated electrons in 450 orbitals, and canonical CCSDTQ calculations on systems with approximately 50 correlated electrons in 115 orbitals, within reach. Finally, we demonstrate the capabilities of the implementation with canonical high-order coupled-cluster calculations that substantially exceed previous literature limits, including CCSDT(Q) calculations of the interaction energies of $\pi$-stacked aromatic and polyene dimers and the CO dissociation energy of a prototypical transition-metal carbonyl complex, as well as CCSDTQ calculations for the Cope rearrangement of semibullvalene.

The remainder of the paper is organized as follows. Section~\ref{sec:shared_mem} presents the shared-memory implementation and its performance. Section~\ref{sec:dis_mem} describes the distributed implementation and its scaling behavior. Section~\ref{sec:practical} provides practical guidance for users. Section~\ref{sec:applications} presents numerical illustrations. Section~\ref{sec:conclusions} provides concluding remarks and an outlook for future work. The theoretical background, notation, and working equations are collected in the Appendix.

\section{Shared memory implementation\label{sec:shared_mem}}

\subsection{Notation and working equations}

We briefly summarize the notation needed in the implementation discussion; the formal definitions and full working equations are given in Appendix~\ref{sec:theory}. Throughout the main text, orbital indices $i,j,k,l,\ldots$, $a,b,c,d,\ldots$, and $p,q,r,s,\ldots$ refer to occupied, virtual, and general molecular orbitals, respectively. In unrestricted cases, a barred index denotes the spin-down channel. $N_{\mathrm{o}}$ and $N_{\mathrm{v}}$ denote the numbers of occupied and virtual spatial orbitals, respectively, and $T_n$ and $R_n$ denote the $n$-fold excitation amplitude and residual tensors.  For closed-shell RCC methods, we use the non-orthogonal spin-free amplitudes $\check{t}$ of Appendix~\ref{sec:nosa}, which obey the paired occupied--virtual column-permutation symmetry; the corresponding paired-index permutation operator, $\mathcal{P}_{(ia)(jb)\cdots}$, is defined in Eq.~\eqref{eq:rcc-perm-op}. Following Ref.~\citenum{matthews2015non}, a check mark on an index denotes spin summation, with the $T_3$ and $T_4$ combinations used below defined in Eqs.~\eqref{eq:triples-spin-sum} and~\eqref{eq:quadruples-spin-sum}.  Unless otherwise stated, all Fock, ERI, and intermediate tensors $F$ and $W$ are formed from the $T_1$-dressed Hamiltonian described in Appendix~\ref{sec:t1dress}. The RCCSDT, RCCSDTQ, and RCCSDT(Q) working equations referenced by the algorithms are given in Appendix~\ref{appendix:rccsdt}, ~\ref{appendix:rccsdtq}, and ~\ref{appendix:rccsdt_q}, respectively.
The UCCSDT equations are collected in Appendix~\ref{appendix:uccsdt}.

\subsection{General considerations}

We implement the RCCSDT, RCCSDT(Q), RCCSDTQ, and UCCSDT methods within the \textsc{PySCF} quantum chemistry package~\cite{sun2018pyscf,sun2020recent,sun2026python} using Python and C. These methods are made available in \textsc{PySCF} 2.13.2. Following \textsc{PySCF}'s general design philosophy, Python and \textsc{NumPy} are utilized for method orchestration, tensor-layout bookkeeping, memory management, and the pre-computation of load-balancing information, while computationally intensive tensor operations are offloaded to specialized C-level routines or external libraries via \texttt{ctypes}. Within C routines, shared-memory parallelization is implemented using OpenMP whenever possible. This design balances flexibility and readability with optimal CPU and memory performance for production-level simulations.

Before discussing the specific implementation details, we outline several key considerations regarding the data storage and manipulation for CC amplitude and Hamiltonian integral tensors, as well as the strategies for their efficient contraction.

\subsubsection{Amplitude tensors}

We utilize full in-memory storage of amplitude, integral, and intermediate tensors throughout the calculation. Disk-based storage is used only optionally, for example, when restarting a CC calculation, loading RCCSDT amplitudes for a subsequent (Q) energy correction, or saving and loading vectors in out-of-core direct inversion in the iterative subspace (DIIS). This choice is motivated by the scaling characteristics of high-order coupled-cluster methods. For example, in RCCSDT, the leading $T_3$ amplitudes require $O(N_{\mathrm{o}}^3 N_{\mathrm{v}}^3)$ memory, whereas the dominant contractions scale as $O(N_{\mathrm{o}}^3 N_{\mathrm{v}}^5)$ for systems with $N_{\mathrm{v}} > N_{\mathrm{o}}$. Thus, large high-order coupled-cluster calculations are limited by both memory and computational cost. While disk-based intermediates could reduce the peak memory footprint, the resulting I/O overhead would substantially increase the wall time. For systems whose amplitudes do not fit in memory, such an approach would generally render the calculation prohibitively slow rather than meaningfully extending the practical system size.

Following the design choice of existing molecular CC modules in \textsc{PySCF}, point-group symmetry is not exploited in our current implementation. We also restrict our implementation to treat only real-valued amplitudes and integrals.

The spin-summation operations that map spin-free amplitudes $\check{t}$ to the spin-summed amplitudes $\check{t}_{\check{i}\cdots}^{\check{a}\cdots}$ [see Eqs.~\eqref{eq:triples-spin-sum} and~\eqref{eq:quadruples-spin-sum}] are implemented as C routines with in-place amplitude modifications following Ref.~\citenum{springer2019spin}. Each routine processes all permutations of the $abc$ (or $abcd$) virtual indices in a single pass using small preallocated intermediate arrays. This avoids large temporary copies of the amplitude tensors and enables efficient multi-threaded execution.

We provide two variants in our shared-memory high-order CC implementations with ``full'' and ``compact'' storage of amplitudes, respectively. The ``full'' variants store the complete $T_3$ or $T_4$ amplitudes and avoid unpacking, whereas the ``compact'' variants store only the hyper-triangular occupied-index sectors and reconstruct symmetry-related blocks on the fly by exploiting the spin-free column-permutation symmetry illustrated in Eq.~\eqref{eq:rccsdt-amps-symm}.
Henceforth, we mainly focus on the implementation variants with ``compact'' amplitude storage, as this is the most relevant for large practical CC calculations.

For RCCSDT, only the $T_3$ components satisfying $i \leq j \leq k$ are stored, reducing the memory footprint by a factor of approximately $6$ in the large-$N_{\mathrm{o}}$ limit. For RCCSDTQ, the $T_3$ amplitudes are stored in full, whereas only the $T_4$ components satisfying $i \leq j \leq k \leq l$ are stored, reducing the $T_4$ memory by a factor of approximately $24$ in the large-$N_{\mathrm{o}}$ limit. This (hyper-)triangular storage is not the most compact possible representation for RCC amplitudes, as it contains three additional sources of symmetry-related redundancies (see Sec.~\ref{sec:symm-puri-ortho} for detailed discussions). While further saving the amplitude storage cost by explicitly removing these redundancies is possible, we do not consider them in our current implementation, as the memory overhead caused by these additional redundancies is negligible at the RCCSDT level and remains minor at the RCCSDTQ level, and storing all virtual indices $abc$ (or $abcd$) contiguously is essential for efficient tensor contractions. Alternatively, one could store only $T_3$ components with $a\leq b \leq c$ while keeping $ijk$ indices contiguous. This is not considered for the RCCSDT implementation in this work, as it can introduce non-negligible computational overhead in the $W^{\mathrm{vv}}_{\mathrm{vv}} \ast T_3$ contraction (see Sec.~\ref{sec:rccsdt-contr-compact} for details), which is the most expensive step when $N_{\mathrm{v}} > N_{\mathrm{o}}$. Here and below, we use $A \ast B$ to denote a generic tensor contraction with the appropriate index summations.

For UCCSDT, the $T_3$ amplitudes are antisymmetric within the occupied ($ijk$) and virtual ($abc$) index groups. The four distinct spin-sector components, $t_{i<j<k}^{a<b<c}$, $t_{\overline{i}<\overline{j}<\overline{k}}^{\overline{a}<\overline{b}<\overline{c}}$, $t_{i<j,\overline{k}}^{a<b,\overline{c}}$, and $t_{i,\overline{j}<\overline{k}}^{a,\overline{b}<\overline{c}}$, are stored separately, which saves approximately a factor of $36$ for same-spin amplitudes and a factor of $4$ for mixed-spin amplitudes. Therefore, for closed-shell systems, the RCCSDT implementation is typically more memory efficient, even compared with a UCCSDT implementation that exploits the closed-shell symmetry allowing $T_3$ to be fully represented by $t_{i<j,\overline{k}}^{a<b,\overline{c}}$.

\subsubsection{Tensor contraction backends}

Our implementation supports switching between different tensor contraction backends (including \texttt{numpy.einsum} and \textsc{pytblis}~\cite{pytblis}) at runtime. For all numerical illustrations shown in this work, we perform tensor contractions using \textsc{pytblis}, a Python wrapper of the TBLIS library that provides multi-threaded tensor contractions for arbitrary index layouts with in-place scaling and accumulation of the output tensor and no memory allocations of large intermediates~\cite{huang2018strassen,matthews2018high,tblis}. This capability is particularly important for high-rank tensors, where avoiding explicit transposed or reordered temporary arrays is essential for preserving the memory savings from compact storage.

\subsection{Tensor contractions for compact amplitudes\label{sec:rccsdt-contr-compact}}

With the compact storage format, the residuals cannot be formed by evaluating each contraction once over the full tensor and subsequently applying the full paired-index permutation (namely, column-permutation of indices) operator [Eq.~\eqref{eq:rcc-perm-op}]. Instead, symmetry-related contributions must be evaluated explicitly, involving up to $3!$ terms for $T_3$/$R_3$ and $4!$ terms for $T_4$/$R_4$. This explicit repetition does not increase the leading-order operation count of the compact-storage algorithm compared to the full tensor storage algorithm, because the residuals themselves obey the same column-permutation symmetry. Thus, for each repeated contraction, only the triangular sectors $i \leq j \leq k$ and $i \leq j \leq k \leq l$ are accumulated, reducing the leading operation count by approximately $1/3!$ for $T_3$/$R_3$ and $1/4!$ for $T_4$/$R_4$ in the large-$N_{\mathrm{o}}$ limit. Here and in the following, the operation count refers to the number of multiply-add pairs, with one multiply-add pair counted as a single operation. In addition, the nominal $3!$ and $4!$ permuted contractions can often be combined or reduced to fewer distinct contractions, typically resulting in a smaller total operation count than the full tensor contraction case, as discussed below.

Consequently, contractions are processed differently according to whether the involved amplitude blocks reside in the same or different storage regions. Some contractions can be evaluated directly via a sequential traversal over stored blocks, provided that all required amplitudes are available from the same stored block. In contrast, other contractions may demand concurrent contributions from multiple distinct blocks; to address this, we develop a tile-based unpacking algorithm that ensures efficient memory access. Examples are given below for RCCSDT and RCCSDTQ, respectively.

\subsubsection{RCCSDT contractions without unpacking\label{sec:rccsdt-wvvvv}}

Consider the most expensive $W^{\mathrm{vv}}_{\mathrm{vv}} \ast T_3$ contraction in the RCCSDT residual equation, Eq.~\eqref{eq:resi-rccsdt}. In terms of the full $T_3$ tensor, this reads
\begin{equation}
    \check{r}_{ijk}^{abc} \leftarrow \mathcal{P}_{(ia)(jb)(kc)} \frac{1}{2}  \sum_{de} W^{ab}_{de} \check{t}_{ijk}^{dec}
\end{equation}
with a dominant operation count of $N_{\mathrm{o}}^3 N_{\mathrm{v}}^5$, where ``$\leftarrow$'' suggests a contribution from the right-hand side expression to the left-hand side. In our implementation with compact amplitude storage, only the $i \leq j \leq k$ components of the $R_3$ amplitude are accumulated, while the permutation operator expands the contractions into three explicit terms considering the column permutation symmetry of both $T_3$ and $W^{\mathrm{vv}}_{\mathrm{vv}}$:
\begin{equation}
\begin{aligned}
    \check{r}_{i\leq j\leq k}^{abc} &\leftarrow \sum_{de} W^{ab}_{de} \check{t}_{ijk}^{dec} + \sum_{de} W^{ac}_{de} \check{t}_{ikj}^{deb} + \sum_{de} W^{bc}_{de} \check{t}_{jki}^{dea}.
\end{aligned}
\end{equation}
For given indices $i, j, k$ (with $i\leq j\leq k$), the permuted amplitudes $\check{t}_{ikj}^{deb}$ and $\check{t}_{jki}^{dea}$ are read from the same compact storage as $\check{t}_{ijk}^{dec}$ after relabeling the paired occupied--virtual columns. The $W^{\mathrm{vv}}_{\mathrm{vv}} \ast T_3$ contraction is thus implemented as a \emph{direct traversal} over $i\le j\le k$ indices. Restricting the occupied sum to the $\frac{1}{6}N_{\mathrm{o}}(N_{\mathrm{o}}+1)(N_{\mathrm{o}}+2)$ triangular index combinations reduces the dominant operation count to $\frac{3}{6}N_{\mathrm{o}}(N_{\mathrm{o}}+1)(N_{\mathrm{o}}+2)N_{\mathrm{v}}^5 \approx \frac{1}{2} N_{\mathrm{o}}^3N_{\mathrm{v}}^5$, a factor of 2 reduction in the large-$N_{\mathrm{o}}$ limit. A further factor-of-two reduction can be achieved by decomposing each contraction into symmetric and antisymmetric components~\cite{scuseria1988efficient}, which is not considered in this work.

\subsubsection{RCCSDT contractions with unpacking}

\begin{figure}
    \centering
    \includegraphics[width=\linewidth]{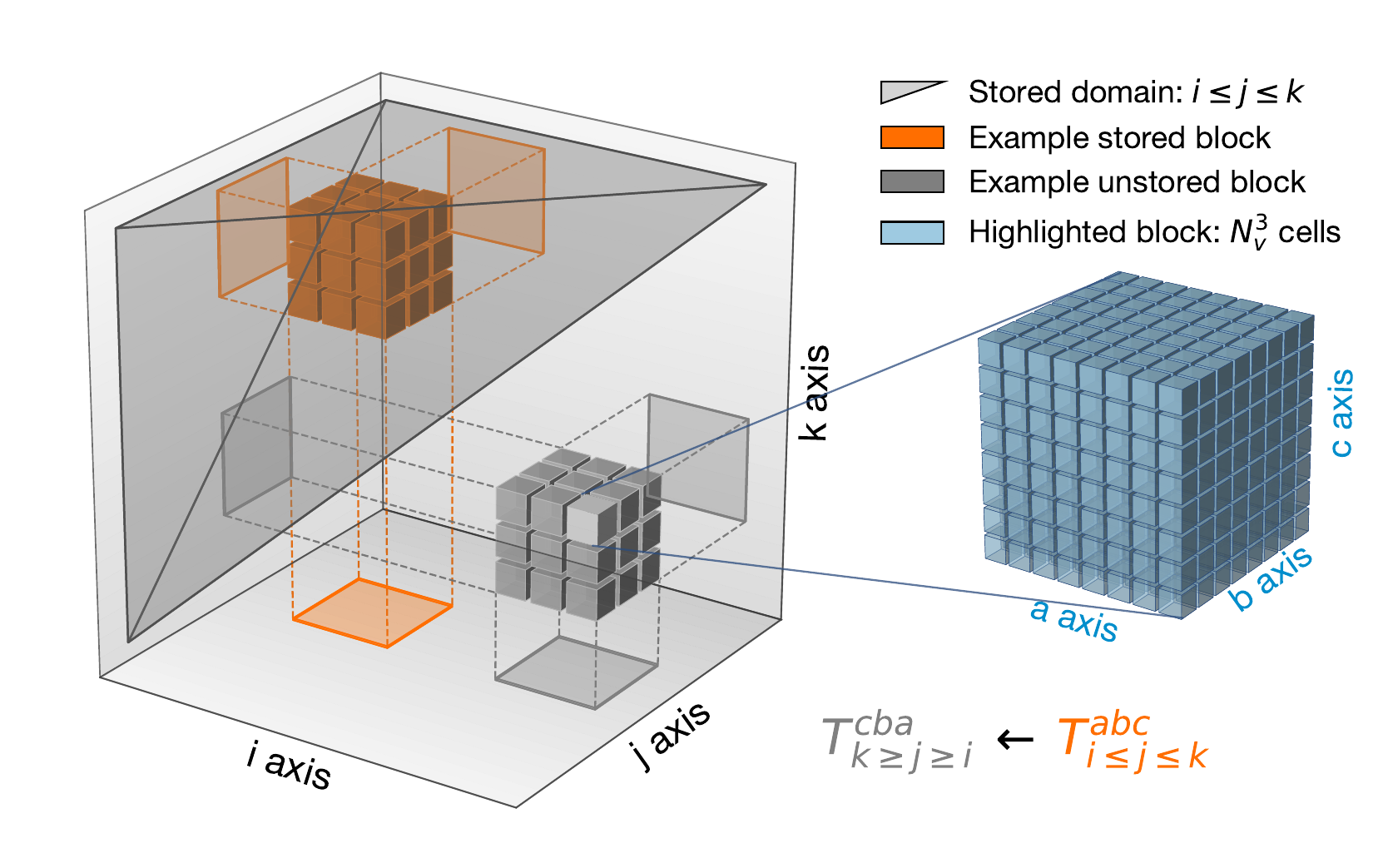}
    \caption{Compact storage of the $T_3$ amplitudes in RCCSDT. The light gray upper-triangular region ($i\leq j\leq k$) is stored explicitly in memory. Off-triangular blocks (dark gray) are not stored but can be reconstructed on the fly from the corresponding stored blocks (orange) using the column permutation symmetry. Each block spans the full $N_{\mathrm{v}}^3$ virtual-index range.}
    \label{fig:t3_perm}
\end{figure}

Other contractions access the occupied indices of $T_3$ more broadly and therefore cannot be evaluated entirely within a single stored $ijk$ block of $T_3$. The $W^{\mathrm{oo}}_{\mathrm{oo}} \ast T_3$ contraction in Eq.~\eqref{eq:resi-rccsdt}, 
\begin{equation}
    \check{r}_{ijk}^{abc} \leftarrow \mathcal{P}_{(ia)(jb)(kc)} \frac{1}{2}  \sum_{lm} W^{lm}_{ij} \check{t}_{lmk}^{abc}
\end{equation}
with an operation count of $N_{\mathrm{o}}^5 N_{\mathrm{v}}^3$,
expands in the compact residual as
\begin{equation}
    \check{r}_{i\leq j\leq k}^{abc} \leftarrow \sum_{lm} W^{lm}_{ij} \check{t}_{lmk}^{abc} + \sum_{lm} W^{lm}_{ik} \check{t}_{lmj}^{acb} + \sum_{lm} W^{lm}_{jk} \check{t}_{lmi}^{bca}
\end{equation}
with an asymptotic operation count of $\frac{1}{2} N_{\mathrm{o}}^5 N_{\mathrm{v}}^3$. Here, $\check{t}_{lmk}^{abc}$ is required for arbitrary $l$ and $m$, which need not satisfy $l \leq m \leq k$. Blocks of $T_3$ outside the stored region are therefore reconstructed from stored blocks using the column permutation symmetry; we refer to this step as \emph{unpacking}. As illustrated in Fig.~\ref{fig:t3_perm}, any off-triangular block can be obtained by permuting a stored block. In practice, unpacking is performed on the fly for a tile of indices $(i,j,k,l,m)$ with an adjustable block size. The block size balances the memory footprint of temporary buffers, the amount of redundant work introduced at tile boundaries, and the efficiency of multi-threaded tensor contractions. The same unpacking strategy is applied to the $W^{\mathrm{ov}}_{\mathrm{ov}} \ast T_3$ and $W^{\mathrm{ov}}_{\mathrm{vo}} \ast T_3$ contractions.

\subsubsection{Total cost of RCCSDT contractions}

Counting only the leading $O(N^8)$ terms, the formal operation count per RCCSDT iteration for the implementation used in this work is $N_{\mathrm{o}}^3N_{\mathrm{v}}^5 + 3N_{\mathrm{o}}^4N_{\mathrm{v}}^4 + N_{\mathrm{o}}^5N_{\mathrm{v}}^3$ with the full $T_3$ tensor and $\frac{1}{2}N_{\mathrm{o}}^3N_{\mathrm{v}}^5 + 2N_{\mathrm{o}}^4N_{\mathrm{v}}^4 + \frac{1}{2}N_{\mathrm{o}}^5N_{\mathrm{v}}^3$ with compact $T_3$ at unit block size in the large-$N_{\mathrm{o}}$ limit. Details on operation count can be found in Table~\ref{tab:rccsdt_operation_count} in the Appendix. Larger blocks slightly increase the compact-storage operation count because edge tiles contain some redundant work, but they also reduce unpacking overhead and improve multi-threaded tensor contraction efficiency.

\subsubsection{RCCSDTQ contractions}

The same direct-traversal/unpacking logic extends to the $T_4$ amplitudes in RCCSDTQ. The $W^{\mathrm{vv}}_{\mathrm{vv}} \ast T_4$ contraction in Eq.~\eqref{eq:resi-rccsdtq} has an operation count of $N_{\mathrm{o}}^4 N_{\mathrm{v}}^6$ with the full tensor,
\begin{equation}
    \check{r}_{ijkl}^{abcd} \leftarrow \mathcal{P}_{(ia)(jb)(kc)(ld)} \frac{1}{4}  \sum_{ef} W^{ab}_{ef} \check{t}_{ijkl}^{efcd}
\end{equation}
and expands in the compact $i \leq j \leq k \leq l$ residual into six terms with $T_4$ all available from the stored block with the same $i \leq j \leq k \leq l$ indices upon permutations:
\begin{equation}
\begin{aligned}
    &\check{r}_{i\leq j\leq k\leq l}^{abcd} \leftarrow \sum_{ef} W^{ab}_{ef} \check{t}_{ijkl}^{efcd} + \sum_{ef} W^{ac}_{ef} \check{t}_{ikjl}^{efbd} + \sum_{ef} W^{ad}_{ef} \check{t}_{iljk}^{efbc} \\
    &\quad + \sum_{ef} W^{bc}_{ef} \check{t}_{jkil}^{efad} + \sum_{ef} W^{bd}_{ef} \check{t}_{jlik}^{efac} + \sum_{ef} W^{cd}_{ef} \check{t}_{klij}^{efab}.
\end{aligned}
\end{equation}
Restricting the occupied sum to the $\frac{1}{24}N_{\mathrm{o}}(N_{\mathrm{o}}+1)(N_{\mathrm{o}}+2)(N_{\mathrm{o}}+3)$ triangular combinations reduces the operation count to $\frac{6}{24}N_{\mathrm{o}}(N_{\mathrm{o}}+1)(N_{\mathrm{o}}+2)(N_{\mathrm{o}}+3)N_{\mathrm{v}}^6 \approx \frac{1}{4} N_{\mathrm{o}}^4 N_{\mathrm{v}}^6$, a factor of 4 reduction in the large $N_{\mathrm{o}}$ limit.

Counting only the leading $O(N^{10})$ terms, the per-iteration total operation count used in this work is $N_{\mathrm{o}}^4N_{\mathrm{v}}^6 + 6N_{\mathrm{o}}^5N_{\mathrm{v}}^5 + 2N_{\mathrm{o}}^6N_{\mathrm{v}}^4$ for the full $T_4$ tensor and $\frac{1}{4}N_{\mathrm{o}}^4N_{\mathrm{v}}^6 + \frac{7}{3}N_{\mathrm{o}}^5N_{\mathrm{v}}^5 + \frac{3}{4}N_{\mathrm{o}}^6N_{\mathrm{v}}^4$ for compact $T_4$ at unit block size. Details on the operation count can be found in Table~\ref{tab:rccsdtq_operation_count} in the Appendix. As for RCCSDT, larger block sizes introduce a modest edge-tile overhead but can improve practical performance by reducing the frequency of unpacking and increasing tensor-contraction granularity.

\subsection{Removing redundancies in RCC amplitudes
\label{sec:symm-puri-ortho}}

After the residual contractions have been accumulated in compact triangular storage, three post-processing steps are applied. These steps enforce the symmetry constraints and remove redundant parameters in the non-orthogonal spin-adapted amplitudes, which would otherwise slow convergence.

\subsubsection{Symmetrization}

The residual must be invariant under the same column permutation symmetry as the amplitudes. For a full, non-compact residual, this condition is imposed directly by applying the paired-index permutation operator, i.e.,
\begin{equation}
    \check{r}_{ijk}^{abc} \leftarrow \mathcal{P}_{(ia)(jb)(kc)} \check{r}_{ijk}^{abc}.
\end{equation}

For compact residuals stored only in the triangular sectors
$i \leq j \leq k$ or $i \leq j \leq k \leq l$, the column-permutation symmetry associated with distinct occupied indices is enforced during the contractions themselves by explicitly evaluating the required column-permuted contributions. However, residual blocks containing two identical occupied indices may retain residual asymmetries among the corresponding virtual indices, primarily due to numerical roundoff. These blocks are therefore symmetrized explicitly. For $R_3$:
\begin{equation}
\begin{aligned}
    \check{r}_{i=j< k}^{abc} \leftarrow&\ \mathcal{P}_{(ia)(jb)} \check{r}_{i=j< k}^{abc}, \\
    \check{r}_{i< j = k}^{abc} \leftarrow&\ \mathcal{P}_{(jb)(kc)} \check{r}_{i< j = k}^{abc},
\end{aligned}
\end{equation}
and for $R_4$:
\begin{equation}
\begin{aligned}
    \check{r}_{i=j < k \leq l}^{abcd}  \leftarrow&\ \mathcal{P}_{(ia)(jb)} \check{r}_{i=j < k \leq l}^{abcd}, \\
    \check{r}_{i< j = k < l}^{abcd} \leftarrow&\ \mathcal{P}_{(jb)(kc)} \check{r}_{i< j = k < l}^{abcd}, \\
    \check{r}_{i\leq j < k = l}^{abcd} \leftarrow&\ \mathcal{P}_{(kc)(ld)} \check{r}_{i\leq j < k = l}^{abcd}.
\end{aligned}
\end{equation}

\subsubsection{Purification}

When three or more occupied indices are equal, the corresponding spin-free residual components must vanish identically because each spatial orbital can accommodate only two electrons or holes. Finite numerical precision can nevertheless produce small nonzero values, which are zeroed explicitly:
\begin{equation}
    \check{r}_{i=j=k}^{abc} = 0, \quad \check{r}_{i=j=k \leq l}^{abcd} = 0, \quad \check{r}_{i\leq j=k=l}^{abcd} = 0.
\end{equation}
While these vanishing conditions could reduce amplitude storage requirements, we do not utilize them because the associated memory costs are marginal in RCCSDT and RCCSDTQ.

\subsubsection{Orthogonalization}

As discussed in Appendix~\ref{sec:nosa} and analyzed in Appendix~\ref{appendix:nosa}, for high-order RCC starting from RCCSDT, the non-orthogonal spin-adapted amplitude tensors contain a null space that arises from the non-orthogonality of the spin-free excited configurations.~\cite{adams1979orthogonally,matthews2013revisitation,matthews2015non} The residual must be projected onto the orthogonal and nonredundant subspace so that subsequent amplitude updates remain in that subspace. This orthogonalization is implemented by removing the null-space component of the residual at the end of each iteration.

For $T_3$, the null space is characterized by the constraint~\cite{wang2018simple}
\begin{equation}
    \sum_{\pi \in S_3} \check{r}_{ijk}^{\pi(abc)} = 0,
\end{equation}
where $S_n$ is the symmetric group on $n$ elements, and $\pi \in S_3$ runs over all permutations of the virtual-index tuple $(a,b,c)$. The projection is applied by subtracting the mean over all $3! = 6$ virtual permutations:
\begin{equation}
    \check{r}_{ijk}^{abc} \leftarrow \check{r}_{ijk}^{abc} - \frac{1}{6} \sum_{\pi \in S_3} \check{r}_{ijk}^{\pi(abc)}.
\end{equation}
For $T_4$, there are 10 independent null-space constraints associated with index permutations. In this work, we impose these constraints explicitly through a projection operator, as described in Appendix~\ref{appendix:nosa}. Alternatively, the same constraints can be enforced by performing spin summation followed by ``de-spin-summation'' at the end of each iteration, as described in Ref.~\citenum{matthews2015non}.

All three post-processing steps are implemented in this work. Symmetrization and purification mainly remove numerical roundoff errors, which can become relevant for very tight convergence thresholds and are less problematic when starting from an initial guess that already satisfies the corresponding constraints, such as the zero initial guess. Orthogonalization, however, is essential for restricting the residual and amplitude updates to the orthogonal nonredundant subspace and is important for robust convergence.

\subsection{RCCSDT(Q) contractions\label{sec:rccsdt_q_impl}}

The CCSDT(Q) energy correction is a non-iterative contraction over four occupied and four virtual indices, requiring dedicated algorithms for efficient evaluation. Here, we consider two algorithms that differ in the summation order of Eq.~\eqref{eq:ener-pt-q}, each suited to different system characteristics.

\subsubsection{IJK-based algorithm}

Because the $T_4$ amplitudes obey fourfold column permutation symmetry, the energy expression Eq.~\eqref{eq:ener-pt-q} can be restricted to the upper-triangular $i \leq j \leq k \leq l$ index set:
\begin{equation}
    E_{(Q)} = \frac{1}{24} \sum_{i\leq j\leq k\leq l} f_{ijkl} \sum_{abcd} z_{ijkl}^{abcd} \check{t}_{\check{i}\check{j}\check{k}\check{l}}^{\check{a}\check{b}\check{c}\check{d}}
\end{equation}
where $f_{ijkl}$ is a degeneracy factor accounting for the multiplicity of each triangular index tuple,
\begin{equation}
f_{ijkl} =
\begin{cases}
    24, &i < j < k < l, \\
    12, &i = j < k < l,\;\; i < j = k < l,\;\; i < j < k = l, \\
    6,  &i = j < k = l, \\
    0,  &\text{otherwise}.
\end{cases}
\end{equation}
We refer to this occupied-tuple traversal as the \emph{IJK-based} algorithm. Its buffer memory requirement is $2N_{\mathrm{v}}^4$ floating point numbers. Since each task fixes one compact occupied tuple $i \leq j \leq k \leq l$, the permutation operator $\mathcal{P}_{(ia)(jb)(kc)(ld)}$ (required for evaluating $\check{t}_{ijkl}^{abcd}$ according to Eq.~\eqref{eq:resi-pt-q}) cannot be applied afterward as a single post-processing step. Instead, each contraction is evaluated explicitly over the required paired-index permutations, up to all $4! = 24$ terms, with reductions to 12 or 6 terms when intermediate symmetry permits.

The intermediates $\overline{W}^{\mathrm{vvv}}_{\mathrm{voo}}$ and $\overline{W}^{\mathrm{vvo}}_{\mathrm{ooo}}$ defined in Eqs.~\eqref{eq:q_w_vvv_voo} and~\eqref{eq:q_w_vvo_ooo}
can become prohibitively large if precomputed and stored for all index combinations in large systems. In our implementation, they are therefore evaluated on the fly for each required index slice. With the intermediates recomputed on the fly, counting only the leading terms, the $IJK$-based operation count in the large-$N_{\mathrm{o}}$ limit is
\begin{equation*}
    N_{\mathrm{o}}^5N_{\mathrm{v}}^4 + \tfrac{3}{2} N_{\mathrm{o}}^4N_{\mathrm{v}}^5 + \tfrac{1}{4} N_{\mathrm{o}}^4 N_{\mathrm{v}}^4 + N_{\mathrm{o}}^6N_{\mathrm{v}}^2 + 2 N_{\mathrm{o}}^5N_{\mathrm{v}}^3.
\end{equation*}
This $IJK$-based implementation is naturally compatible with the compact $i \leq j \leq k$ storage of $T_3$, because the virtual indices remain contiguous in memory.

\subsubsection{ABC-based algorithm}

\begin{figure}
    \centering
    \includegraphics[width=\linewidth]{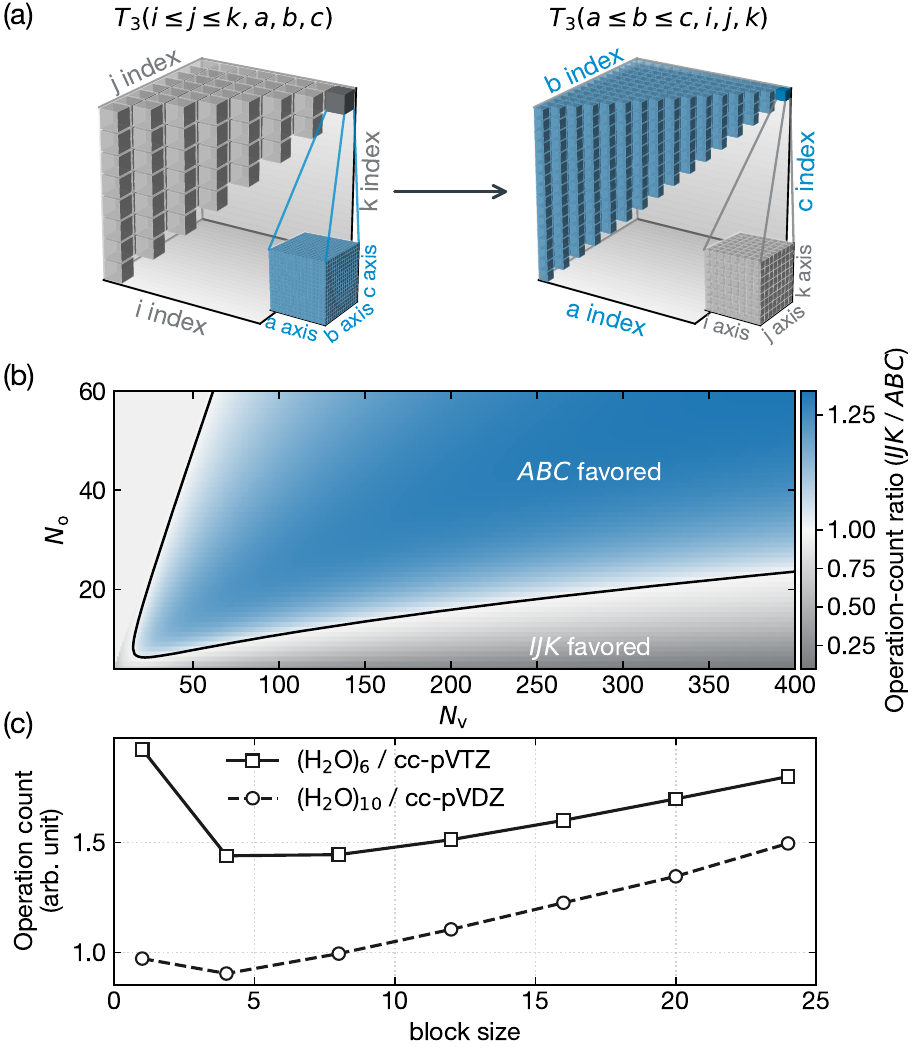}
    \caption{Transposition of compact $T_3$ amplitudes and operation-count analysis for the RCCSDT(Q) correction. (a) The $T_3$ amplitudes are transposed from the occupied-triangular representation, $T_3(i\leq j\leq k,a,b,c)$, where each stored $(i,j,k)$ block contains $N_{\mathrm{v}}^3$ contiguous virtual-index elements, to the virtual-triangular representation, $T_3(a\leq b\leq c,i,j,k)$, where each stored $(a,b,c)$ block contains $N_{\mathrm{o}}^3$ contiguous occupied-index elements, as required by the $ABC$-based algorithm. (b) Ratio of the formal operation counts of the $IJK$-based and $ABC$-based algorithms as a function of $N_{\mathrm{o}}$ and $N_{\mathrm{v}}$, assuming a unit block size. The black contour marks equal operation counts. (c) Formal operation count of the $ABC$-based implementation as a function of virtual block size for (H$_2$O)$_{10}$/cc-pVDZ and (H$_2$O)$_6$/cc-pVTZ.}
    \label{fig:q_operation_counts}
\end{figure}

An alternative \emph{ABC-based} algorithm reverses the summation order in Eq.~\eqref{eq:ener-pt-q} and loops over virtual index quadruples:
\begin{equation}
    E_{(Q)} = \frac{1}{24} \sum_{a\leq b\leq c\leq d} f_{abcd} \sum_{ijkl} z_{ijkl}^{abcd} \check{t}_{\check{i}\check{j}\check{k}\check{l}}^{\check{a}\check{b}\check{c}\check{d}},
    \label{eq:ener-pt-q-abc}
\end{equation}
with the leading operation count
\begin{equation*}
    N_{\mathrm{o}}^4N_{\mathrm{v}}^5 + \tfrac{3}{2} N_{\mathrm{o}}^5N_{\mathrm{v}}^4 + \tfrac{1}{4} N_{\mathrm{o}}^4N_{\mathrm{v}}^4 + N_{\mathrm{o}}^2N_{\mathrm{v}}^6 + 2 N_{\mathrm{o}}^3N_{\mathrm{v}}^5,
\end{equation*}
using the modified intermediates defined in Eq.~\eqref{eq:q_w_alter}. The $ABC$-based approach requires the $T_3$ amplitudes to be stored with the virtual indices triangularized ($a \leq b \leq c$) and the occupied indices contiguous in memory, rather than the occupied-triangular layout used in the iterative RCCSDT calculation. This one-time transposition of the compact $T_3$ representation, illustrated in Fig.~\ref{fig:q_operation_counts}(a), is performed before the (Q) evaluation.

A comparison of the formal operation counts for the two algorithms is shown in Fig.~\ref{fig:q_operation_counts}(b). The $ABC$-based approach is favored in the typical regime with $N_{\mathrm{v}} > N_{\mathrm{o}}$, unless $N_{\mathrm{v}} \gtrsim 20 N_{\mathrm{o}}$, at which point the $N_{\mathrm{o}}^2N_{\mathrm{v}}^6$ term dominates the overall cost, and thus the $IJK$-based approach is favored.

Compared to the $IJK$-based algorithm, the $ABC$-based approach further reduces the buffer memory requirement from $2N_{\mathrm{v}}^4$ to $2N_{\mathrm{o}}^4$ tensor elements,
which is particularly beneficial when $N_{\mathrm{v}}$ is large. In our implementation, the virtual indices $abcd$ are tiled into adjustable blocks of size $N^{\mathrm{blk}}_{\mathrm{v}}$ to improve tensor-contraction efficiency while requiring only a manageable increase in memory. As shown in Fig.~\ref{fig:q_operation_counts}(c), using a moderate block size also reduces the actual operation count relative to the unit-block-size estimate. This reduction arises from the effective caching and reuse of slices of the intermediates $\overline{W}^{\mathrm{vvv}}_{\mathrm{voo}}$ and $\overline{W}^{\mathrm{vvo}}_{\mathrm{ooo}}$ within each $abcd$ tile.

\subsection{Performance}

\begin{figure}
    \centering
    \includegraphics[width=\linewidth]{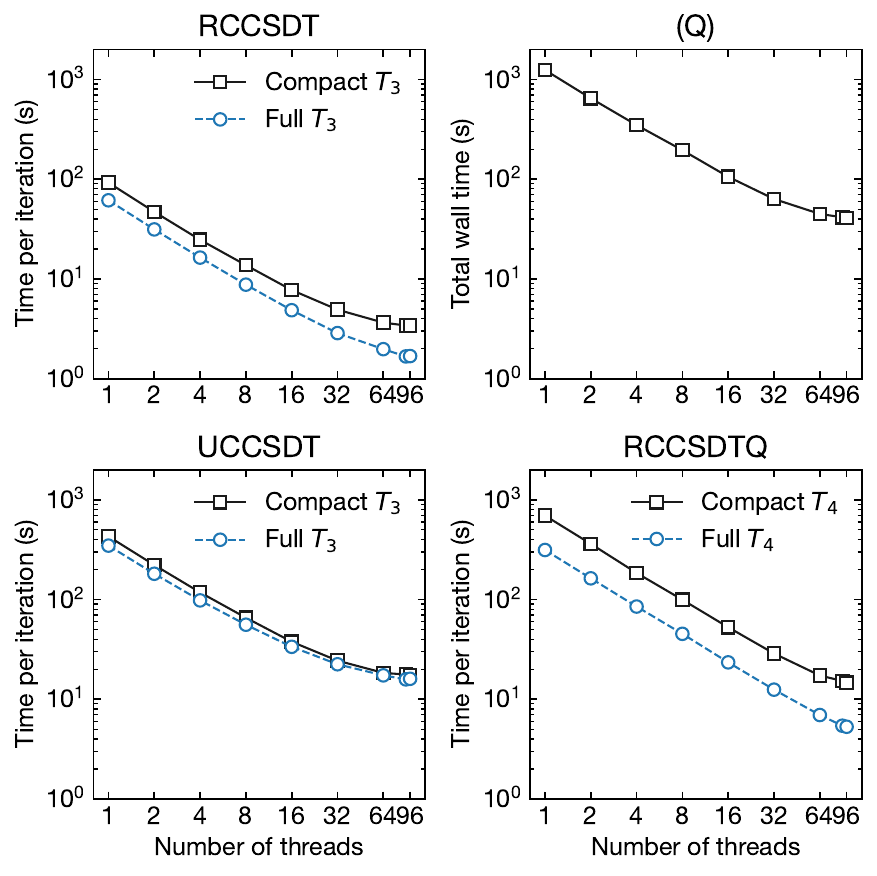}
    \caption{Average per-iteration wall time as a function of thread count for RCCSDT, RCCSDT(Q), UCCSDT, and RCCSDTQ applied to hydrogen thioperoxide (HSOH) with the frozen-core approximation. The cc-pVTZ basis is used for RCCSDT, RCCSDT(Q), and UCCSDT ($N_{\mathrm{o}} = 7$, $N_{\mathrm{v}} = 79$), and the cc-pVDZ basis is used for RCCSDTQ ($N_{\mathrm{o}} = 7$, $N_{\mathrm{v}} = 29$). ``Full'' denotes implementations that store the complete $T_3$ (RCCSDT, UCCSDT) or $T_4$ (RCCSDTQ) tensor; ``Compact'' denotes implementations using triangular storage together with on-the-fly unpacking. Benchmarks were run on a single 96-core AMD Genoa node using \textsc{pytblis}~\cite{pytblis} (a Python wrapper for \textsc{TBLIS}~\cite{tblis}) for tensor contractions. 
    }
    \label{fig:threads_scaling}
\end{figure}

We first benchmark the thread scaling of the shared-memory implementation on a single AMD Genoa node using up to 96 threads. Each Genoa node is equipped with two 48-core AMD EPYC 9474F processors, totaling 96 physical CPU cores across two NUMA domains, with 1.5 TB of memory. Figure~\ref{fig:threads_scaling} reports the average per-iteration wall time for RCCSDT, UCCSDT, and RCCSDTQ and the total wall time for RCCSDT(Q) calculations on hydrogen thioperoxide (HSOH). The cc-pVTZ basis was used for RCCSDT, RCCSDT(Q), and UCCSDT, whereas the cc-pVDZ basis was used for RCCSDTQ.~\cite{dunning1989gaussian,woon1993gaussian} The frozen-core approximation is used throughout this work. All methods exhibit nearly linear thread scaling up to 90 threads, reflecting the efficiency of the multi-threaded \textsc{pytblis}/\textsc{TBLIS} tensor contractions,~\cite{matthews2018high} as well as the parallel C implementations of spin summation and amplitude unpacking. For the small benchmark systems considered here, calculations using the full amplitudes implementation are consistently faster than their compact-amplitude counterparts, despite their formally larger floating-point operation counts. This behavior arises because contractions over large, contiguous full-amplitude tensors are highly optimized, whereas the compact formulation requires contractions over many smaller unpacked tensor blocks, which are less favorable for cache utilization and tensor-contraction efficiency. In addition, the unpacking operations can become memory-bandwidth limited, particularly when the subsequent tensor contractions are relatively inexpensive. For larger production calculations, however, the compact implementation is expected to become advantageous in both memory footprint and overall computational cost.

\begin{figure}
    \centering
    \includegraphics[width=\linewidth]{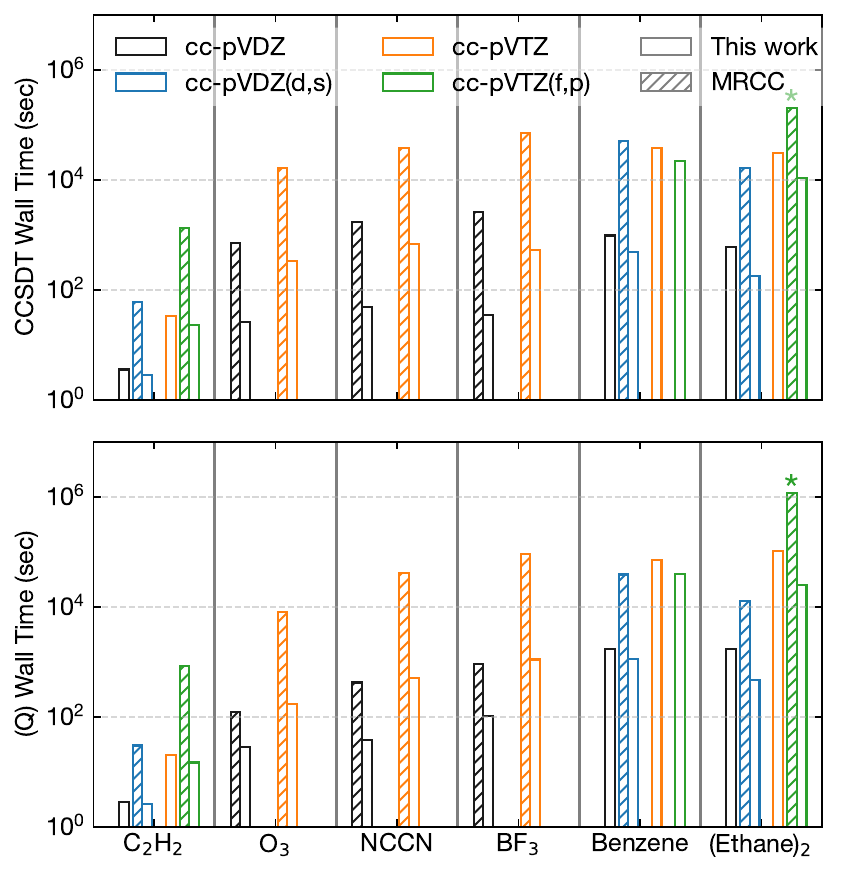}
    \caption{Wall-time comparison between the present compact amplitudes implementation (This work) and \textsc{MRCC}~\cite{fishman2026development} for CCSDT and CCSDT(Q) applied to C$_2$H$_4$, O$_3$, NCCN, BF$_3$, benzene, and (ethane)$_2$ in the cc-pVDZ and cc-pVTZ basis sets and the truncated cc-pVDZ(d,s) and cc-pVTZ(f,p) variants. ${}^\star$Point-group symmetry was used in the \textsc{MRCC} calculation for (ethane)$_2$/cc-pVTZ(f,p).
    }
    \label{fig:comp_mrcc}
\end{figure}

Figure~\ref{fig:comp_mrcc} compares the performance of the present implementation using compact amplitudes with recent \textsc{MRCC} timing data reported in Ref.~\citenum{fishman2026development} for CCSDT and CCSDT(Q) calculations on a representative set of molecules, including C$_2$H$_4$, O$_3$, NCCN, BF$_3$, benzene, and (ethane)$_2$. The comparison includes the standard cc-pVDZ and cc-pVTZ basis sets, as well as the truncated cc-pVDZ(d,s) and cc-pVTZ(f,p) variants used in Ref.~\citenum{fishman2026development}, in which the highest-angular-momentum polarization functions on hydrogen are removed. The point group symmetry was not exploited in \textsc{MRCC} calculations except for the (ethane)$_2$ in cc-pVTZ(f,p) basis set. For this benchmark set, the present implementation is approximately one to two orders of magnitude faster than \textsc{MRCC}, based on the timings reported in Ref.~\citenum{fishman2026development}. This improvement arises from several compounding factors: (i) the spin-free formulation reduces the formal operation count relative to spin-integrated implementations, (ii) the implementation exhibits near-linear thread scaling up to 90 cores, and (iii) the fully in-memory design avoids disk-I/O overhead. The \textsc{MRCC} reference timings were obtained on Intel Xeon Gold 6240R CPUs (2.40~GHz, with 380~GB of memory and 3.6~TB of SSD storage), using 16 cores per job, with which the CCSDT timings were reported to reach peak performance.~\cite{fishman2026development}

Recent \textsc{CFOUR} timing data for CCSDTQ calculations provide an additional point of comparison.~\cite{barman2026new} For linear N$_2$O in the cc-pVTZ(d,p) basis, where the highest-angular-momentum functions are truncated, \textsc{MRCC} required 4,367~seconds per iteration, whereas \textsc{CFOUR} required 236~seconds per iteration.~\cite{barman2026new} These calculations employed point-group symmetry to reduce the computational cost and were performed on dual-socket Intel Xeon Gold 5320 CPUs operating at 2.20~GHz, with 16 threads; only marginal speedup was observed with additional threads. In comparison, the present RCCSDTQ implementation, which does not currently exploit point-group symmetry, requires 493~seconds per iteration using compact amplitudes and 227~seconds per iteration using full amplitudes on an AMD Genoa node using 90 threads. Thus, despite not using point-group symmetry, the present implementation achieves performance comparable to the state-of-the-art \textsc{CFOUR} RCCSDTQ implementation, primarily owing to efficient multithreading and near-linear scaling to large core counts.

\begin{figure*}
    \centering
    \includegraphics[width=0.8\linewidth]{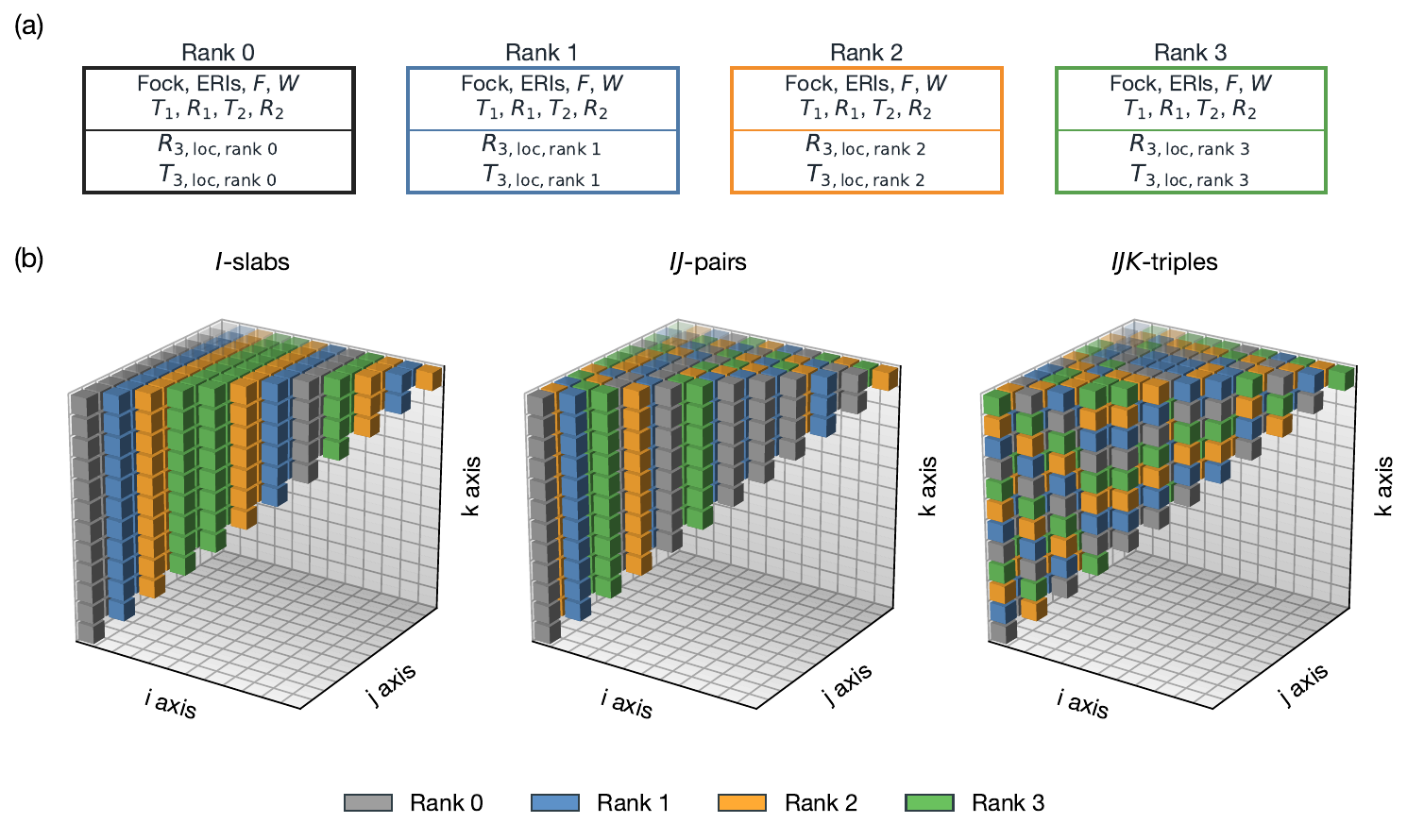}
    \caption{Data distribution for the distributed RCCSDT implementation. (a) The Fock matrix, ERIs, intermediates $F$ and $W$, and the lower-order amplitudes and residuals ($T_1$, $R_1$, $T_2$, $R_2$) are replicated on every rank, while the $T_3$ amplitudes and residual $R_3$ are distributed. (b) Data distribution pattern is illustrated for three strategies for distributing the triangular $T_3$ and $R_3$ tensors: $I$-slabs, $IJ$-pairs, and $IJK$-triples. The $IJK$-triples strategy is adopted for best load balance, at the cost of reduced data locality.}
    \label{fig:t3_dis}
\end{figure*}

The maximal capacity of the shared-memory implementation for treating large systems can be estimated from the dominant simultaneous memory storage of the highest-order amplitude tensor $T_n$ and its residual $R_n$. On a node with 1.5~TB of memory, each can occupy roughly 600~GB in compact triangular form while leaving space for lower-rank amplitudes, ERIs, and intermediates. For RCCSDT and RCCSDT(Q), this accommodates a cluster of 10 water molecules in the cc-pVDZ basis ($N_{\mathrm{o}} = 40$, $N_{\mathrm{v}} = 190$; 587~GB for $T_3$/$R_3$) or 6 water molecules in the cc-pVTZ basis ($N_{\mathrm{o}} = 24$, $N_{\mathrm{v}} = 318$; 623~GB for $T_3$/$R_3$). For RCCSDTQ, whose memory cost scales as $O(N_{\mathrm{o}}^4 N_{\mathrm{v}}^4)$ even in compact storage, the feasible system size is more limited: $N_{\mathrm{o}} = 15$, $N_{\mathrm{v}} = 70$ (547~GB for $T_4$/$R_4$). For UCCSDT, the dominant sectors are $t_{i<j,\bar{k}}^{a<b,\bar{c}}$ and $t_{i,\bar{j}<\bar{k}}^{a,\bar{b}<\bar{c}}$; representative limit is 8 water molecules in cc-pVDZ ($N_{\mathrm{o}} = 32$, $N_{\mathrm{v}} = 152$; 455~GB for all $T_3$/$R_3$). Larger systems require a distributed implementation, as discussed in the next section.

\section{Distributed implementation\label{sec:dis_mem}}

The distributed implementation extends the compact triangular $T_3$ representation described in Sec.~\ref{sec:shared_mem} for multi-node parallel RCCSDT, RCCSDT(Q), and RCCSDTQ calculations. To maximize CPU and memory efficiency while minimizing communication overhead, our distributed implementation employs a hybrid OpenMP--MPI strategy: OpenMP threads provide the shared-memory intra-node parallelism described in Sec.~\ref{sec:shared_mem}, while MPI ranks distribute the memory and work across nodes.  We use the naphthalene dimer in the cc-pVDZ basis with a frozen core ($N_{\mathrm{o}} = 48$, $N_{\mathrm{v}} = 292$; compact $T_3/R_3$ size of 3.6 TB) as the primary large-system example for contextualizing our strategy choices for the distributed algorithms for RCCSDT and (Q) energy correction. The calculations on the naphthalene dimer are performed on Intel Ice Lake nodes, each equipped with two 32-core Intel Xeon Platinum 8362 processors, totaling 64 physical CPU cores across two NUMA domains, with 1 TB of memory.

In this work, we use \textsc{Open MPI} 4.1.6~\cite{gabriel04:_open_mpi} as the MPI library implementation and \textsc{mpi4py} 3.1.5~\cite{dalcin2005mpi,dalcin2021computing} as the Python interface for MPI. We also use a few special environment variables tested on our computing platform, such as \texttt{OMPI\_MCA\_coll\_tuned\_use\_dynamic\_rules=true},  \texttt{OMPI\_MCA\_coll\_tuned\_allgatherv\_algorithm=3}, and \texttt{OMPI\_MCA\_mpi\_thread\_multiple=1}. We use \texttt{mpirun} options including \texttt{--mca orte\_keepalive\_timeout 300} and \texttt{--mca orte\_base\_help\_aggregate 0}.

\subsection{Distributed RCCSDT}

The development of an efficient distributed RCCSDT implementation requires carefully balancing the memory footprint, computational workload, and communication overhead. In this section, we first discuss different data distribution strategies of $T_3$ amplitudes and the corresponding computational procedure. We then elaborate on the communication strategies employed to alleviate data transfer bottlenecks. The efficiency of the distributed RCCSDT implementation is demonstrated through scaling benchmarks on water clusters. Finally, we discuss the distributed convergence acceleration algorithm required for studying highly correlated systems.

\subsubsection{Data distribution strategies}

\begin{figure}
    \centering
    \includegraphics[width=0.95\linewidth]{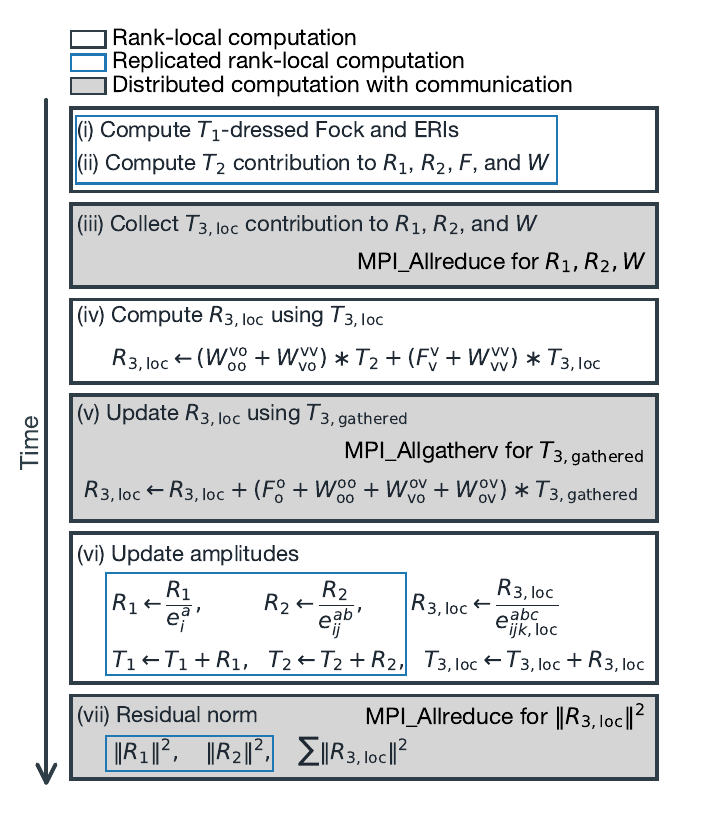}
    \caption{Workflow of one distributed RCCSDT iteration. White boxes: steps local to each rank. Blue boxes: computations replicated on every rank. Gray-filled boxes: steps involving inter-rank communication.
    }
    \label{fig:mpit_workflow}
\end{figure}

The memory bottleneck in RCCSDT is the compact $T_3$ amplitude tensor and its residual $R_3$. All other data, including the Fock matrix, ERIs, intermediates $F$ and $W$, and the lower-order amplitudes and residuals $T_1$, $R_1$, $T_2$, and $R_2$, are comparatively small. For the naphthalene dimer benchmark, the compact $T_3$ occupies 3.6~TB, so storing both $T_3$ and $R_3$ requires at least 7.2~TB, whereas the ERIs require only approximately 100~GB. It is therefore sufficient to distribute only $T_3$ and $R_3$ across MPI ranks and replicate all lower-order quantities on every rank, as illustrated in Fig.~\ref{fig:t3_dis}(a).

Because $T_3$ and $R_3$ are stored in compact triangular form, the distributed tensor is naturally indexed by the upper-triangular occupied triple $i \leq j \leq k$, with the virtual indices $a, b, c$ kept contiguous within each block. Three distribution strategies were considered, as shown in Fig.~\ref{fig:t3_dis}(b): $I$-slabs (partition by the smallest index $i$), $IJ$-pairs (partition by the pair $i,j$), and $IJK$-triples (partition by individual triples $i,j,k$). Coarser partitions preserve more data locality but can be severely imbalanced, whereas finer partitions improve load balance at the cost of locality. The data distribution strategy can also significantly impact communication load balancing, particularly when performing the $W^{\mathrm{oo}}_{\mathrm{oo}} \ast T_3$ contractions, where $T_3$ blocks need to be gathered across MPI ranks. We adopt the $IJK$-triples strategy because maintaining load balance is the primary concern at large node counts.

\subsubsection{Computational procedure}

Under this data distribution, the computational procedure within each iteration is divided into seven steps (shown in Fig.~\ref{fig:mpit_workflow}): (i)~update the $T_1$-dressed Fock matrix and ERIs; (ii)~compute the $T_2$ contributions to $R_1$, $R_2$, and the intermediates $F$ and $W$; (iii)~accumulate the $T_3$ contributions to $R_1$, $R_2$, and $W$ through an inexpensive \texttt{MPI\_Allreduce} over all ranks; (iv)~form the local residual fragment $R_{3,\text{loc}}$ from contractions involving $T_2$ and the local $T_3$ block,
\begin{equation*}
    R_{3,\text{loc}} \leftarrow \bigl(W^{\mathrm{vv}}_{\mathrm{vo}} + W^{\mathrm{vo}}_{\mathrm{oo}}\bigr) \ast T_2 + \bigl(F^{\mathrm{v}}_{\mathrm{v}} + W^{\mathrm{vv}}_{\mathrm{vv}}\bigr) \ast T_{3,\text{loc}},
\end{equation*}
which requires no inter-rank communication; (v)~update $R_{3,\text{loc}}$ using $T_3$ blocks gathered from other ranks,
\begin{equation*}
    R_{3,\text{loc}} \leftarrow R_{3,\text{loc}} + \bigl(F^{\mathrm{o}}_{\mathrm{o}} + W^{\mathrm{oo}}_{\mathrm{oo}} + W^{\mathrm{ov}}_{\mathrm{ov}} + W^{\mathrm{ov}}_{\mathrm{vo}}\bigr) \ast T_{3,\text{gathered}},
\end{equation*}
which is the dominant communication step; (vi)~update $T_1$, $T_2$, and $T_{3,\text{loc}}$ from the computed residuals; and (vii)~evaluate the residual norms $\|R_1\|^2$, $\|R_2\|^2$, and $\|R_3\|^2$, where the global $\|R_3\|^2$ is assembled from local contributions through \texttt{MPI\_Allreduce}.

The workload of all $T_3$-dependent computations is distributed across ranks. The only work duplicated on every rank is the updates involving low-order quantities in steps~(i), (ii), and part of (vi) and (vii): the $T_1$-dressed Fock and ERI transformation, the $T_2$-only contributions to $R_1$ and $R_2$, and the $T_1$ and $T_2$ updates. This replicated CCSD-like cost is negligible compared to the $T_3$ workload unless the calculation is spread over a very large number of nodes. The only communication-intensive part of an RCCSDT iteration is step~(v), an effective all-to-all exchange of $T_3$ blocks among ranks, implemented using \texttt{MPI\_Allgatherv} as described below.

\begin{figure*}
    \centering
    \includegraphics[width=0.9\linewidth]{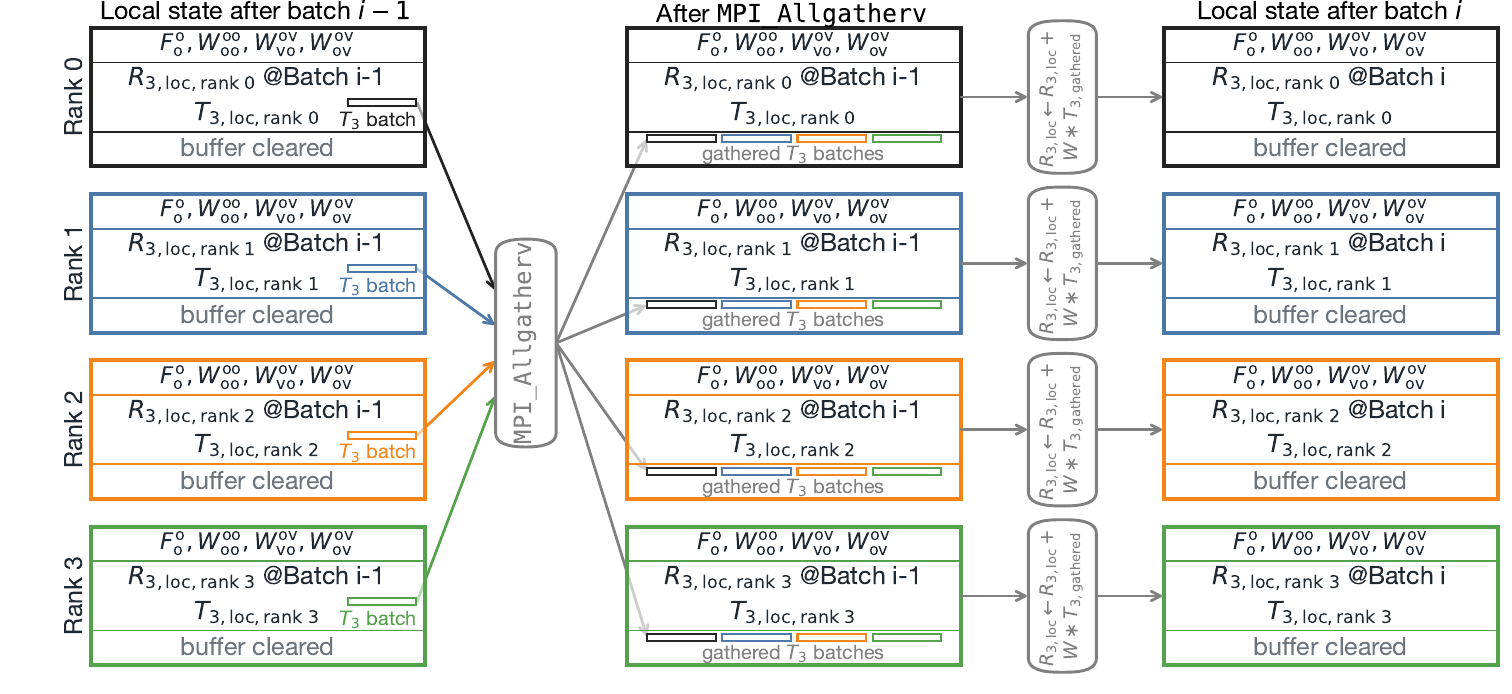}
    \caption{One communication--computation batch for the $\left(F^{\mathrm{o}}_{\mathrm{o}} + W^{\mathrm{oo}}_{\mathrm{oo}} + W^{\mathrm{ov}}_{\mathrm{vo}} + W^{\mathrm{ov}}_{\mathrm{ov}} \right) \ast T_3$ step in distributed RCCSDT. \texttt{MPI\_Allgatherv} assembles a batch of $T_3$ fragments from their owning ranks and broadcasts the result to all ranks; each rank then performs local contractions to update $R_{3,\text{loc}}$.
    }
    \label{fig:mpit_allgatherv}
\end{figure*}

\subsubsection{Communication strategies}

The all-to-all exchange in step~(v) is implemented using a static dispatching protocol based on \texttt{MPI\_Allgatherv}, in which the $T_3$ communication is organized into batches containing $N_{IJK}^{\mathrm{batch}}$ ordered $IJK$ triples, as illustrated in Fig.~\ref{fig:mpit_allgatherv}. For each batch, the required $T_3$ fragments are identified and gathered from their owning ranks. The assembled $T_3$ block is made available to all ranks and determines the local contractions performed on each rank. Thus, although every rank receives the same gathered $T_3$ data for a given batch, each rank contracts this data only for its own local residual block. Load balance is maintained by pre-estimating the contraction cost of each $IJK$ triplet and grouping triples so that each batch carries approximately equal work among all ranks. With this design, each $T_3$ element is communicated only once per cycle, which is essential for calculations involving TB-scale $T_3$ amplitudes. One source of redundancy remains: the $W^{\mathrm{ov}}_{\mathrm{vo}} \ast T_3$ contraction requires all three spin-summation permutations of the gathered block, $\check{t}_{\check{i}jk}^{\check{a}bc}$, $\check{t}_{i\check{j}k}^{a\check{b}c}$, and $\check{t}_{ij\check{k}}^{ab\check{c}}$, which are currently recomputed independently on every rank. This duplicated work introduces a constant overhead that leads to a slight degradation from ideal strong scaling, as demonstrated below.

\begin{figure}
    \centering
    \includegraphics[width=\linewidth]{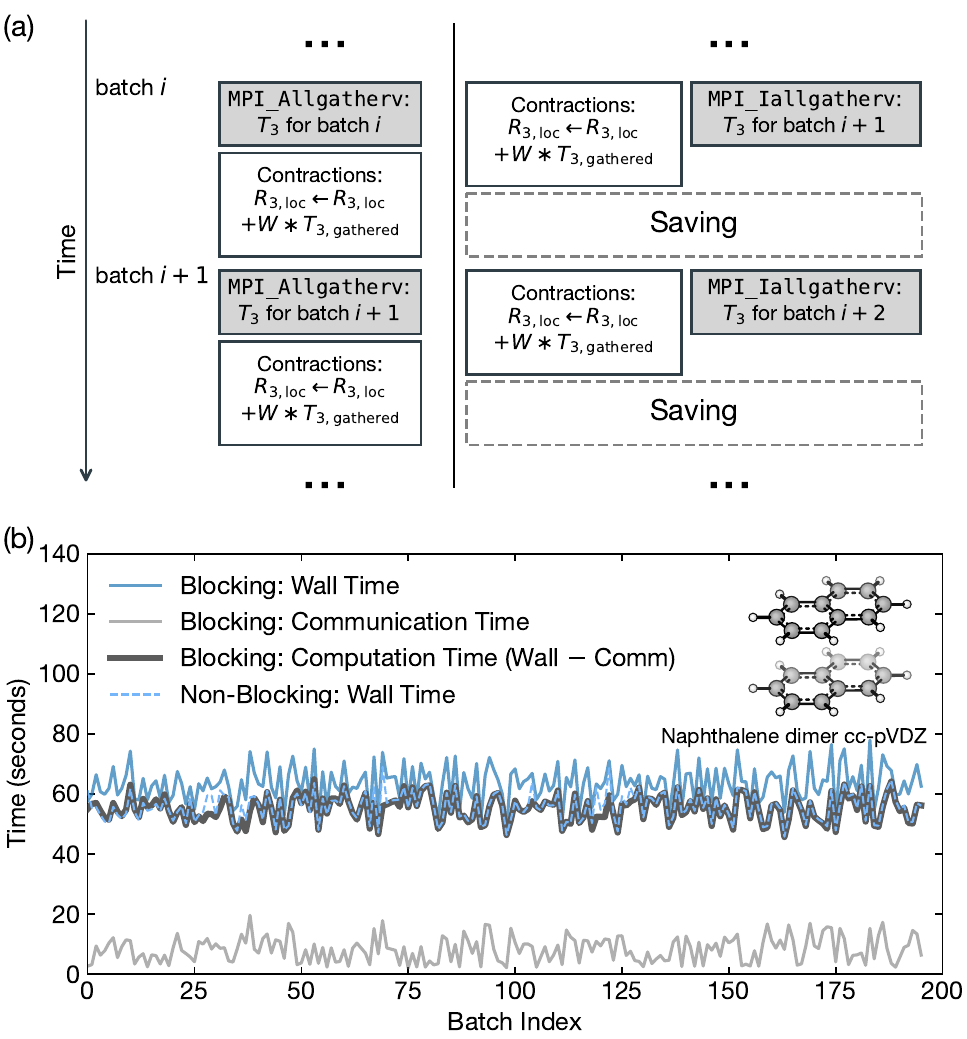}
    \caption{Blocking and non-blocking communication strategies for the $\left(F^{\mathrm{o}}_{\mathrm{o}} + W^{\mathrm{oo}}_{\mathrm{oo}} + W^{\mathrm{ov}}_{\mathrm{vo}} + W^{\mathrm{ov}}_{\mathrm{ov}} \right) \ast T_3$ step in distributed RCCSDT. (a) In the blocking implementation (with \texttt{MPI\_Allgatherv}), communication and computation proceed sequentially. In the non-blocking implementation (with \texttt{MPI\_Iallgatherv}), communication for batch $i+1$ overlaps with computation for batch $i$, hiding the inter-node latency. (b) Per-batch wall time of this step for the naphthalene dimer on 16 nodes with 64 Ice Lake cores each (batch size 100; 196 batches). The non-blocking wall time matches the pure computation time (total wall time minus communication time) of the blocking version, confirming full latency hiding. Here and throughout, molecules are rendered with xyzrender~\cite{goodfellow2026xyzgraph}. 
    }
    \label{fig:mpit_nonblocking}
\end{figure}

To hide communication latency, step~(v) uses non-blocking \texttt{MPI\_Iallgatherv} in place of the blocking \texttt{MPI\_Allgatherv}: while each rank performs the contraction for batch $i$, the gathering of $T_3$ fragments for batch $i+1$ proceeds in the background, as illustrated in Fig.~\ref{fig:mpit_nonblocking}(a). This requires only one additional buffer for receiving the next batch
in a double-buffer scheme. Fig.~\ref{fig:mpit_nonblocking}(b) shows per-batch wall times for step~(v) for the naphthalene dimer benchmark using 16 nodes with 64 Ice Lake cores each. A batch size of 100 (in terms of the number of $IJK$ triples included in each batch) is used in this calculation, which gives 196 batches in total. The non-blocking wall time per batch matches the pure computation time (i.e., total wall time minus communication time) of the blocking version, confirming that communication is fully overlapped. This overlap eliminates 12\% of the step~(v) total wall time for this system. The small batch-to-batch fluctuations confirm that the pre-estimated workload achieves good load balance within each batch.

\subsubsection{Parallel performance and scaling}

\begin{figure}
    \centering
    \includegraphics[width=\linewidth]{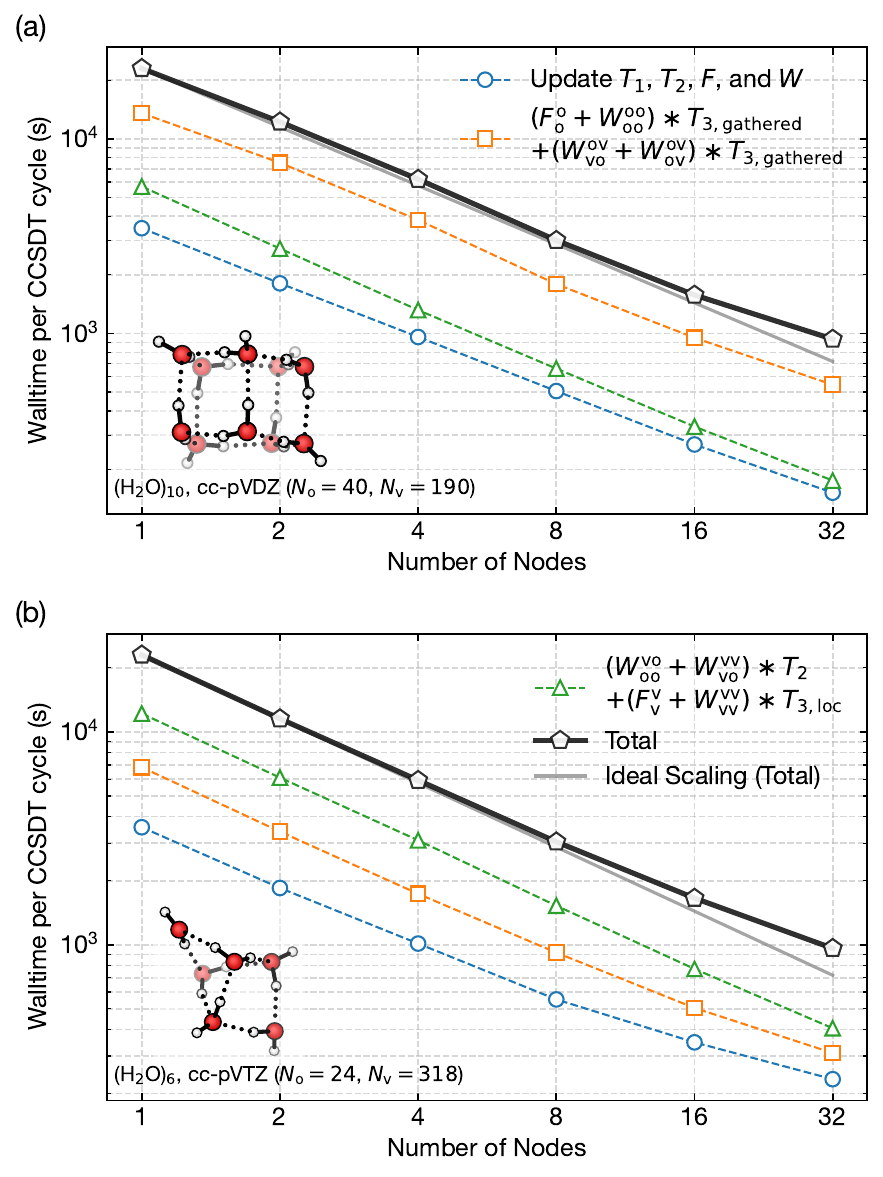}
    \caption{Per-cycle wall time for distributed RCCSDT as a function of node count for (a) (H$_2$O)$_{10}$/cc-pVDZ and (b) (H$_2$O)$_{6}$/cc-pVTZ. Total and step-decomposed wall times are shown. Benchmarks were run on 96-core AMD Genoa nodes. Water cluster geometries are from Ref.~\citenum{wales1998global}.
    }
    \label{fig:mpit_scaling}
\end{figure}

To benchmark the scaling of the distributed RCCSDT code, we performed calculations on the (H$_2$O)$_{10}$/cc-pVDZ and (H$_2$O)$_6$/cc-pVTZ systems on 96-core AMD Genoa nodes. These systems are used because they represent the largest systems that are feasible on a single node. Figure~\ref{fig:mpit_scaling} shows the per-cycle wall time as a function of node count. The rank-local contractions in step~(iv), $\big(W^{\mathrm{vv}}_{\mathrm{vo}} + W^{\mathrm{vo}}_{\mathrm{oo}}\big) \ast T_2 + \big(F^{\mathrm{v}}_{\mathrm{v}} + W^{\mathrm{vv}}_{\mathrm{vv}}\big) \ast T_{3,\text{loc}}$, exhibit ideal strong scaling. The updates of $T_1$, $T_2$, $F$, and $W$ in steps~(i)--(iii) show slight deviations from ideality because steps (i) and (ii) include a constant replicated CCSD-like computation and step (iii) is not load-balanced among ranks; however, this portion of the total cost is negligible. The dominant source of residual scaling loss is step~(v), $R_{3,\text{loc}} \leftarrow R_{3,\text{loc}} + \bigl(F^{\mathrm{o}}_{\mathrm{o}} + W^{\mathrm{oo}}_{\mathrm{oo}} + W^{\mathrm{ov}}_{\mathrm{ov}} + W^{\mathrm{ov}}_{\mathrm{vo}}\bigr) \ast T_{3,\text{gathered}}$, where the duplicated spin-summation permutations of the gathered $T_3$ blocks act as a constant overhead. Overall, the implementation achieves close to ideal strong scaling up to 32 nodes, delivering a 24-fold speedup on 32 nodes.

\subsubsection{Convergence acceleration}
DIIS~\cite{pulay1980convergence,scuseria1986accelerating} is implemented for distributed RCCSDT to accelerate the convergence of CC iterations. In the $m$th iteration (counting from 1), the full $T_1^{(m)}$ and $T_2^{(m)}$ amplitudes are duplicated on all MPI ranks, while rank $p$ (counting from 0) stores only its local block of the $T_3$ amplitudes, denoted by $T_{3,p}^{(m)}$. The corresponding flattened DIIS error vectors are denoted by $\mathbf{e}_1^{(m)}$, $\mathbf{e}_2^{(m)}$, and $\mathbf{e}_{3,p}^{(m)}$. The amplitudes and error vectors from the last $M$ iterations are kept in memory for DIIS extrapolation, and the DIIS overlap matrix is assembled as
\begin{equation}
    B_{mn} = \mathbf{e}_1^{(m)} \cdot \mathbf{e}_1^{(n)} + \mathbf{e}_2^{(m)} \cdot \mathbf{e}_2^{(n)} + \sum_p \mathbf{e}_{3,p}^{(m)} \cdot \mathbf{e}_{3,p}^{(n)}
\end{equation}
where the summation over MPI ranks is performed using a negligible-cost \texttt{MPI\_Allreduce} over the $B_{mn}$ coefficients. The DIIS coefficients are obtained by solving
\begin{equation}
\begin{bmatrix}
    B_{11} & \cdots & B_{1M} & 1 \\
    \vdots & \ddots & \vdots & \vdots \\
    B_{M1} & \cdots & B_{MM} & 1 \\
    1 & \cdots & 1 & 0
\end{bmatrix}
\begin{bmatrix}
    c_1 \\ \vdots \\ c_M \\ \lambda
\end{bmatrix}
    =
\begin{bmatrix}
    0 \\ \vdots \\ 0 \\ 1
\end{bmatrix},
\end{equation}
which is solved redundantly on all MPI ranks. Each rank then updates its amplitudes as
\begin{equation}
    T_{1/2}^{\mathrm{DIIS}} = \sum_{m=1}^{M} c_m T_{1/2}^{(m)},\quad T_{3,p}^{\mathrm{DIIS}} = \sum_{m=1}^{M} c_m T_{3,p}^{(m)} .
\end{equation}
Thus, DIIS requires communication only to construct the global error-overlap matrix; the extrapolation itself is performed locally on each MPI rank.

To further reduce memory usage, DIIS for the triples amplitude can be performed in a reduced orbital subspace, for example, by retaining only the components associated with low-energy virtual orbitals. Since the storage of $T_3$ scales as $O(N_{\mathrm{o}}^3N_{\mathrm{v}}^3)$, this subspace DIIS strategy can significantly reduce the memory overhead for storing DIIS amplitude and error vectors while retaining the main convergence-acceleration effect. Alternatively, DIIS can be applied only to $T_1$ and $T_2$, which is already sufficient for easy-to-converge systems.~\cite{matthews2015accelerating,matthews2020accelerating}

\subsection{Distributed (Q) energy correction}

\begin{figure*}
    \centering
    \includegraphics[width=0.9\linewidth]{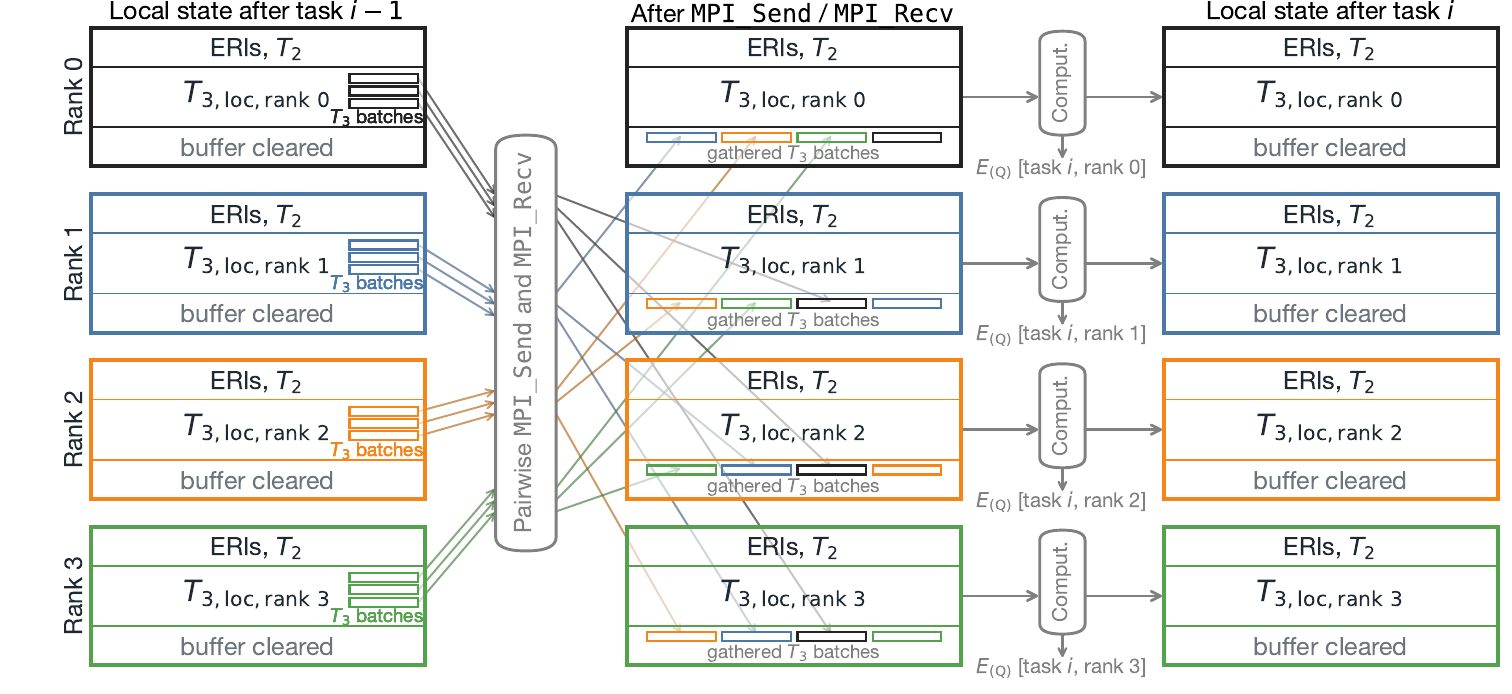}
    \caption{One distributed (Q) energy correction computational task. Pairwise \texttt{MPI\_Send} and \texttt{MPI\_Recv} exchanges deliver the required $T_3$ fragments to every rank; each rank then computes its local contribution to the (Q) correction.
    }
    \label{fig:mpi_q_comm}
\end{figure*}

The distributed (Q) energy correction implementation uses the $ABC$-based algorithm introduced in Sec.~\ref{sec:rccsdt_q_impl}. This formulation is well-suited for distributed execution because the total (Q) correction decomposes naturally into independent virtual-index tasks, each requiring only a subset of the $T_3$ data and producing a modest communication volume. The $(N_{\mathrm{v}}^{\mathrm{blk}})^4 N_{\mathrm{o}}^4$ buffer associated with one virtual-index tile is also manageable compared to the $N_{\mathrm{v}}^4$ buffer required by the $IJK$-based algorithm, especially for the large systems targeted by the MPI implementation. The virtual block size $N_{\mathrm{v}}^{\mathrm{blk}}$ therefore controls both the task granularity and the peak memory requirement. Given that the tasks are independent, they can also be distributed across multiple asynchronous MPI jobs, offering an alternative path to strong scaling on large supercomputers.

Before the (Q) correction begins, the distributed $T_3$ amplitudes must be transposed from the occupied-triangular layout used in distributed RCCSDT, stored as $(i\leq j\leq k,a,b,c)$ and partitioned by $IJK$ triples, into the virtual-triangular layout $(a\leq b\leq c,i,j,k)$ required by the $ABC$-based algorithm, as shown in Fig.~\ref{fig:q_operation_counts}(a). This global $T_3$ transposition step is implemented as a distributed algorithm and executed one-time before the (Q) computation. 

The distributed (Q) energy correction algorithm operates in two phases: planning and execution. In the planning phase, every rank simulates a dry-run of the contractions to precompute a deterministic, global communication schedule. During execution, each rank follows this schedule to exchange $T_3$ fragments via pairwise \texttt{MPI\_Send} and \texttt{MPI\_Recv} operations (Fig.~\ref{fig:mpi_q_comm})
, and subsequently accumulate their local contributions to the (Q) correction. The total communication volume across all tasks is substantially larger than the size of $T_3$ itself, making this step more communication-intensive than one distributed RCCSDT iteration. However, the $O(N^9)$ computation per task dominates over the $O(N^6)$ communication cost by a wide margin.

Tasks are ordered and distributed across ranks so that adjacent tasks on the same rank partially share the required $T_3$ data, reducing total communication by approximately a factor of two. To see this, consider a task labeled $(A,B,C,D)$ where each letter denotes a range of virtual indices $(a_0, a_1)$, $(b_0,b_1)$, $(c_0,c_1)$, $(d_0,d_1)$, where $a, b, c, d$ represents the four virtual-orbital indices involved in the summation in Eq.~\eqref{eq:ener-pt-q-abc}. It requires six $T_3$ fragments (with the superscript $:$ denoting all third-virtual blocks):
\begin{equation*}
    \check{t}^{AB:},\ \check{t}^{AC:},\ \check{t}^{AD:},\ \check{t}^{BC:},\ \check{t}^{BD:},\ 
    \check{t}^{CD:},
\end{equation*}
for the contractions shown in Eq.~\eqref{eq:q_t3_terms}. The immediately following task $(A, B, C, D')$, which differs only in the last index, needs
\begin{equation*}
    \check{t}^{AB:},\ \check{t}^{AC:},\ \check{t}^{AD':},\ \check{t}^{BC:},\ \check{t}^{BD':},\ \check{t}^{CD':},
\end{equation*}
so the three fragments $\check{t}^{AB:}$, $\check{t}^{AC:}$, and $\check{t}^{BC:}$ are reused without additional communication. This reuse lowers the communication volume and reduces network congestion.

\begin{figure}
    \centering
    \includegraphics[width=\linewidth]{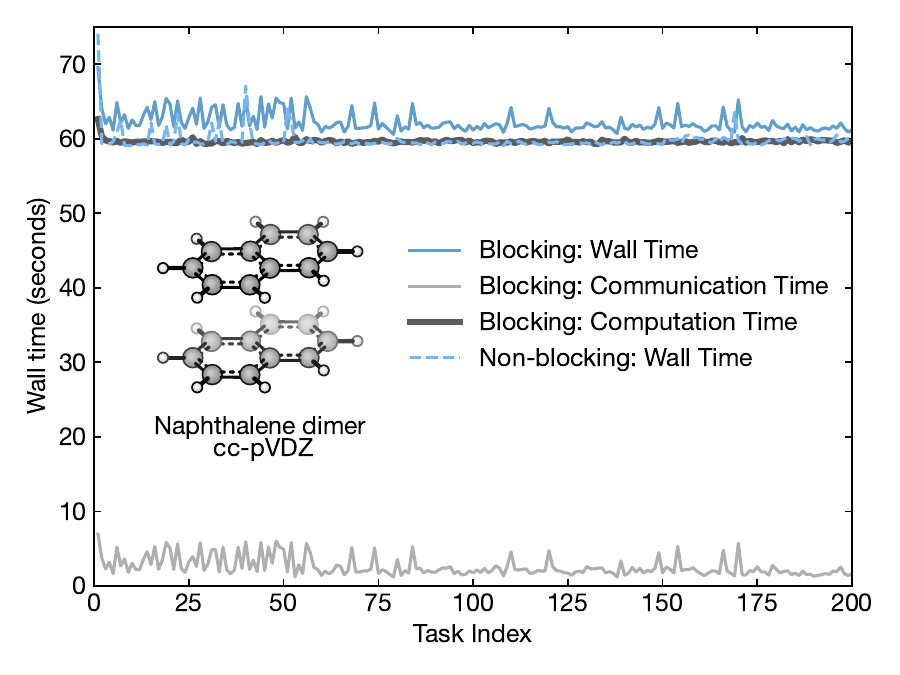}
    \caption{Per-task wall time for 200 representative tasks from the distributed (Q) calculation of the cc-pVDZ naphthalene dimer on 16 Intel Ice Lake nodes, comparing blocking and nonblocking communication. The non-blocking wall time matches the pure computation time of the blocking version almost perfectly (blocking wall time minus communication time), confirming near-complete latency hiding.
    }
    \label{fig:mpi_q_nonblocking}
\end{figure}

As in distributed RCCSDT, non-blocking communication is used to overlap data transfer with computation: while task $i$ is being computed, the $T_3$ fragments needed for task $i+1$ are gathered in the background. A synchronization barrier is inserted every 50 tasks to prevent rank desynchronization from accumulating due to per-task imbalances in communication volume or hardware variability. Figure~\ref{fig:mpi_q_nonblocking} compares the per-task wall times of the blocking and non-blocking implementations for the naphthalene dimer using 16 Ice Lake nodes. The calculation uses a block size of 6, giving 270,725 tasks in total. The plotted window contains 3,200 tasks, or 200 tasks per node. The blocking version shows noticeably uneven communication times across tasks; the non-blocking version eliminates this spiking behavior by absorbing the latency into the computation of the preceding task.
The communication overhead in the blocking case is already small, only $\sim3\%$ of the per-task wall time, as the computation has a much larger computational complexity.

\begin{figure}
    \centering
    \includegraphics[width=\linewidth]{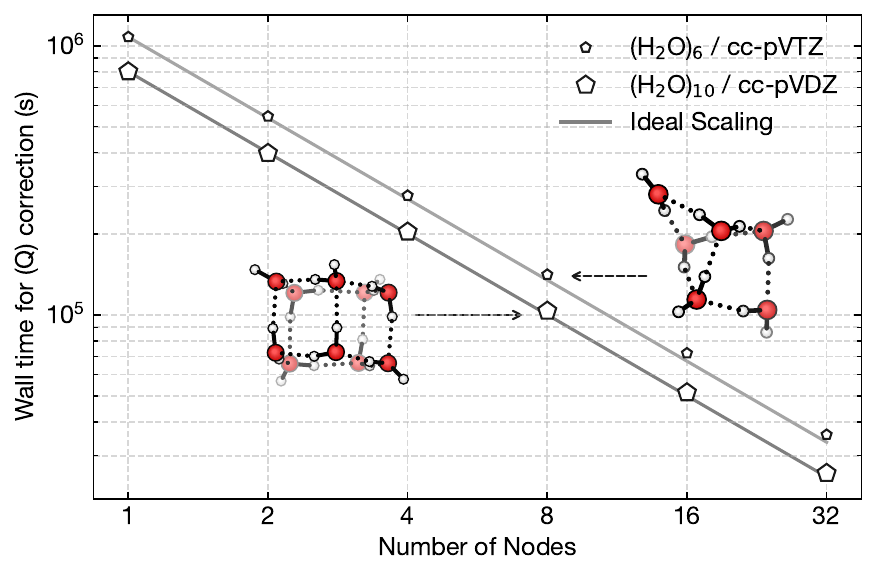}
    \caption{Wall time for the CCSDT(Q) correction as a function of node count for (H$_2$O)$_{10}$/cc-pVDZ and (H$_2$O)$_{6}$/cc-pVTZ. Virtual block sizes of 9 and 12 give 12{,}650 and 27{,}405 tasks, respectively. Total wall times are estimated from 256 representative tasks scaled proportionally. Benchmarks were run on 96-core AMD Genoa nodes.}
    \label{fig:mpi_q_scaling}
\end{figure}

We again use the $(\text{H}_2\text{O})_{10}$/cc-pVDZ and $(\text{H}_2\text{O})_6$/cc-pVTZ systems on AMD Genoa nodes to test the scaling of the distributed (Q) code. Figure~\ref{fig:mpi_q_scaling} shows the total wall time for the (Q) correction as a function of node count. The total wall time is estimated from the measured wall time of 256 representative tasks [block sizes of 9 and 12, giving 12{,}650 and 27{,}405 tasks in total for $(\text{H}_2\text{O})_{10}$/cc-pVDZ and $(\text{H}_2\text{O})_6$/cc-pVTZ, respectively]. Nearly ideal strong scaling is observed, with a $30\times$ speedup on 32 nodes, consistent with the computation-dominated task complexity, hidden communication, and absence of duplicated work.

\section{Distributed RCCSDTQ}

We also implemented a distributed-memory RCCSDTQ algorithm following the same design principles as the distributed RCCSDT implementation. The compact $T_4$ amplitudes and residuals are distributed over occupied-index quadruples $IJKL$, while the lower-order amplitudes and intermediates are replicated on each MPI rank. All $T_4$-dependent contractions are distributed across MPI ranks, with communication performed using nonblocking \texttt{MPI\_Iallgatherv} and convergence accelerated by distributed DIIS in a reduced orbital subspace.

\begin{figure}
    \centering
    \includegraphics[width=\linewidth]{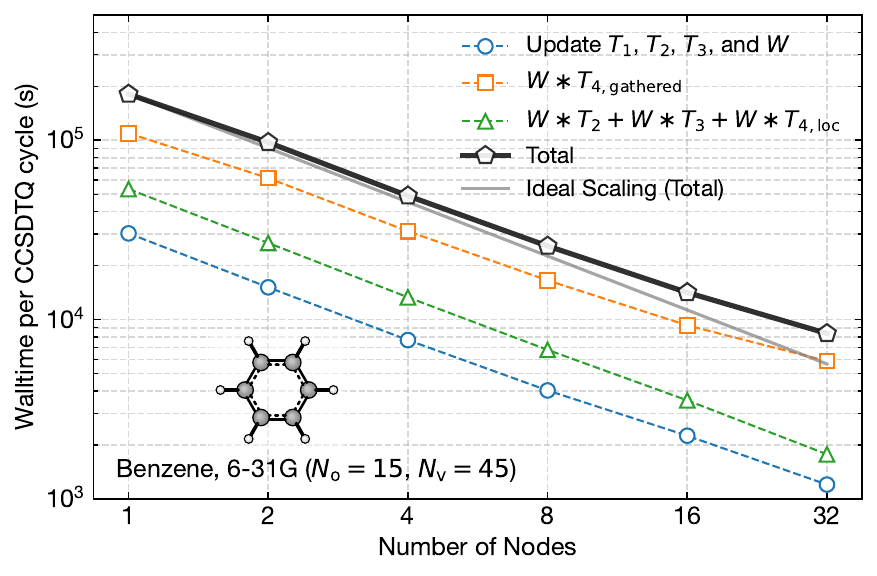}
    \caption{Per-cycle wall time for distributed RCCSDTQ as a function of node count for benzene/6-31G. Total and step-decomposed wall times are shown. Benchmarks were run on 64-core Intel Ice Lake nodes.}
    \label{fig:mpi_ccsdtq_scaling}
\end{figure}

We use the benzene/6-31G system on Ice Lake nodes to assess the strong-scaling performance of the distributed RCCSDTQ implementation. Figure~\ref{fig:mpi_ccsdtq_scaling} shows the per-cycle wall time as a function of the number of nodes. Similar to the distributed RCCSDT case, the dominant source of scaling loss is the step in which the local $R_4$ residual is updated through contractions between the intermediates $W$ and the gathered $T_4$ blocks. In this step, the duplicated spin-summation permutations of the gathered $T_4$ blocks introduce an approximately constant overhead. Overall, the implementation exhibits close-to-ideal strong scaling up to 32 nodes, achieving a 20-fold speedup relative to the single-node calculation.

\section{Practical considerations\label{sec:practical}}

In this section, we summarize several practical considerations for algorithm selection and parameter tuning to maximize performance for production applications.

The present algorithms are designed for the conventional molecular regime in which $N_{\mathrm{v}} > N_{\mathrm{o}}$, as is typical for standard Gaussian basis sets. The opposite regime, $N_{\mathrm{v}} \lesssim N_{\mathrm{o}}$, can arise in active-space~\cite{zhai2026classical} or model calculations and is not the focus here. The MPI implementation additionally assumes that the $N_{\mathrm{v}}/N_{\mathrm{o}}$ ratio is not too large (about 10 or smaller). Otherwise, the memory cost of the ERIs, which scales as $O[\left(N_{\mathrm{o}} + N_{\mathrm{v}}\right)^4]$, will be another bottleneck when the system size is large.

With our current implementation, memory cost remains the primary limiting factor for most high-order CC applications considered in this work. The overall memory requirement for high-order CC can be reasonably estimated from the highest-order amplitude and residual tensors, given that ERIs, intermediates, and lower-order amplitudes are comparatively small. For RCCSDT and RCCSDT(Q), $T_3$ and $R_3$ have identical storage footprints, so the minimum high-order memory requirement is twice the size of the chosen $T_3$ representation. For RCCSDTQ, the same consideration applies to $T_4$ and $R_4$, making the single-node memory limit much more restrictive. For systems that are difficult to converge without DIIS acceleration, DIIS can be restricted to a selected subspace of the high-order amplitudes, which should avoid any substantial additional memory overhead. 

The appropriate implementation should then be chosen according to this memory estimate. For small systems whose full-amplitude tensors fit within a shared-memory node, the shared-memory code without compact storage is the simplest and avoids unpacking overhead. For larger systems whose full tensors do not fit but whose compact triangular $T_3$ (or $T_4$) tensor and residual still fit on a single node, the shared-memory code with compact storage should be used. For RCCSDT, the recommended unpacking block size is approximately $N_{\mathrm{o}}/2$, which may be further reduced when the memory budget is tight. For RCCSDTQ, smaller unpacking blocks are usually preferable because the temporary buffers grow more rapidly with system size.

For systems whose compact highest-order amplitudes and residuals together exceed the single-node memory limit, the MPI-distributed implementation is required. In distributed RCCSDT (RCCSDTQ), the batch size in step~(v) controls the amount of $T_3$ ($T_4$) data gathered at once and should be chosen to avoid network congestion while keeping each batch sufficiently large to provide enough tensor-contraction work on each rank for efficient multithreaded execution. In distributed RCCSDT(Q), the virtual block size controls both the buffer memory and the amount of computation per task. In practice, both parameters should be tuned to keep the per-batch or per-task communication volume below roughly 20~GB while maintaining enough computation per batch or task, for example, on the order of 60 seconds in the naphthalene dimer benchmarks shown in Figs.~\ref{fig:mpit_nonblocking} and~\ref{fig:mpi_q_nonblocking}.

\section{Applications\label{sec:applications}}

\begin{figure}
    \centering
    \includegraphics[width=\linewidth]{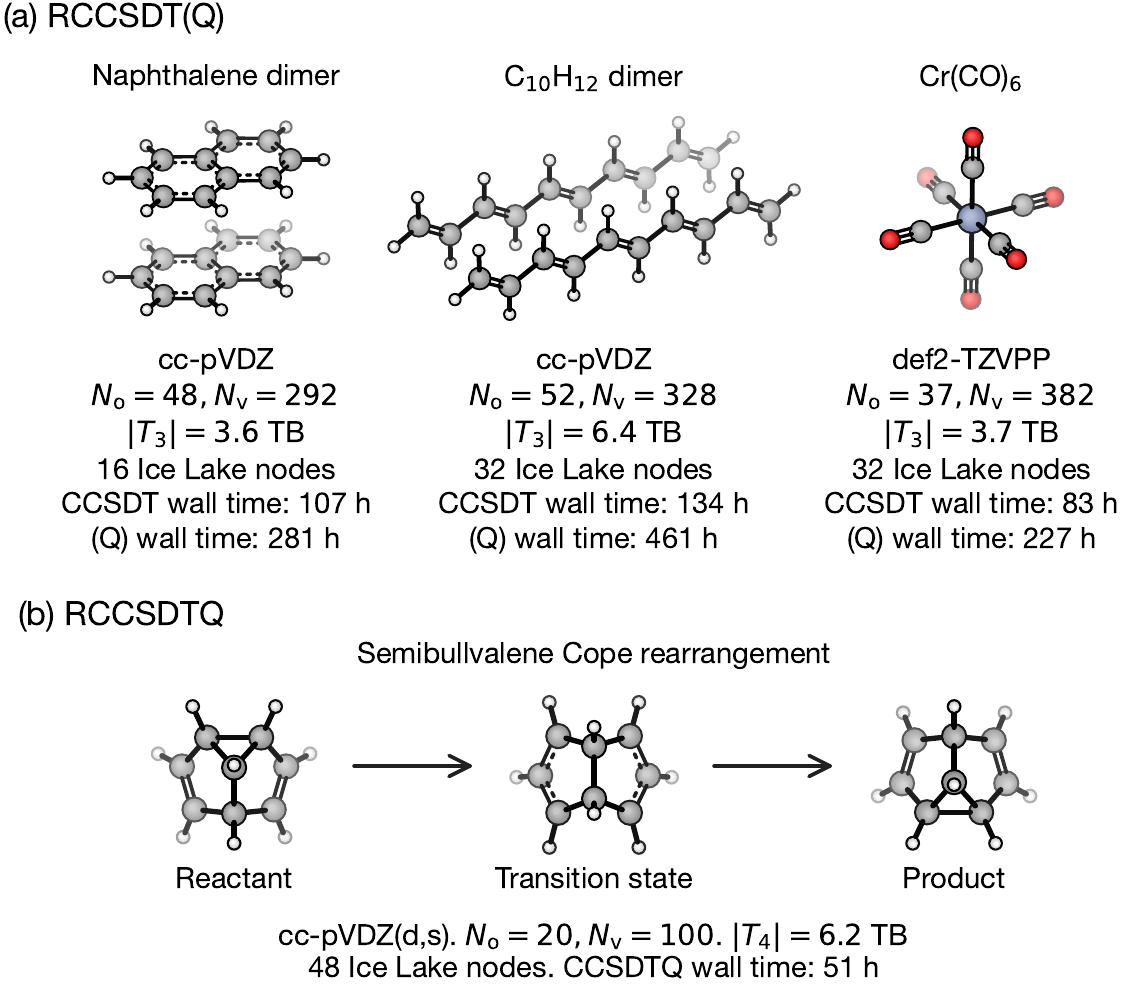}
    \caption{System specifications for the (a) three largest CCSDT(Q) calculations and (b) one largest CCSDTQ calculation in this work: molecular structures, basis sets, numbers of occupied and virtual orbitals, compact $T_3$/$T_4$ sizes, and computational resources. Geometries for the naphthalene dimer and (C$_{10}$H$_{12}$)$_2$ are from Ref.~\citenum{fishman2025another}; the Cr(CO)$_6$ geometry is from Ref.~\citenum{dohm2018comprehensive}. The Semibullvalene geometry is from Ref.~\citenum{karton2020cope}.
    }
    \label{fig:structures}
\end{figure}

We now demonstrate the reach of our implementation on chemically relevant systems for which post-CCSD(T) contributions are expected to matter. The examples include two types of noncovalent $\pi$-stacking interactions, namely aromatic dimers and conjugated polyene dimers, as well as a transition-metal bond-dissociation reaction and an organic rearrangement barrier height. The three largest CCSDT(Q) calculations and one largest CCSDTQ calculation are summarized in Fig.~\ref{fig:structures}, including the computational resources used and the total calculation time. The converged energies of all calculations are given in Table~\ref{tab:raw_data} in the Appendix. Although the benchmark systems possess point group symmetry, our implementation does not exploit it. Therefore, the reported timings reflect the baseline performance for any general system of equivalent size.

We first computed CCSDT and CCSDT(Q) interaction energies for sandwiched benzene and naphthalene dimers using geometries from Ref.~\citenum{fishman2025another} with counterpoise (CP) corrections. Other arrangements with the same basis and electron count, such as parallel-displaced and T-shaped dimers, would have comparable cost. For the naphthalene dimer in the cc-pVDZ basis with a frozen core ($N_{\mathrm{o}} = 48$, $N_{\mathrm{v}} = 292$; compact $T_3$ size of 3.6~TB), the CCSDT calculation on 16 Intel Ice Lake nodes required 107~hours (5.6~hours per cycle), and the (Q) energy correction required 281~hours. The resulting interaction energies are reported in Table~\ref{tab:benzene_int}. The iterative-triples correction ($\delta$CCSDT) and the perturbative-quadruples correction [$\delta$CCSDT(Q)] are small on an absolute scale, but they grow substantially from benzene to naphthalene and are comparable to the final interaction energies in the cc-pVDZ basis. Accurate benchmarks for these weak interactions, therefore, require both iterative triples and perturbative quadruples. The present $\delta$CCSDT contributions of $+0.51$ and $+1.34$~kJ/mol for benzene and naphthalene dimers, and $\delta$CCSDT(Q) contributions of $-0.31$ and $-0.79$~kJ/mol, are close to the previous rank-reduced estimates of $+0.82 \pm 0.02$ and $+1.57 \pm 0.04$~kJ/mol for $\delta$CCSDT and $-0.5 \pm 0.2$ and $-0.9 \pm 0.4$~kJ/mol for $\delta$CCSDT(Q)~\cite{fishman2025another,lesiuk2022gold}. This agreement independently verifies the reliability of the rank-reduced methodology.

\begin{table}[ht]
\caption{Interaction energies (kJ/mol) for sandwiched benzene and naphthalene dimers in the cc-pVDZ basis. Geometries are from Ref.~\citenum{fishman2025another}. Starting from MP2, each row reports the incremental correction relative to the preceding level of theory; for example, $\delta$CCSDT $\equiv$ CCSDT $-$ CCSD(T). The total interaction energy is obtained as the sum of the HF contribution and all subsequent increments.
}
\label{tab:benzene_int}
\centering
\begin{tabular}{lrr}
\hline \hline
& Benzene dimer & Naphthalene dimer \\
\hline
HF               & $+15.67$ & $+28.68$ \\
$\delta$MP2      & $-20.04$ & $-44.20$ \\
$\delta$CCSD     & $+7.04$  & $+17.36$ \\
$\delta$CCSD(T)  & $-2.28$  & $-5.11$ \\
$\delta$CCSDT    & $+0.51$  & $+1.34$  \\
$\delta$CCSDT(Q) & $-0.31$  & $-0.79$  \\
Total            & $+0.58$  & $-2.73$ \\
\hline \hline
\end{tabular}
\end{table}

We next considered a series of conjugated polyene dimers of increasing size, (C$_2$H$_4$)$_2$, (C$_4$H$_6$)$_2$, (C$_6$H$_8$)$_2$, (C$_8$H$_{10}$)$_2$, and (C$_{10}$H$_{12}$)$_2$, in the cc-pVDZ basis with frozen cores and relaxed geometries from Ref.~\citenum{fishman2025another}. CP corrections were included in the interaction energy calculations. This series provides a controlled test of how post-CCSD(T) corrections evolve with increasing $\pi$-system size. The compact $T_3$ sizes grow from 379~GB for (C$_6$H$_8$)$_2$ ($N_{\mathrm{o}} = 32$, $N_{\mathrm{v}} = 204$) to 1.8~TB for (C$_8$H$_{10}$)$_2$ ($N_{\mathrm{o}} = 42$, $N_{\mathrm{v}} = 266$) and 6.4~TB for the decapentaene dimer (C$_{10}$H$_{12}$)$_2$ ($N_{\mathrm{o}} = 52$, $N_{\mathrm{v}} = 328$). For the largest calculation performed on 32 Ice Lake nodes, the CCSDT calculation took 134~hours (7.1~hours per cycle), and the CCSDT(Q) correction took 461~hours. Prior to this work, the largest reported calculations, despite already exploiting $\mathrm{D_{2d}}$ point-group symmetry to reduce computational and memory costs, were limited to (C$_6$H$_8$)$_2$ at the CCSDT level and to (C$_4$H$_6$)$_2$ for the (Q) correction.~\cite{fishman2025another} Interaction energies for the full series are compiled in Table~\ref{tab:polyene_int}. In contrast to the aromatic dimers, the $\delta$CCSDT and $\delta$CCSDT(Q) contributions in the polyene series have opposite signs and largely cancel. Both corrections grow with system size, but they remain a modest fraction of the total interaction energy over the range studied here. These calculations show that exact CCSDT(Q) benchmarks for extended non-covalent systems are already possible, and that combining the present implementation with frozen natural orbital (FNO) or local natural orbital (LNO) approximations should further extend this reach.

\begin{table}[ht]
\caption{Interaction energies (kJ/mol) for the polyene dimer series in the cc-pVDZ basis. Geometries are from Ref.~\citenum{fishman2025another}. Starting from MP2, each row reports the incremental correction relative to the preceding level of theory; for example, $\delta$CCSDT $\equiv$ CCSDT $-$ CCSD(T). The total interaction energy is obtained as the sum of the HF contribution and all subsequent increments.}
\label{tab:polyene_int}
\centering
\begin{tabular}{lrrrrr}
\hline \hline
& (C$_2$H$_4$)$_2$ & (C$_4$H$_6$)$_2$ & (C$_6$H$_8$)$_2$ & (C$_8$H$_{10}$)$_2$ & (C$_{10}$H$_{12}$)$_2$ \\
\hline
HF                & $+2.38$ & $+8.20$ & $+13.64$ & $+19.53$ & $+24.83$ \\
$\delta$MP2       & $-1.92$ & $-8.45$ & $-15.89$ & $-24.23$ & $-32.29$ \\
$\delta$CCSD      & $+0.33$ & $+2.12$ & $+4.53$  & $+7.37$  & $+10.22$ \\
$\delta$CCSD(T)   & $-0.18$ & $-0.91$ & $-1.78$  & $-2.77$  & $-3.74$  \\
$\delta$CCSDT     & $+0.02$ & $+0.13$ & $+0.30$  & $+0.51$  & $+0.73$  \\
$\delta$CCSDT(Q)  & $-0.02$ & $-0.15$ & $-0.33$  & $-0.54$  &  $-0.75$        \\
Total             & $+0.60$ & $+0.93$ & $+0.46$  & $-0.13$  & $-0.99$       \\
\hline \hline
\end{tabular}
\end{table}

As a representative transition-metal application, we computed the CO association energy of chromium hexacarbonyl, Cr(CO)$_5$ + CO $\to$ Cr(CO)$_6$, using the def2-TZVPP basis set~\cite{weigend2005balanced,weigend2006accurate} with a frozen core [$N_{\mathrm{o}} = 37$, $N_{\mathrm{v}} = 382$; compact $T_3$ size of 3.7~TB for Cr(CO)$_6$] and the geometry from Ref.~\citenum{dohm2018comprehensive}. Following the same reference, CP corrections were not included. This system is more challenging than the non-covalent dimers: convergence of the CCSDT iterations required DIIS acceleration on the $T_3$ amplitudes. To limit the additional memory overhead, DIIS extrapolation for $T_3$ was performed in a reduced subspace with 90 virtual orbitals, which was sufficient for convergence. Using 32 Ice Lake nodes, the CCSDT calculation took 83~hours (3.9~hours per cycle), and the CCSDT(Q) correction required 262~hours. The results are given in Table~\ref{tab:crco6_reac}. The $\delta$CCSDT contribution of $+8.00$~kJ/mol is substantial and shifts the total reaction energy toward the experimental value of $-162.3 \pm 23.0$~kJ/mol~\cite{dohm2018comprehensive}, whereas the (Q) contribution is small for this particular reaction. This example illustrates the importance of exact post-CCSD(T) corrections for transition-metal chemistry and the value of CCSDT(Q)-level data for validating lower-cost electronic-structure methods, including new exchange-correlation functionals and benchmark sets such as MOR41~\cite{dohm2018comprehensive}.

As a representative application of CCSDTQ, we compute the barrier height of the Cope rearrangement of semibullvalene using the cc-pVDZ(d,s) basis set with the frozen-core approximation ($N_{\mathrm{o}} = 20$, $N_{\mathrm{v}} = 100$; compact $T_4$ size of 6.2~TB). Using 48 Ice Lake nodes, the CCSDTQ calculation required 51 hours, corresponding to 4.2 hours per iteration. The results are reported in Table~\ref{tab:cope}. Although the iterative-quadruples correction ($\delta$CCSDTQ) is relatively small, its magnitude is comparable to those of the $\delta$CCSDT and $\delta$CCSDT(Q) contributions, indicating that post-perturbative quadruple effects remain relevant for this barrier height. To the best of our knowledge, previous high-order coupled-cluster benchmarks for this system have been limited to perturbative quadruples, (Q), and employed a truncated cc-pVDZ basis set with the $d$-type orbitals removed from the carbon atoms.~\cite{karton2020cope} This example demonstrates that the present study enables canonical iterative-quadruple calculations for chemically relevant systems beyond the reach of previous implementations.

\begin{table}[ht]
\caption{Reaction energy (kJ/mol) for Cr(CO)$_5$ + CO $\to$ Cr(CO)$_6$  in the def2-TZVPP basis set. Geometries are from Ref.~\citenum{dohm2018comprehensive}. Starting from MP2, each row reports the incremental correction relative to the preceding level of theory; for example, $\delta$CCSDT $\equiv$ CCSDT $-$ CCSD(T). The total reaction energy is obtained as the sum of the HF contribution and all subsequent increments.}
\label{tab:crco6_reac}
\centering
\begin{tabular}{lr}
\hline \hline
           & Reaction energy (kJ/mol) \\
\hline
HF               & $-78.91$\\
$\delta$MP2      & $-182.26$ \\
$\delta$CCSD     & $+103.49$ \\
$\delta$CCSD(T)  & $-23.60$ \\
$\delta$CCSDT    & $+8.00$ \\
$\delta$CCSDT(Q) & $-0.77$ \\
Total            & $-174.05$ \\
Expt.~\cite{dohm2018comprehensive} & $-162.3 \pm 23.0$ \\
\hline \hline
\end{tabular}
\end{table}

Across these MPI calculations, the CPU core utilization exceeds 80\%, and the sustained floating-point throughput reaches 30--50\% of the single-node MKL dense matrix multiplication (DGEMM) performance on the Ice Lake nodes. MKL DGEMM is used here as a practical reference for highly optimized dense matrix multiplication with regular memory access. The reduced performance in the high-order CC calculations is expected, because the relevant tensor contractions are over multi-dimensional arrays rather than a single large DGEMM. Although these contractions are evaluated by \textsc{pytblis} without explicit index permutations, their tensor strides, effective block sizes, and packing overhead are generally less favorable than those of dense matrix multiplication, leading to lower cache reuse and reduced floating-point efficiency. 

To the best of our knowledge, the applications presented here are the largest canonical CCSDT(Q) and CCSDTQ calculations reported to date. They demonstrate that the current distributed implementations extend canonical high-order coupled-cluster calculations to regimes that were previously inaccessible. Based on the demonstrated memory distribution and parallel performance, compact highest-order cluster amplitudes at the 10~TB scale are within practical reach, corresponding to RCCSDT and RCCSDT(Q) calculations for systems with approximately 100 correlated electrons in 450 orbitals, and RCCSDTQ calculations for systems with approximately 50 correlated electrons in 115 orbitals.

\begin{table}[ht]
\caption{Barrier height (kJ/mol) for the Cope rearrangements reaction of Semibullvalene in the cc-pVDZ(d,s) basis set. Geometries are from Ref.~\citenum{karton2020cope}. Starting from MP2, each row reports the incremental correction relative to the preceding level of theory; for example, $\delta$CCSDT $\equiv$ CCSDT $-$ CCSD(T). The total barrier height is obtained as the sum of the HF contribution and all subsequent increments.}
\label{tab:cope}
\centering
\begin{tabular}{lr}
\hline \hline
           & Barrier height (kJ/mol) \\
\hline
HF               & $+71.57$\\
$\delta$MP2      & $-58.43$ \\
$\delta$CCSD     & $+36.23$ \\
$\delta$CCSD(T)  & $-16.02$ \\
$\delta$CCSDT    & $+1.30$ \\
$\delta$CCSDT(Q) & $-2.20$ \\
$\delta$CCSDTQ   & $+0.43$ \\
Total            & $+32.89$ \\
\hline \hline
\end{tabular}
\end{table}

\section{Conclusions and future work\label{sec:conclusions}}

We have presented efficient Python implementations of high-order coupled-cluster methods RCCSDT, RCCSDT(Q), RCCSDTQ, and UCCSDT within the \textsc{PySCF} quantum chemistry package. The central objective of this work is to make exact iterative triples and perturbative or iterative quadruples calculations practical beyond the small-molecule regime where they have traditionally been confined. The resulting implementation addresses both sides of the bottleneck: the high formal operation counts of the residual equations and the large memory footprint of the highest-order amplitude and residual tensors.

On a shared-memory node, the implementation combines the full and compact high-rank amplitude representations and multi-threaded tensor contractions through \textsc{pytblis}/\textsc{TBLIS}. The compact variants store only the triangular occupied-index sectors of $T_3$ or $T_4$ and reconstruct symmetry-related blocks only when needed, preserving the memory savings while retaining a reduced effective leading operation count. Together, these features provide near-ideal thread scaling and make our single-node RCCSDT, RCCSDT(Q), RCCSDTQ, and UCCSDT implementations faster than, or competitive with, existing alternatives.

The distributed-memory layer extends this strategy to systems for which the compact $T_3$ or $T_4$ amplitudes and residuals exceed the memory capacity of a single node. In distributed RCCSDT and RCCSDTQ, the highest-order amplitudes and residuals, $T_3/R_3$ and $T_4/R_4$, respectively, are partitioned by occupied-index triples and quadruples. Cost-balanced batches are used to avoid load imbalance, while asynchronous collective communication overlaps the transfer of high-order amplitude blocks with local contractions. For the distributed (Q) energy correction, we instead decompose the calculation over tiled virtual-index quadruples $(A,B,C,D)$; each independent tile is evaluated using pairwise nonblocking communication to retrieve the required $T_3$ fragments. Together, these algorithms achieve close-to-ideal strong scaling up to 32 nodes and bring CCSDT(Q) calculations on systems with approximately 100 correlated electrons in 450 orbitals, as well as CCSDTQ calculations on systems with approximately 50 correlated electrons in 115 orbitals, within reach.

The application calculations demonstrate our method's capacity to tackle realistic chemical systems. For $\pi$-stacked aromatic dimers, exact iterative-triples and perturbative-quadruples corrections validate prior rank-reduced estimates and show that post-CCSD(T) effects can be comparable to the total interaction energy. For the polyene dimer series, the iterative-triples and perturbative-quadruples corrections grow with system size but largely cancel, providing a different benchmark trend for extended non-covalent systems. For the CO association energy of Cr(CO)$_6$, the sizeable iterative-triples contribution shifts the reaction energy toward experiment and highlights the importance of exact post-CCSD(T) data for transition-metal chemistry. For the barrier height of the Cope rearrangement of semibullvalene, the iterative-quadruples correction beyond perturbative quadruples is modest but not negligible, suggesting that post-perturbative quadruple effects remain relevant for accurately determining the barrier height. To the best of our knowledge, these examples include the largest canonical CCSDT(Q) calculation reported to date ($N_{\mathrm{o}} = 52$, $N_{\mathrm{v}} = 328$; compact $T_3$ size of 6.4~TB) and the largest canonical CCSDTQ calculation reported to date ($N_{\mathrm{o}} = 20$, $N_{\mathrm{v}} = 100$; compact $T_4$ size of 6.2~TB). In addition to demonstrating the computational reach of the present implementation, these calculations provide reference data for assessing reduced-scaling, local, and stochastic approaches, as well as density-functional theory calculations. The relatively small basis sets used here, however, should be carefully considered when making such comparisons.

Several directions follow naturally from this work. A distributed spin-unrestricted CCSDT, CCSDT(Q), and CCSDTQ implementation would extend the same strategy to open-shell and strongly correlated systems, enabling benchmark calculations for thermochemical atomization energies as well as enzyme- and transition-metal-catalyzed reactions~\cite{zhai2026classical}. Generalizing the implementation to $\Lambda$ equations, density matrices, and response properties would broaden its scope beyond total energies~\cite{salter1989analytic,gauss2002analytic,matthews2020analytic}, while additional approximate quadruple corrections could provide lower-cost routes to post-CCSDT accuracy.~\cite{kallay2005approximate,eriksen2015communication} Periodic extensions with $k$-point sampling~\cite{hirata2004coupled,mcclain2017gaussian,zhang2019coupled,gruber2018applying} would bring exact high-order CC benchmarks to solids. Finally, combining the present implementation with frozen natural orbital (FNO)~\cite{manisha2025frozen}, local natural orbital (LNO)~\cite{fishman2026development,ye2023ab,ye2024adsorption,ye2024periodic}, rank-reduced~\cite{lesiuk2022gold}, or stochastic~\cite{sun2026stochastic,martin2025development} approximations would extend the accessible length scales further while retaining canonical calculations as calibration points.

\begin{acknowledgments}
The Flatiron Institute is a division of the Simons Foundation. CH was supported by a fellowship from the Molecular Sciences Software Institute under NSF award No.~CHE-2136142. The computations in this work were run at facilities supported by the Scientific Computing Core at the Flatiron Institute.
\end{acknowledgments}

\section*{Data Availability Statement}

The code and example scripts are open source and available in \href{https://github.com/pyscf}{https://github.com/pyscf} and \href{https://github.com/jinyuchem/Distributed-CC}{https://github.com/jinyuchem/Distributed-CC}. The data supporting the findings of this study are available in the article and the Appendix.

\appendix
\section{Theory\label{sec:theory}}

This appendix revisits established coupled-cluster (CC) formalisms relevant to the present work. Beginning with standard CC theory in spin-orbital notation, we introduce the spin-integrated formalism with spin-orbital excitation operators and spin-unrestricted (UCC) amplitudes, and the non-orthogonal spin-free formalism with spin-free excitation operators and spin-free (RCC) amplitudes. We present the explicit RCCSDT, RCCSDTQ, and RCCSDT(Q) equations that underlie the implementations described below within the spin-free formalism. We also include the UCCSDT equations for completeness.

\subsection{Coupled-cluster theory\label{sec:cc}}

CC theory parameterizes the many-electron wavefunction by applying an exponential excitation operator to a reference Slater determinant $|\Phi_0 \rangle$,
\begin{equation}
    | \Psi_{\mathrm{CC}} \rangle = \mathrm{e}^{\hat{T}} | \Phi_0 \rangle ,
\end{equation}
where the cluster operator is expanded as a sum of excitation operators of different ranks,
\begin{equation}
    \hat{T} = \sum_{n=1}^{N} \hat{T}_n .
\end{equation}
Here, $\hat{T}_n$ generates all $n$-fold particle-hole excitations from the reference determinant:
\begin{equation}
    \hat{T}_n = \dfrac{1}{(n!)^2} \sum_{\substack{a_1 \cdots a_n \\ i_1 \cdots i_n}} t_{i_1 \cdots i_n}^{a_1 \cdots a_n} \hat{T}_{i_1 \cdots i_n}^{a_1 \cdots a_n},
\end{equation}
with
\begin{equation}
    \hat{T}_{i_1 \cdots i_n}^{a_1 \cdots a_n} = \hat{a}_{a_1}^{\dagger} \cdots \hat{a}_{a_n}^{\dagger} \hat{a}_{i_n} \cdots \hat{a}_{i_1}.
\end{equation}
where the indices $i,j,k,\ldots$ denote occupied spin-orbitals, $a,b,c,\ldots$ denote virtual spin-orbitals, and $n=1,\cdots N$ denotes excitation orders. We may use $T_n$ to denote the amplitude tensor at excitation order $n$ with elements $t_{i_1 \cdots i_n}^{a_1 \cdots a_n}$. The exponential ansatz is size extensive and, through the disconnected products generated by $\mathrm{e}^{\hat{T}}$, includes contributions from higher-order excited configurations even when the cluster operator is truncated at a finite excitation rank $N < n_{\mathrm{elec}}$. The value of $N$ defines the corresponding CC approximation: $N=2$ gives CCSD, $N=3$ gives CCSDT, $N=4$ gives CCSDTQ, and so forth.

The CC energy is obtained from the reference projection of the similarity-transformed Hamiltonian,
\begin{equation}
    E_{\mathrm{CC}}
    =
    \big\langle \Phi_0 \big\lvert
    \mathrm{e}^{-\hat{T}} \hat{H} \mathrm{e}^{\hat{T}}
    \big\rvert \Phi_0 \big\rangle.
\end{equation}
The cluster amplitudes $t_{i_1 \cdots i_k}^{a_1 \cdots a_k}$ are determined by requiring that the similarity-transformed Hamiltonian have zero projection onto all excited determinants included in the truncation,
\begin{equation}
    0 = r_{i_1 \cdots i_n}^{a_1 \cdots a_n} = \big\langle \Phi_{i_1 \cdots i_n}^{a_1 \cdots a_n} \big\lvert \mathrm{e}^{-\hat{T}} \hat{H} \mathrm{e}^{\hat{T}} \big\rvert \Phi_0 \big\rangle ,
\end{equation}
for \(n = 1, \ldots, N\) and for all occupied and virtual index combinations. $R_n$ with elements $ r_{i_1 \cdots i_n}^{a_1 \cdots a_n}$ is called the residual tensor. The corresponding excited determinants are
\begin{equation}
    \big \langle \Phi_{i_1 \cdots i_n}^{a_1 \cdots a_n} \big\lvert = \big\langle \Phi_0 \big| \hat{a}_{a_1} \cdots \hat{a}_{a_n} \hat{a}_{i_n}^{\dagger} \cdots \hat{a}_{i_1}^{\dagger}.
\end{equation}

In practical implementations, the nonlinear amplitude equations are solved iteratively. A basic Jacobi-like update has the form
\begin{equation}
    t_{i_1 \cdots i_n}^{a_1 \cdots a_n} \leftarrow t_{i_1 \cdots i_n}^{a_1 \cdots a_n} + r_{i_1 \cdots i_n}^{a_1 \cdots a_n} \big/ \epsilon_{i_1 \cdots i_n}^{a_1 \cdots a_n},
\end{equation}
where the orbital-energy denominator is
\begin{equation}
    \epsilon_{i_1 \cdots i_n}^{a_1 \cdots a_n} = \epsilon_{i_1} + \cdots +  \epsilon_{i_n} - \epsilon_{a_1} - \cdots - \epsilon_{a_n},
    \label{eq:orb-energy-deno}
\end{equation}
and $\epsilon_p$ denotes the orbital energy of the spin-orbital $p$.

The spin-orbital formalism is general, but it carries implicit spin labels on every orbital index. For spin-unrestricted reference states, it is often more convenient to use a spin-integrated notation, in which spatial orbitals are retained, and the spin labels are summed explicitly. This formulation applies to both open-shell and closed-shell references. Henceforth, we use the indices $i, j, k, \ldots$ to denote occupied spatial orbitals, and $a, b, c, \ldots$ to denote virtual spatial orbitals. Each spatial orbital index carries a spin label $\sigma,\tau,\ldots$ summed over $\{\alpha,\beta\}$, and only terms with matching spin labels between occupied and virtual orbitals survive:
\begin{equation}
    \hat{T}_n = \frac{1}{(n!)^2} \sum_{\substack{\sigma_1 \cdots \sigma_n}} \sum_{\substack{a_1\cdots a_n\\i_1 \cdots i_n}}  t_{i_{1,\sigma_1}\cdots i_{n,\sigma_n}}^{a_{1,\sigma_1} \cdots a_{n,\sigma_n}} \hat{T}_{i_{1,\sigma_1}\cdots i_{n,\sigma_n}}^{a_{1,\sigma_1}\cdots a_{n,\sigma_n}}
\end{equation}
With this notation, the spin-unrestricted amplitudes are antisymmetric/symmetric under independent odd/even permutations of occupied and virtual indices with the same spin index. Taking the $T_3$ amplitudes as an example,
\begin{subequations}
\begin{equation}
\begin{aligned}
    t_{ijk}^{abc} &= - t_{ikj}^{abc} = - t_{jik}^{abc} = - t_{kji}^{abc} =  t_{jki}^{abc} =  t_{kij}^{abc} \\
    &= - t_{ijk}^{acb} = - t_{ijk}^{bac} = - t_{ijk}^{cba} =  t_{ijk}^{bca} = t_{ijk}^{cab},
\end{aligned}
\end{equation}
\vspace{-10pt}
\begin{equation}
    t_{ij\overline{k}}^{ab\overline{c}} = - t_{ji\overline{k}}^{ab\overline{c}} = - t_{ij\overline{k}}^{ba\overline{c}} =   t_{ji\overline{k}}^{ba\overline{c}}.
\end{equation}
\end{subequations}
Similar relations follow for $t_{\overline{i}\overline{j}k}^{\overline{a}\overline{b}c}$ and $t_{\overline{i}\overline{j}\overline{k}}^{\overline{a}\overline{b}\overline{c}}$. Here, indices $i, j, k, \ldots, a,b,c,\ldots$ with/without an overline denote spatial orbitals with alpha/beta spin.

\subsection{Spin-free formalism\label{sec:nosa}}

For spin-restricted closed-shell systems, we can further exploit SU(2) spin symmetry and reduce the computational cost~\cite{matthews2013revisitation,matthews2015non} by re-expressing the cluster operator in terms of spin-free excitation operators and amplitudes as

\begin{equation}
    \hat{T}_n = \frac{1}{n!} \sum_{\substack{a_1\cdots a_n\\i_1 \cdots i_n}} \check{t}_{i_1\cdots i_n}^{a_1 \cdots a_n} \hat{E}_{i_1\cdots i_n}^{a_1\cdots a_n},
\end{equation}
where the spin-free excitation operator is
\begin{equation}
\label{eq:spinfree_e}
    \hat{E}_{i_1\cdots i_n}^{a_1\cdots a_n} = \sum_{\substack{\sigma_1 \cdots \sigma_n}} \hat{T}_{i_{1,\sigma_1}\cdots i_{n,\sigma_n}}^{a_{1,\sigma_1}\cdots a_{n,\sigma_n}}.
\end{equation}
The spin-free amplitudes $\check{t}$ satisfy a column permutation symmetry: simultaneously permuting an occupied index and the paired virtual index leaves the amplitude invariant. Taking the $T_3$ amplitudes as an example:
\begin{equation}
    \check{t}_{ijk}^{abc} = \check{t}_{ikj}^{acb} = \check{t}_{jik}^{bac} = \check{t}_{jki}^{bca} = \check{t}_{kij}^{cab} = \check{t}_{kji}^{cba}.
    \label{eq:rccsdt-amps-symm}
\end{equation}

The spin-unrestricted amplitudes $t_{i\cdots}^{a\cdots}$ can be recovered from antisymmetrized sums of the spin-free amplitudes. For $T_2$ amplitudes,
\begin{equation}
    t_{ij}^{ab} = t_{\overline{i}\overline{j}}^{\overline{a}\overline{b}} = \mathcal{A}^{ab} \check{t}_{ij}^{ab} = \mathcal{A}_{ij} \check{t}_{ij}^{ab}, \quad t_{i\overline{j}}^{a\overline{b}} = \check{t}_{ij}^{ab}
\end{equation}
while for $T_3$ amplitudes,
\begin{equation}
\begin{aligned}
    t_{ijk}^{abc} &= t_{\overline{i}\overline{j}\overline{k}}^{\overline{a}\overline{b}\overline{c}} = \mathcal{A}^{abc} \check{t}_{ijk}^{abc} = \mathcal{A}_{ijk} \check{t}_{ijk}^{abc}, \\
    t_{ij\overline{k}}^{ab\overline{c}} &= t_{\overline{i}\overline{j}k}^{\overline{a}\overline{b}c} = \mathcal{A}^{ab} \check{t}_{ijk}^{abc} = \mathcal{A}_{ij} \check{t}_{ijk}^{abc},
\end{aligned}
\end{equation}
and for $T_4$ amplitudes,
\begin{equation}
\begin{aligned}
    t_{ijkl}^{abcd} &= t_{\overline{i}\overline{j}\overline{k}\overline{l}}^{\overline{a}\overline{b}\overline{c}\overline{d}} = \mathcal{A}^{abcd} \check{t}_{ijkl}^{abcd} = \mathcal{A}_{ijkl} \check{t}_{ijkl}^{abcd}, \\
    t_{ijk\overline{l}}^{abc\overline{d}} &= t_{\overline{i}\overline{j}\overline{k}l}^{\overline{a}\overline{b}\overline{c}d} = \mathcal{A}^{abc} \check{t}_{ijkl}^{abcd} = \mathcal{A}_{ijk} \check{t}_{ijkl}^{abcd}, \\
    t_{ij\overline{k}\overline{l}}^{ab\overline{c}\overline{d}} &= t_{\overline{i}\overline{j}kl}^{\overline{a}\overline{b}cd} = \mathcal{A}^{ab} \mathcal{A}^{cd} \check{t}_{ijkl}^{abcd} = \mathcal{A}_{ij} \mathcal{A}_{kl} \check{t}_{ijkl}^{abcd}.
\end{aligned}
\end{equation}
Here, the antisymmetrizers $\mathcal{A}$ over virtual and occupied indices are defined as
\begin{subequations}
\begin{align}
    \mathcal{A}^{a_1\cdots a_n} X^{a_1\cdots a_n}_{i_1\cdots i_n} &= \sum_{\pi\in S_n} \mathrm{sgn}(\pi) X^{a_{\pi(1)}\cdots a_{\pi(n)}}_{i_1\cdots i_n}, \\
    \mathcal{A}_{i_1\cdots i_n} X^{a_1\cdots a_n}_{i_1\cdots i_n} &= \sum_{\pi\in S_n} \mathrm{sgn}(\pi) X_{i_{\pi(1)}\cdots i_{\pi(n)}}^{a_1\cdots a_n},
\end{align}
\end{subequations}
where $S_n$ denotes the symmetric group of degree $n$, i.e., the set of all permutations of $n$ elements. For $\pi \in S_n$, $\mathrm{sgn}(\pi)$ denotes the parity of the permutation, equal to $+1$ for even permutations and $-1$ for odd permutations.

Finally, we note that the excited configurations generated by the spin-free excitation operators $\hat{E}_{i_1\cdots i_n}^{a_1\cdots a_n}$
are not mutually orthogonal: their overlap matrix has nonzero off-diagonal elements, which gives rise to a null space corresponding to linear dependencies among the spin-free amplitudes. Consequently, the spatial amplitude tensors $\check{t}_{ijk}^{abc}$ and $\check{t}_{ijkl}^{abcd}$ are overcomplete parameterizations of RCCSDT and RCCSDTQ: they contain 6 and 24 index-permutation-symmetry independent components, respectively, whereas only 5 (for $T_3$) and 14 (for $T_4$) are linearly independent for spin-$\tfrac{1}{2}$ particles. Working with this non-orthogonal spin-free formalism, this redundancy must be controlled during the iterative solution, either by a spin-summation/de-spin-summation procedure~\cite{matthews2013revisitation,matthews2015non} or by explicitly projecting out permutation components incompatible with spin-$\tfrac{1}{2}$ systems (see Appendix~\ref{appendix:nosa}).

\subsection{Hamiltonian and integrals\label{sec:hamil}}

The standard electronic Hamiltonian in the closed-shell molecular orbital basis is given by 
\begin{equation}
    \hat{H} = \sum_{pq} h^p_q \sum_{\sigma} \hat{a}_{p\sigma}^{\dagger} \hat{a}_{q\sigma} + \frac{1}{2} \sum_{pqrs} v^{pq}_{rs} \sum_{\sigma\tau} \hat{a}_{p\sigma}^{\dagger} \hat{a}_{q\tau}^{\dagger} \hat{a}_{s\tau} \hat{a}_{r\sigma},
\end{equation}
which can be rewritten in spin-free operators:
\begin{equation}
    \hat{H} = \sum_{pq} h^p_q \hat{E}^p_q + \frac{1}{2} \sum_{pqrs} v^{pq}_{rs} \left( \hat{E}^p_r \hat{E}^q_s - \delta_{qr} \hat{E}^p_s \right),
\end{equation}
where the electron-repulsive integrals (ERIs) are written in physicists' notation as $v^{pq}_{rs} \equiv \langle pq | rs \rangle$, and the Fock matrix elements $f^r_s$ are
\begin{equation}
    f^r_s = h^r_s + \sum_i \left( 2 v^{ri}_{si} - v^{ri}_{is} \right)
\end{equation}
where we use the indices $p, q, r, \ldots$ to denote arbitrary (occupied or virtual) spatial orbitals.

\subsection{$T_1$-dressed formalism\label{sec:t1dress}}

The $T_1$-dressed formalism~\cite{koch1994direct} absorbs the singles amplitudes into a partially similarity-transformed Hamiltonian,
\begin{equation}
    \widetilde{H} = \mathrm{e}^{-\hat{T}_1} \hat{H} \mathrm{e}^{\hat{T}_1},
\end{equation}
so that the remaining amplitude equations can be written without explicit $T_1$-containing terms. 

In practice, the transformation is implemented through the matrix $\mathbf{t}_1$ with the shape $(N_{\mathrm{o}} + N_{\mathrm{v}})\times(N_{\mathrm{o}} + N_{\mathrm{v}})$ (with $N_{\mathrm{o}}$ and $N_{\mathrm{v}}$ denote the number of occupied and virtual orbitals, respectively), which embeds the singles amplitudes $[\check{\mathbf{t}}_1^{\mathrm{vo}}]_i^a = \check{t}_i^a$ into the full one-particle orbital space:
\begin{equation}
    \mathbf{t}_1 = \left[ \begin{matrix}
        \mathbf{0}^{\mathrm{oo}} & \mathbf{0}^{\mathrm{ov}} \\
        \check{\mathbf{t}}_1^{\mathrm{vo}} & \mathbf{0}^{\mathrm{vv}} \\
    \end{matrix} \right],
\end{equation}
The associated left and right orbital transformation matrices are
\begin{equation}
    \mathbf{x} = \mathbf{I} - \mathbf{t}_1, \quad \mathbf{y} = \mathbf{I} + \mathbf{t}_1^T.
\end{equation}
The dressed ERIs and Fock matrix elements are then obtained by contracting the bare quantities with $\mathbf{x}$ and $\mathbf{y}$:
\begin{equation}
    \tilde{v}^{pq}_{rs} = \sum_{tuvw} x_{pt} x_{qu} v^{tu}_{vw} y_{rv} y_{sw} 
\end{equation}
\begin{equation}
    \tilde{f}^{p}_{q} = \sum_{rs} x_{pr} \bigg[ f^{r}_{s} + \sum_{ia} \left( 2 v^{ri}_{sa} - v^{ri}_{as} \right) \check{t}_i^a \bigg] y_{qs}
\end{equation}

Unlike standard ERIs, the dressed ERIs possess only a two-fold symmetry and do not, in general, satisfy the full eight-fold permutation symmetry for real wavefunctions:
\begin{equation*}
    \tilde{v}^{pq}_{rs} = \tilde{v}^{qp}_{sr}, \quad \tilde{v}^{pq}_{rs} \neq \tilde{v}^{ps}_{rq}.
\end{equation*}

\subsection{Coupled-cluster amplitude equations}
\label{sec:amp_eq}

The RCCSD, RCCSDT, and RCCSDTQ residual equations are written below in terms of the spin-free amplitudes $\check{t}$ and the $T_1$-dressed integrals $\tilde{v}$ and $\tilde{f}$. 

We first introduce two notational conventions used throughout these expressions. Following Refs.~\citenum{matthews2013revisitation,matthews2015non}, a check mark on an index label denotes spin summation over that index pair. For example, the spin-summed two-electron integrals are
\begin{equation}
\begin{aligned}
    v^{\check{p}q}_{\check{r}s} &= \left( 2 - P_r^s \right) v^{pq}_{rs} = 2 v^{pq}_{rs} - v^{pq}_{sr} \\
    &= \left( 2 - P_p^q \right) v^{pq}_{rs} = 2 v^{pq}_{rs} - v^{qp}_{rs},
\end{aligned}
\end{equation}
where $P_r^s$ exchanges the two indices $r$ and $s$. The CC correlation energy can therefore be written as
\begin{equation}
    E_{\mathrm{CC}} = \sum_{ijab} v^{\check{i}j}_{\check{a}b} \left( \check{t}_{ij}^{ab} + \check{t}_i^a \check{t}_j^b \right) + 2 \sum_{ia} f^{i}_{a} \check{t}_i^a.
\end{equation}

We also define the full paired-index permutation operator,
\begin{equation}
    \mathcal{P}_{(i_1a_1)\cdots(i_na_n)} X_{i_1\cdots i_n}^{a_1\cdots a_n} = \sum_{\pi\in S_n} X_{i_{\pi(1)}\cdots i_{\pi(n)}}^{a_{\pi(1)}\cdots a_{\pi(n)}}.
    \label{eq:rcc-perm-op}
\end{equation}
For example, $\mathcal{P}_{(ia)(jb)(kc)(ld)}$ generates all $4!$ simultaneous permutations of the occupied--virtual index pairs $(ia)$, $(jb)$, $(kc)$, and $(ld)$.

\subsubsection{RCCSD}

Adapting the non-orthogonal spin-adapted CC equations of Ref.~\citenum{matthews2015non} to the $T_1$-dressed formalism, the RCCSD singles residual is
\begin{equation}
\label{eq:ccsd_r1}
    \check{r}^{a}_{i} (\mathrm{RCCSD}) = \tilde{f}^{a}_{i} + \sum_{kc} \tilde{f}^{k}_{c}  \check{t}_{\check{i}k}^{\check{a}c} + \sum_{kcd} \tilde{v}^{ak}_{cd} \check{t}_{\check{i}k}^{\check{c}d} - \sum_{klc} \tilde{v}^{kl}_{ic} \check{t}_{\check{k}l}^{\check{a}c},
\end{equation}
and the doubles residual is
\begin{equation}
\label{eq:ccsd_r2}
\begin{aligned}
    &\check{r}^{ab}_{ij} (\mathrm{RCCSD}) = \mathcal{P}_{(ia)(jb)}  \bigg( \frac{1}{2} \tilde{v}^{ab}_{ij} + \sum_c F^{b}_{c} \check{t}_{ij}^{ac} - \sum_k F^{k}_{j} \check{t}_{ik}^{ab}  \\
    &+ \frac{1}{2} \sum_{cd} \tilde{v}^{ab}_{cd} \check{t}_{ij}^{cd} + \frac{1}{2} \sum_{kl} W^{kl}_{ij} \check{t}_{kl}^{ab} + \frac{1}{2} \sum_{kc} W^{ka}_{ci} \check{t}_{\check{k}j}^{\check{c}b} \\
    & - \frac{1}{2} \sum_{kc} W^{ka}_{ic} \check{t}_{jk}^{cb} - \sum_{kc} W^{kb}_{ic} \check{t}_{jk}^{ca} \bigg).
\end{aligned}
\end{equation}
The spin-summed double-amplitude combinations appearing in these residuals are
\begin{equation}
\begin{aligned}
    \check{t}_{\check{i}j}^{\check{a}b} &= \big(2 - P_a^b \big) \check{t}_{ij}^{ab} = \big(2 - P_i^j \big) \check{t}_{ij}^{ab} = 2 \check{t}_{ij}^{ab} - \check{t}_{ij}^{ba},
\end{aligned}
\end{equation}
and the RCCSD intermediates are
\begin{subequations}
\begin{align}
    F^b_c &= \tilde{f}^b_c - \sum_{kld} \tilde{v}^{\check{k}l}_{\check{d}c} \check{t}_{kl}^{db}, \\
    F^k_j &= \tilde{f}^k_j + \sum_{lcd} \tilde{v}^{\check{l}k}_{\check{c}d} \check{t}_{lj}^{cd}, \\
    W^{kl}_{ij} &= \tilde{v}^{kl}_{ij} + \sum_{cd} \tilde{v}^{kl}_{cd} \check{t}_{ij}^{cd}, \\
    W^{ka}_{ci} &= \tilde{v}^{\check{k}a}_{\check{c}i} + \frac{1}{2} \sum_{ld} \tilde{v}^{\check{l}k}_{\check{d}c} \check{t}^{\check{d}a}_{\check{l}i}, \\
    W^{ka}_{ic} &= \tilde{v}^{ka}_{ic} - \frac{1}{2} \sum_{ld} \tilde{v}^{lk}_{cd} \check{t}^{da}_{il}.
\end{align}
\end{subequations}

\subsubsection{RCCSDT\label{appendix:rccsdt}}

Including the triples operator $\hat{T}_3$ gives RCCSDT. The singles residual then acquires a $T_3$ correction:
\begin{equation}
    \check{r}_i^a (\mathrm{RCCSDT}) = \check{r}_i^a (\mathrm{RCCSD}) + \frac{1}{2} \sum_{jkbc} \tilde{v}^{jk}_{bc} \check{t}_{\check{j}\check{k}i}^{\check{b}\check{c}a}.
\end{equation}
The doubles residual is augmented in a similar way:
\begin{equation}
\begin{aligned}
    &\check{r}_{ij}^{ab}(\mathrm{RCCSDT}) = \check{r}_{ij}^{ab}(\mathrm{RCCSD}) \\
    &+ \mathcal{P}_{(ia)(jb)} \bigg(\frac{1}{2} \tilde{f}^{k}_{c} \check{t}_{\check{k}ij}^{\check{c}ab} + \sum_{kcd} \tilde{v}^{bk}_{cd} \check{t}_{\check{k}ij}^{\check{d}ac} - \sum_{klc} \tilde{v}^{kl}_{jc} \check{t}_{\check{l}ik}^{\check{c}ab}\bigg).
\end{aligned}
\end{equation}
The corresponding triples residual is
\begin{equation}
\begin{aligned}
    &\check{r}_{ijk}^{abc}(\mathrm{RCCSDT}) = \mathcal{P}_{(ia)(jb)(kc)} \\
    & \bigg( \sum_d W_{dj}^{ab} \check{t}_{ik}^{dc} - \sum_l W_{ij}^{al} \check{t}_{lk}^{bc} + \frac{1}{2} \sum_d F_d^a \check{t}_{ijk}^{dbc} - \frac{1}{2} \sum_l F_i^l \check{t}_{ljk}^{abc} \\
    &+ \frac{1}{4} \sum_{ld} \overline{W}^{la}_{di} \check{t}_{\check{l}jk}^{\check{d}bc} - \frac{1}{2} \sum_{ld} \overline{W}^{la}_{id} \check{t}_{jlk}^{dbc} - \sum_{ld} \overline{W}^{lb}_{id} \check{t}_{jlk}^{dac} \\
    &+ \frac{1}{2} \sum_{lm} W_{ij}^{lm} \check{t}_{lmk}^{abc} + \frac{1}{2} \sum_{de} W_{de}^{ab} \check{t}_{ijk}^{dec} \bigg).
    \label{eq:resi-rccsdt}
\end{aligned}
\end{equation}
The spin-summed triples amplitude combinations entering the residuals are
\begin{equation}
\begin{aligned}
    \check{t}_{\check{i}\check{j}k}^{\check{a}\check{b}c} &= \big( 2 - P_b^c \big) \check{t}_{\check{i}jk}^{\check{a}bc} = \big(2 - P_b^c \big) \big( 2 - P_a^b - P_a^c \big) \check{t}_{ijk}^{abc} \\
    &= \big( 2 - P_j^k \big) \check{t}_{\check{i}jk}^{\check{a}bc} = \big(2 - P_j^k \big) \big( 2 - P_i^j - P_i^k \big) \check{t}_{ijk}^{abc}. \\
    \label{eq:triples-spin-sum}
\end{aligned}
\end{equation}
The additional intermediates needed for RCCSDT are
\begin{subequations}
    \begin{equation}
    \begin{aligned}
        W_{dj}^{ab} =&\ \tilde{v}^{ab}_{dj} + \frac{1}{2} \sum_{el} \tilde{v}^{\check{l}a}_{\check{e}d} \check{t}_{\check{l}j}^{\check{e}b} - \frac{1}{2} \sum_{el} \tilde{v}^{la}_{de} \check{t}_{jl}^{eb} \\
        &- \sum_{el} \tilde{v}^{lb}_{de} \check{t}_{jl}^{ea} + \sum_{lm} \tilde{v}^{lm}_{dj} \check{t}_{lm}^{ab} - \sum_{lme} \tilde{v}^{lm}_{de} \check{t}_{\check{m}jl}^{\check{e}ba},
    \end{aligned}
    \end{equation}
    \vspace{-10pt}
    \begin{equation}
    \begin{aligned}
    W_{ij}^{al} =&\ \tilde{v}_{ij}^{al} + \sum_d \tilde{f}^{l}_{d} \check{t}_{ij}^{ad} + \frac{1}{2} \sum_{dm} \tilde{v}^{\check{m}l}_{\check{d}j} \check{t}_{\check{m}i}^{\check{d}a} - \frac{1}{2} \sum_{dm} \tilde{v}^{ml}_{jd} \check{t}_{im}^{da}, \\
    &- \sum_{dm} \tilde{v}^{ml}_{id} \check{t}_{jm}^{da} + \sum_{de} \tilde{v}^{al}_{de} \check{t}_{ij}^{de} + \sum_{mde} \tilde{v}^{lm}_{de} \check{t}_{\check{m}ij}^{\check{e}ad},
    \end{aligned}
    \end{equation}
    \vspace{-20pt}
    \begin{align}
    \overline{W}^{la}_{di} &= W^{la}_{di} + \frac{1}{2} \sum_{me} \tilde{v}^{\check{m}l}_{\check{e}d} \check{t}_{\check{m}i}^{\check{e}a}, \\
    \overline{W}^{la}_{id} &= W^{la}_{id} - \frac{1}{2} \sum_{me} \tilde{v}^{ml}_{de} \check{t}_{im}^{ea}, \\
    W_{de}^{ab} &= \tilde{v}^{ab}_{de} + \sum_{lm} \tilde{v}^{lm}_{de} \check{t}_{lm}^{ab}.
\end{align}
\end{subequations}

\subsubsection{RCCSDTQ\label{appendix:rccsdtq}}

Including the quadruples operator $\hat{T}_4$ gives RCCSDTQ. The doubles residual receives an additional quadruples-driven term:
\begin{equation}
\begin{aligned}
    \check{r}_{ij}^{ab}(\mathrm{RCCSDTQ}) =&\ \check{r}_{ij}^{ab}(\mathrm{RCCSDT}) \\
    &+ \mathcal{P}_{(ia)(jb)} \bigg( \frac{1}{4} \sum_{efmn} \tilde{v}_{ef}^{mn} \check{t}_{\check{m}\check{n}ij}^{\check{e}\check{f}ab} \bigg).
\end{aligned}
\end{equation}
The triples residual also receives a $T_4$ correction:
\begin{equation}
\begin{aligned}
    & \check{r}_{ijk}^{abc}(\mathrm{RCCSDTQ}) = \check{r}_{ijk}^{abc}(\mathrm{RCCSDT}) + \mathcal{P}_{(ia)(jb)(kc)} \\
    &\times \bigg( \frac{1}{6} \sum_{em} \tilde{f}_{e}^m \check{t}_{\check{m}ijk}^{\check{e}abc} + \frac{1}{2} \sum_{efm} \tilde{v}_{ef}^{am} \check{t}_{\check{m}ijk}^{\check{f}ebc} - \frac{1}{2} \sum_{emn} \tilde{v}_{ej}^{mn} \check{t}_{\check{m}ink}^{\check{e}abc} \bigg).
\end{aligned}
\end{equation}
The new quadruples residual is
\begin{equation}
\begin{aligned}
    &\check{r}_{ijkl}^{abcd}(\mathrm{RCCSDTQ}) = \mathcal{P}_{(ia)(jb)(kc)(ld)} \\
    &\times \bigg( \frac{1}{2} \sum_e \overline{W}_{ej}^{ab} \check{t}_{ikl}^{ecd} -\frac{1}{2} \sum_m W_{ij}^{am} \check{t}_{mkl}^{bcd} + \frac{1}{6} \sum_e F_e^a \check{t}_{ijkl}^{ebcd} \\
    &- \frac{1}{6} \sum_m F_{i}^m \check{t}_{mjkl}^{abcd}  + \frac{1}{12} \sum_{em} \overline{W}_{ei}^{ma} \check{t}_{\check{m}jkl}^{\check{e}bcd} \\
    &- \frac{1}{4} \sum_{em} \overline{W}_{ie}^{ma} \check{t}_{jmkl}^{ebcd} - \frac{1}{2} \sum_{em} \overline{W}_{ie}^{mb} \check{t}_{jmkl}^{eacd} + \frac{1}{4} \sum_{mn} W_{ij}^{mn} \check{t}_{mnkl}^{abcd} \\
    &+ \frac{1}{4} \sum_{ef} W_{ef}^{ab} \check{t}_{ijkl}^{efcd} + \frac{1}{8} \sum_{em} W_{eij}^{mab} \check{t}_{\check{m}kl}^{\check{e}cd} \\
    &- \frac{1}{2} \sum_{em} W_{iej}^{mab} \check{t}_{kml}^{ecd} - \sum_{em} W_{iej}^{mcb} \check{t}_{kml}^{ead} + \frac{1}{2} \sum_{mn} W_{ijk}^{amn} \check{t}_{mnl}^{bcd} \\
    &- \frac{1}{2} \sum_m W_{ijk}^{abm} \check{t}_{ml}^{cd} + \frac{1}{2} \sum_e W_{ejk}^{abc} \check{t}_{il}^{ed} \bigg).
    \label{eq:resi-rccsdtq}
\end{aligned}
\end{equation}
The spin-summed quadruples amplitude combinations used above are
\begin{equation}
\begin{aligned}
    \check{t}_{\check{i}\check{j}kl}^{\check{a}\check{b}cd} &= \big(2 - P_b^c - P_b^d \big) \check{t}_{\check{i}jkl}^{\check{a}bcd}, \\
    \check{t}_{\check{i}jkl}^{\check{a}bcd} &= \big(2 - P_a^b - P_a^c - P_a^d \big) \check{t}_{ijkl}^{abcd}. \\
    \label{eq:quadruples-spin-sum}
\end{aligned}
\end{equation}
The additional intermediates needed for RCCSDTQ are
\begin{subequations}
\begin{align}
    \overline{W}_{ej}^{ab} &= W_{ej}^{ab} - \sum_m \tilde{f}_e^m \check{t}_{mj}^{ab} \\
    W_{iej}^{mab} &= \sum_f \tilde{v}_{fe}^{ma} \check{t}_{ji}^{bf} - \sum_n \tilde{v}_{ie}^{mn} \check{t}_{nj}^{ab} - \frac{1}{2} \sum_{fn} \tilde{v}_{ef}^{nm} \check{t}_{inj}^{fab} \\
    W_{ijk}^{amn} &= \mathcal{P}_{(mj)(nk)} \bigg( \sum_e \tilde{v}_{ek}^{mn} \check{t}_{ij}^{ae} + \frac{1}{2} \sum_{ef} \tilde{v}_{ef}^{mn} \check{t}_{ijk}^{aef} \bigg)
\end{align}
\vspace{-15pt}
\begin{equation}
    \begin{aligned}
        W_{eij}^{mab}
        =&\ \mathcal{P}_{(ia)(jb)} \\
        &\times \bigg( \sum_f \tilde{v}_{\check{e}f}^{\check{m}a} \check{t}_{ji}^{bf} - \sum_n \tilde{v}_{\check{e}i}^{\check{m}n} \check{t}_{nj}^{ab} + \frac{1}{4} \sum_{fn} \tilde{v}_{\check{f}e}^{\check{n}m} \check{t}_{\check{n}ij}^{\check{f}ab} \bigg)
    \end{aligned}
    \end{equation}
    \begin{align}
        W_{ejk}^{abc} &= \mathcal{P}_{(jb)(kc)} \bigg( \frac{1}{2} \sum_f W_{ef}^{ab} \check{t}_{jk}^{fc} - \frac{1}{2} \sum_{fmn} \tilde{v}_{ef}^{mn} \check{t}_{\check{n}mjk}^{\check{f}abc} \bigg)
    \end{align}
    \vspace{-15pt}
    \begin{equation}
    \begin{aligned}
        W_{ijk}^{abm} =&\ \mathcal{P}_{(ia)(jb)} \bigg( \sum_{ef} \tilde{v}_{ef}^{am} \check{t}_{ijk}^{ebf} + \sum_e \widetilde{W}_{ei}^{ma} \check{t}_{jk}^{be} \\
        &+ \sum_e W_{ke}^{ma} \check{t}_{ji}^{be} - \frac{1}{2} \sum_n W_{ki}^{mn} \check{t}_{nj}^{ab} + \frac{1}{2} \sum_{efn} \tilde{v}_{ef}^{mn} \check{t}_{\check{n}ijk}^{\check{f}abe} \bigg)
    \end{aligned}
    \end{equation}
    \vspace{-15pt}
    \begin{align}
        \widetilde{W}_{ei}^{ma} &= \tilde{v}^{ma}_{ei} + \frac{1}{2} \sum_{nf} \tilde{v}^{\check{n}m}_{\check{f}e} \check{t}_{\check{n}i}^{\check{f}a} - \frac{1}{2} \sum_{nf} \tilde{v}^{nm}_{ef} \check{t}_{in}^{fa}
    \end{align}
\end{subequations}

\subsection{(Q) energy correction\label{appendix:rccsdt_q}}

Within the non-orthogonal spin-adapted CC framework, the RCCSDT(Q) perturbative correction is evaluated from the converged RCCSDT amplitudes without iterating the quadruples equations. The energy correction is
\begin{equation}
    E_{\text{(Q)}} = \frac{1}{24} \sum_{ijkl} \sum_{abcd} z_{ijkl}^{abcd} \check{t}_{\check{i}\check{j}\check{k}\check{l}}^{\check{a}\check{b}\check{c}\check{d}}.
    \label{eq:ener-pt-q}
\end{equation}
Here, the non-iterative quadruples amplitudes $\check{t}_{ijkl}^{abcd}$ are obtained by dividing the residual $\check{r}_{ijkl}^{abcd}$ by the corresponding orbital-energy denominator [see Eq.~\eqref{eq:orb-energy-deno}], i.e., $\check{t}_{ijkl}^{abcd} = \check{r}_{ijkl}^{abcd} / \epsilon_{ijkl}^{abcd}$, where the residual $\check{r}_{ijkl}^{abcd}$ is
\begin{equation}
    \check{r}_{ijkl}^{abcd} = \mathcal{P}_{(ia)(jb)(kc)(ld)} \big( X_{ijkl}^{abcd} + Y_{ijkl}^{abcd} \big)
    \label{eq:resi-pt-q}
\end{equation}
The fully spin-summed amplitude entering the energy expression is obtained from the spin-free amplitudes as
\begin{equation}
\begin{aligned}
    \check{t}_{\check{i}\check{j}\check{k}\check{l}}^{\check{a}\check{b}\check{c}\check{d}} &= 2  \check{t}_{\check{i}\check{j}\check{k}l}^{\check{a}\check{b}\check{c}d} = 2 \big(2 - P_c^d \big) \check{t}_{\check{i}\check{j}kl}^{\check{a}\check{b}cd} \\
\end{aligned}
\end{equation}
The energy-weighting quantity $z_{ijkl}^{abcd}$ is
\begin{equation}
    z_{ijkl}^{abcd} = \mathcal{P}_{(ia)(jb)(kc)(ld)} \big( X_{ijkl}^{abcd} + V_{ijkl}^{abcd} \big).
\end{equation}
The required intermediates are
\begin{subequations}
\begin{align}
    X_{ijkl}^{abcd} &= \frac{1}{2} \sum_e v^{ab}_{ej} \check{t}_{ikl}^{ecd} - \frac{1}{2} \sum_m v^{am}_{ij} \check{t}_{mkl}^{bcd}, \label{eq:q_t3_terms} \\
    Y_{ijkl}^{abcd} &= \frac{1}{2} \sum_e \overline{W}_{ejk}^{abc} \check{t}_{il}^{ed} - \frac{1}{2} \sum_m \overline{W}_{ijk}^{abm} \check{t}_{ml}^{cd}, \\
    V_{ijkl}^{abcd} &= \frac{1}{4} v^{ab}_{ij} \check{t}_{kl}^{cd}, \\
    \overline{W}_{ejk}^{abc} &= \mathcal{P}_{(jb)(kc)} \bigg( \frac{1}{2} \sum_f v_{ef}^{ab} \check{t}_{jk}^{fc} \bigg),
    \label{eq:q_w_vvv_voo}
\end{align}
\vspace{-20pt}
\begin{equation}
\begin{aligned}
    \overline{W}_{ijk}^{abm} =&\ \mathcal{P}_{(ia)(jb)} \\& \bigg( \sum_e v_{ei}^{ma} \check{t}_{jk}^{be} + \sum_e v_{ke}^{ma} \check{t}_{ji}^{be} - \frac{1}{2} \sum_n v_{ki}^{mn} \check{t}_{nj}^{ab} \bigg).
    \label{eq:q_w_vvo_ooo}
\end{aligned}
\end{equation}
\end{subequations}
Here, $v$ denotes the bare ERIs rather than the $T_1$-dressed integrals $\tilde{v}$ used in the iterative residual equations. The last two intermediates can be defined in an alternative way:
\begin{subequations}
\begin{equation}
\begin{aligned}
    \overline{W}_{ejk}^{abc} &= \mathcal{P}_{(jb)(kc)} \\& \bigg(- \sum_m v_{ej}^{mb} \check{t}_{mk}^{ac} - \sum_m v_{je}^{ma} \check{t}_{mk}^{bc} + \frac{1}{2} \sum_f v_{ef}^{ab} \check{t}_{jk}^{fc} \bigg),
\end{aligned}
\end{equation}
\vspace{-20pt}
\begin{equation}
\begin{aligned}
   \overline{W}_{ijk}^{abm} &= \mathcal{P}_{(ia)(jb)} \bigg( - \frac{1}{2} \sum_n v_{ki}^{mn} \check{t}_{nj}^{ab} \bigg).
\end{aligned}
\end{equation}
\label{eq:q_w_alter}
\end{subequations}

\subsection{UCCSDT equations\label{appendix:uccsdt}}

This section collects the $T_1$-dressed residual equations for UCCSDT. Unbarred indices $i,j,k,\ldots$ ($a,b,c,\ldots$) denote $\alpha$-spin occupied (virtual) orbitals; barred indices $\bar{i},\bar{j},\bar{k},\ldots$ ($\bar{a},\bar{b},\bar{c},\ldots$) denote $\beta$-spin orbitals. For each spin channel $\sigma$, the singles amplitude matrix $\mathbf{t}_{1\sigma}$ and the associated orbital rotation matrices $\mathbf{x}_\sigma$, $\mathbf{y}_\sigma$ are defined as in Appendix~\ref{sec:t1dress}: $\mathbf{t}_{1\sigma}$ is the matrix with the following form:
\begin{equation}
    \mathbf{t}_{1\alpha} = \begin{bmatrix}
        \mathbf{0}^{\mathrm{oo}} & \mathbf{0}^{\mathrm{ov}} \\
        \mathbf{t}_{1\alpha}^{\mathrm{vo}} & \mathbf{0}^{\mathrm{vv}} \\
    \end{bmatrix},\quad
    \mathbf{t}_{1\beta} = \begin{bmatrix}
        \mathbf{0}^{\overline{\mathrm{oo}}} & \mathbf{0}^{\overline{\mathrm{ov}}} \\
        \mathbf{t}_{1\beta}^{\overline{\mathrm{vo}}} & \mathbf{0}^{\overline{\mathrm{vv}}} \\
    \end{bmatrix},
\end{equation}
together with
\begin{equation}
    \mathbf{x}_\sigma = \mathbf{I} - \mathbf{t}_{1\sigma}, \quad \mathbf{y}_\sigma = \mathbf{I} + \mathbf{t}_{1\sigma}^T.
\end{equation}

The same-spin antisymmetrized ERIs and the opposite-spin ERIs ($\sigma \neq \sigma'$) are, respectively,
\begin{equation}
    v^{p_{\sigma} q_{\sigma}}_{r_{\sigma} s_{\sigma}} \equiv \langle p_{\sigma} q_{\sigma} \| r_{\sigma} s_{\sigma} \rangle
\end{equation}
and
\begin{equation}
    v^{p_{\sigma} q_{\sigma'}}_{r_{\sigma} s_{\sigma'}} \equiv \langle p_{\sigma} q_{\sigma'} | r_{\sigma} s_{\sigma'} \rangle.
\end{equation}
The $T_1$-dressed ERIs and Fock matrix for spin channel $\sigma$ are
\begin{equation}
    \tilde{v}^{p_{\sigma} q_{\sigma'}}_{r_{\sigma} s_{\sigma'}} = \sum_{tuvw} x_{pt,\sigma} x_{qu,\sigma'} v^{t_{\sigma} u_{\sigma'}}_{v_{\sigma} w_{\sigma'}} y_{rv,\sigma} y_{sw,\sigma'} 
\end{equation}
and
\begin{equation}
\begin{aligned}
    \tilde{f}^{p_\sigma}_{q_\sigma} =& \sum_{rs} x_{pr,\sigma} \bigg[ f^{r_\sigma}_{s_\sigma} + \sum_{ia} \left( v^{r_\sigma i_\sigma}_{s_\sigma a_\sigma} - v^{r_\sigma i_\sigma}_{a_\sigma s_\sigma} \right) t_{i_\sigma}^{a_\sigma} \\
    & + \sum_{ia} v^{r_\sigma i_{\sigma'}}_{s_\sigma a_{\sigma'}}  t_{i_{\sigma'}}^{a_{\sigma'}} \bigg] y_{qs,\sigma},
\end{aligned}
\end{equation}
where the bare Fock matrix elements are
\begin{equation}
    f^{r_\sigma}_{s_\sigma} = h^{r_\sigma}_{s_\sigma} + \sum_i v^{r_\sigma i_\sigma}_{s_\sigma i_\sigma} + \sum_i v^{r_\sigma i_{\sigma'}}_{s_\sigma i_{\sigma'}} - \sum_i v^{r_\sigma i_\sigma}_{i_\sigma s_\sigma}.
\end{equation}

The spin-unrestricted correlation energy is
\begin{equation}
\begin{aligned}
    \Delta E & = \frac{1}{4} \sum_{ijab} v^{ij}_{ab} \left( t_{ij}^{ab} + t_i^a t_j^b - t_i^b t_j^a \right) \\
    &+ \frac{1}{4} \sum_{\overline{i}\overline{j}\overline{a}\overline{b}} v^{\overline{i}\overline{j}}_{\overline{a}\overline{b}} \left( t_{\overline{i}\overline{j}}^{\overline{a}\overline{b}} + t_{\overline{i}}^{\overline{a}} t_{\overline{j}}^{\overline{b}} - t_{\overline{i}}^{\overline{b}} t_{\overline{j}}^{\overline{a}} \right) \\
    &+ \sum_{i\overline{j}a\overline{b}} v^{i\overline{j}}_{a\overline{b}} \left( t_{i\overline{j}}^{a\overline{b}} + t_{i}^{a} t_{\overline{j}}^{\overline{b}} \right) + \sum_{ia} f^i_{a} t_i^a + \sum_{\overline{i}\overline{a}} f^{\overline{i}}_{\overline{a}} t_{\overline{i}}^{\overline{a}} \\
\end{aligned}
\end{equation}

The $\alpha$-spin singles residual is
\begin{equation}
\begin{aligned}
    & r_i^a (\mathrm{CCSD}) = \tilde{f}^a_i + \sum_{kc} \tilde{f}^k_{c} t_{ik}^{ac} + \sum_{\overline{k}\overline{c}} \tilde{f}^{\overline{k}}_{\overline{c}} t_{i\overline{k}}^{a\overline{c}} \\
    &+ \frac{1}{2} \sum_{kcd} \tilde{v}^{ak}_{cd} t_{ik}^{cd} + \sum_{\overline{k}c\overline{d}} \tilde{v}^{a\overline{k}}_{c\overline{d}} t_{i\overline{k}}^{c\overline{d}} - \frac{1}{2} \sum_{klc} \tilde{v}^{kl}_{ic} t_{kl}^{ac} - \sum_{k\overline{l}\overline{c}} \tilde{v}^{k\overline{l}}_{i\overline{c}}  t_{k\overline{l}}^{a\overline{c}}
\end{aligned}
\end{equation}
The expression for the $\beta$-spin singles residual $r_{\overline{i}}^{\overline{a}}$ is obtained by interchanging barred and unbarred indices throughout the expression above.

The $\alpha\alpha$ doubles residual is
\begin{equation}
\begin{aligned}
    & r_{ij}^{ab} (\mathrm{CCSD}) = \mathcal{A}^{ab} \mathcal{A}_{ij} \bigg(\frac{1}{4} \tilde{v}^{ab}_{ij} + \frac{1}{2} \sum_c F^{b}_{c} t_{ij}^{ac} \\
    &- \frac{1}{2} \sum_k F^k_j t_{ik}^{ab} + \frac{1}{8} \sum_{kl} W^{kl}_{ij} t_{kl}^{ab} + \frac{1}{8} \sum_{cd} \tilde{v}^{ab}_{cd} t_{ij}^{cd} \\
    &+ \sum_{kc} W^{bk}_{jc} t_{ik}^{ac} + \sum_{\overline{k}\overline{c}} W^{b\overline{k}}_{j\overline{c}} t_{i\overline{k}}^{a\overline{c}} \bigg)
\end{aligned}
\end{equation}
where the $\alpha\alpha$ intermediates are defined below:
\begin{subequations}
\begin{align}
    F^b_c &= \tilde{f}^b_c- \frac{1}{2} \sum_{kld} \tilde{v}^{kl}_{cd} t_{kl}^{bd} - \sum_{l\overline{k}\overline{d}} \tilde{v}^{l\overline{k}}_{c\overline{d}}  t_{l\overline{k}}^{b\overline{d}}, \\
    F^k_{j} &= \tilde{f}^k_{j} + \frac{1}{2} \sum_{lcd} \tilde{v}^{kl}_{cd} t_{jl}^{cd} + \sum_{d\overline{l}\overline{c}} \tilde{v}^{k\overline{l}}_{d\overline{c}} t_{j\overline{l}}^{d\overline{c}}, \\
    W^{kl}_{ij} &= \tilde{v}^{kl}_{ij} + \frac{1}{2} \sum_{cd} \tilde{v}^{kl}_{cd} t_{ij}^{cd}, \\
    W^{bk}_{jc} &= \tilde{v}^{bk}_{jc} + \frac{1}{2} \sum_{ld} \tilde{v}^{kl}_{cd} t_{jl}^{bd} + \frac{1}{2} \sum_{\overline{l}\overline{d}} \tilde{v}^{k\overline{l}}_{c\overline{d}} t_{j\overline{l}}^{b\overline{d}}, \\
    W^{b\overline{k}}_{j\overline{c}} &= \tilde{v}^{b\overline{k}}_{j\overline{c}} + \frac{1}{2} \sum_{ld} \tilde{v}^{l\overline{k}}_{d\overline{c}} t_{jl}^{bd}  + \frac{1}{2} \sum_{\overline{l}\overline{d}} \tilde{v}^{\overline{k}\overline{l}}_{\overline{c}\overline{d}} t_{j\overline{l}}^{b\overline{d}}.
\end{align}
\end{subequations}

The expressions for the $\beta\beta$ doubles residual $r_{\overline{i}\overline{j}}^{\overline{a}\overline{b}}$ and the corresponding intermediates are obtained by interchanging barred and unbarred indices throughout the $\alpha\alpha$ expressions. The $\alpha\beta$ mixed-spin doubles residual is
\begin{equation}
\begin{aligned}
    &r^{a\overline{b}}_{i\overline{j}} (\mathrm{CCSD}) = \tilde{v}^{a\overline{b}}_{i\overline{j}} + \sum_c F^a_{c} t_{i\overline{j}}^{c\overline{b}} + \sum_{\overline{c}} F^{\overline{b}}_{\overline{c}} t_{i\overline{j}}^{a\overline{c}}\\
    & - \sum_k F^k_i t_{k\overline{j}}^{a\overline{b}}  - \sum_{\overline{k}} F^{\overline{k}}_{\overline{j}} t_{i\overline{k}}^{a\overline{b}} + \sum_{c\overline{d}} \tilde{v}^{a\overline{b}}_{c\overline{d}} t_{i\overline{j}}^{c\overline{d}} + \sum_{k\overline{l}} W^{k\overline{l}}_{i\overline{j}} t_{k\overline{l}}^{a\overline{b}} \\
    &- \sum_{kc} W^{ak}_{ci} t_{k\overline{j}}^{c\overline{b}} - \sum_{\overline{k}c} W^{a\overline{k}}_{c\overline{j}} t_{i\overline{k}}^{c\overline{b}} + \sum_{\overline{k}\overline{c}} W^{a\overline{k}}_{i\overline{c}} t_{\overline{k}\overline{j}}^{\overline{c}\overline{b}} \\
    & - \sum_{\overline{k}\overline{c}} W^{\overline{b}\overline{k}}_{\overline{c}\overline{j}} t_{i\overline{k}}^{a\overline{c}} - \sum_{k\overline{c}} W^{k\overline{b}}_{i\overline{c}} t_{k\overline{j}}^{a\overline{c}} + \sum_{kc} W^{k\overline{b}}_{c\overline{j}} t_{ik}^{ac}
\end{aligned}
\end{equation}
where the additional mixed-spin intermediates are
\begin{subequations}
\begin{align}
    W^{k\overline{l}}_{i\overline{j}} &= \tilde{v}^{k \overline{l}}_{i\overline{j}} + \sum_{c\overline{d}} \tilde{v}^{k \overline{l}}_{c \overline{d}} t_{i\overline{j}}^{c\overline{d}} \\
    W^{ak}_{ci} &= \tilde{v}^{ak}_{ci} - \frac{1}{2} \sum_{ld} \tilde{v}^{kl}_{cd} t_{il}^{ad} - \frac{1}{2} \sum_{\overline{l}\overline{d}} \tilde{v}^{k \overline{l}}_{c \overline{d}} t_{i\overline{l}}^{a\overline{d}} \\
    W^{a\overline{k}}_{c\overline{j}} &= \tilde{v}^{a \overline{k}}_{c\overline{j}} - \frac{1}{2} \sum_{l\overline{d}} \tilde{v}^{l\overline{k}}_{c\overline{d}}  t_{l\overline{j}}^{a\overline{d}} \\
    W^{\overline{b}\overline{k}}_{\overline{c}\overline{j}} &=  \tilde{v}^{\overline{b}\overline{k}}_{\overline{c}\overline{j}} - \frac{1}{2} \sum_{\overline{l}\overline{d}} \tilde{v}^{\overline{k}\overline{l}}_{\overline{c}\overline{d}} t_{\overline{j}\overline{l}}^{\overline{b}\overline{d}} - \frac{1}{2} \sum_{ld} \tilde{v}^{l \overline{k}}_{d \overline{c}} t_{l\overline{j}}^{d\overline{b}} \\
    W^{k\overline{b}}_{i\overline{c}} &= \tilde{v}^{k\overline{b}}_{i\overline{c}} - \frac{1}{2} \sum_{\overline{l}d} \tilde{v}^{k\overline{l}}_{d \overline{c}} t_{i\overline{l}}^{d\overline{b}} 
\end{align}
\end{subequations}

The CCSDT corrections add $T_3$-dependent terms to the singles, doubles, and triples residuals. The correction to the singles residual is
\begin{equation}
\begin{aligned}
    &r_i^a (\mathrm{CCSDT}) = r_i^a (\mathrm{CCSD}) + \frac{1}{4} \sum_{mnef} \tilde{v}^{mn}_{ef} t_{imn}^{aef} \\
    &+ \sum_{m\overline{n}e\overline{f}} \tilde{v}^{m\overline{n}}_{e\overline{f}} t_{im\overline{n}}^{ae\overline{f}} + \frac{1}{4} \sum_{\overline{m}\overline{n}\overline{e}\overline{f}} \tilde{v}^{\overline{m}\overline{n}}_{\overline{e}\overline{f}} t_{i\overline{m}\overline{n}}^{a\overline{e}\overline{f}}.
\end{aligned}
\end{equation}
The corrections to the doubles residual are
\begin{equation}
\begin{aligned}
    &r_{ij}^{ab} (\mathrm{CCSDT}) = r_{ij}^{ab} (\mathrm{CCSD}) + \mathcal{A}^{ab} \mathcal{A}_{ij} \left( \frac{1}{4} \sum_{me} \tilde{f}^m_{e} t_{ijm}^{abe} \right.\\
    &+ \frac{1}{4} \sum_{\overline{me}} \tilde{f}^{\overline{m}}_{\overline{e}} t_{ij\overline{m}}^{ab\overline{e}} + \frac{1}{4} \sum_{mef} \tilde{v}^{bm}_{ef} t_{ijm}^{aef} + \frac{1}{2} \sum_{\overline{m}e\overline{f}} \tilde{v}^{b\overline{m}}_{e\overline{f}} t_{ij\overline{m}}^{ae\overline{f}} \\
    & \left. - \frac{1}{4} \sum_{mne} \tilde{v}^{mn}_{je} t_{imn}^{abe} - \frac{1}{2} \sum_{m\overline{ne}} \tilde{v}^{m\overline{n}}_{j\overline{e}} t_{im\overline{n}}^{ab\overline{e}} \right) \\
\end{aligned}
\end{equation}
and
\begin{equation}
\begin{aligned}
    &r_{i\overline{j}}^{a\overline{b}} (\mathrm{CCSDT}) = r_{i\overline{j}}^{a\overline{b}} (\mathrm{CCSD}) + \sum_{me} \tilde{f}^m_{e} t_{im\overline{j}}^{ae\overline{b}} + \sum_{\overline{m}\overline{e}} \tilde{f}^{\overline{m}}_{\overline{e}} t_{i\overline{j}\overline{m}}^{a\overline{b}\overline{e}} \\
    &+ \sum_{m\overline{e}f} \tilde{v}^{m\overline{b}}_{f\overline{e}} t_{im\overline{j}}^{af\overline{e}} + \frac{1}{2} \sum_{\overline{m}\overline{e}\overline{f}} \tilde{v}^{\overline{b}\overline{m}}_{\overline{e}\overline{f}} t_{i\overline{j}\overline{m}}^{a\overline{e}\overline{f}} + \frac{1}{2} \sum_{mef} \tilde{v}^{am}_{ef} t_{im\overline{j}}^{ef\overline{b}} \\
    &+ \sum_{\overline{m}e\overline{f}} \tilde{v}^{a\overline{m}}_{e \overline{f}} t_{i\overline{j}\overline{m}}^{e\overline{b}\overline{f}} - \sum_{\overline{m}ne} \tilde{v}^{n\overline{m}}_{e\overline{j}} t_{in\overline{m}}^{ae\overline{b}} - \frac{1}{2} \sum_{\overline{m}\overline{n}\overline{e}} \tilde{v}^{\overline{m}\overline{n}}_{\overline{j}\overline{e}} t_{i\overline{m}\overline{n}}^{a\overline{b}\overline{e}} \\
    &- \frac{1}{2} \sum_{mne} \tilde{v}^{mn}_{ie} t_{mn\overline{j}}^{ae\overline{b}} - \sum_{m\overline{n}\overline{e}} \tilde{v}^{m\overline{n}}_{i \overline{e}} t_{m\overline{j}\overline{n}}^{a\overline{b}\overline{e}}. \\
\end{aligned}
\end{equation}
The $\alpha\alpha\alpha$ same-spin triples residual is
\begin{equation}
\begin{aligned}
    &r^{abc}_{ijk} (\mathrm{CCSDT}) = \mathcal{A}^{abc} \mathcal{A}_{ijk} \bigg( \frac{1}{4} \sum_d W^{bc}_{dk} t_{ij}^{ad} - \frac{1}{4} \sum_l W^{lc}_{jk} t_{il}^{ab} \\
    &+ \frac{1}{12} \sum_d F^c_d t_{ijk}^{abd} - \frac{1}{12} \sum_l F^l_k t_{ijl}^{abc} + \frac{1}{24} \sum_{de} W^{ab}_{de} t_{ijk}^{dec} \\
    &+ \frac{1}{24} \sum_{lm} W^{lm}_{ij} t_{lmk}^{abc} + \frac{1}{4} \sum_{ld} \overline{W}^{al}_{id} t_{ljk}^{dbc} + \frac{1}{4} \sum_{\overline{ld}} \overline{W}^{a\overline{l}}_{i\overline{d}} t_{jk\overline{l}}^{bc\overline{d}} \bigg)
\end{aligned}
\end{equation}
where the $\alpha\alpha\alpha$ intermediates are
\begin{subequations}
\begin{align}
    W^{ab}_{de} &= \tilde{v}^{ab}_{de} + \frac{1}{2} \sum_{lm} \tilde{v}^{lm}_{de} t^{ab}_{lm},\\
    \overline{W}^{al}_{id} &= W^{al}_{id} + \frac{1}{2} \sum_{me} \tilde{v}^{ml}_{ed} t_{im}^{ae} + \frac{1}{2} \sum_{\overline{m}\overline{e}} \tilde{v}^{l\overline{m}}_{d\overline{e}} t_{i\overline{m}}^{a\overline{e}}, \\
    \overline{W}^{a\overline{l}}_{i\overline{d}} &= W^{a\overline{l}}_{i\overline{d}} + \frac{1}{2} \sum_{me} \tilde{v}^{m\overline{l}}_{e\overline{d}} t_{im}^{ae} + \frac{1}{2} \sum_{\overline{me}} \tilde{v}^{\overline{m}\overline{l}}_{\overline{e}\overline{d}} t_{i\overline{m}}^{a\overline{e}}
    \end{align}
    \vspace{-20pt}
    \begin{equation}
    \begin{aligned}
        W^{bc}_{dk} =&\ \tilde{v}^{bc}_{dk} + 2 \sum_{el} \tilde{v}^{lb}_{ed} t_{kl}^{ce} + 2 \sum_{\overline{el}} \tilde{v}^{b\overline{l}}_{d\overline{e}}  t_{k\overline{l}}^{c\overline{e}} \\
        &+ \frac{1}{2} \sum_{lm} \tilde{v}^{lm}_{dk} t_{lm}^{bc} - \frac{1}{2} \sum_{lme} \tilde{v}^{lm}_{de} t_{lmk}^{bec} - \sum_{l\overline{m}\overline{e}} \tilde{v}^{l\overline{m}}_{d\overline{e}}  t_{lk\overline{m}}^{bc\overline{e}}
    \end{aligned}
    \end{equation}
    \begin{equation}
    \begin{aligned}
        W^{lc}_{jk} =&\ \tilde{v}^{lc}_{jk} + \sum_{d} \tilde{f}^l_{d} t_{jk}^{dc} + 2 \sum_{dm} \tilde{v}^{ml}_{dj} t_{km}^{cd} + 2 \sum_{\overline{d}\overline{m}} \tilde{v}^{l\overline{m}}_{j\overline{d}} t_{k\overline{m}}^{c\overline{d}} \\
        &+ \frac{1}{2} \sum_{de} \tilde{v}^{lc}_{de} t_{jk}^{de} + \frac{1}{2} \sum_{mde} \tilde{v}^{lm}_{de} t_{jmk}^{dec} + \sum_{\overline{m}d\overline{e}} \tilde{v}^{l\overline{m}}_{d\overline{e}} t_{jk\overline{m}}^{dc\overline{e}}
    \end{aligned}
    \end{equation}
\end{subequations}
The expressions for the all-$\beta$ triples residual $r_{\overline{i}\overline{j}\overline{k}}^{\overline{a}\overline{b}\overline{c}}$ and the corresponding intermediates are obtained by interchanging barred and unbarred indices throughout the all-$\alpha$ expressions. The $\alpha\alpha\beta$ mixed-spin triples residual is
\begin{equation}
\begin{aligned}
    &r_{ij\overline{k}}^{ab\overline{c}} (\mathrm{CCSDT}) = \mathcal{A}^{ab} \mathcal{A}_{ij} \bigg( \frac{1}{2} \sum_d W^{b\overline{c}}_{d\overline{k}} t_{ij}^{ad} + \sum_{\overline{d}} W^{b\overline{c}}_{j\overline{d}} t_{i\overline{k}}^{a\overline{d}} \\
    &- \frac{1}{2} \sum_d W^{ab}_{di} t_{j\overline{k}}^{d\overline{c}} - \frac{1}{2} \sum_l W^{l\overline{c}}_{j\overline{k}} t_{il}^{ab} + \sum_{\overline{l}} W^{a\overline{l}}_{j\overline{k}} t_{i\overline{l}}^{b\overline{c}} \\
    &+ \frac{1}{2} \sum_l W^{la}_{ij} t_{l\overline{k}}^{b\overline{c}} + \frac{1}{4} \sum_{\overline{d}} F^{\overline{c}}_{\overline{d}} t_{ij\overline{k}}^{ab\overline{d}} - \frac{1}{2} \sum_d F^a_d t_{ij\overline{k}}^{bd\overline{c}} \\
    &- \frac{1}{4} \sum_l F^{\overline{l}}_{\overline{k}} t_{ij\overline{l}}^{ab\overline{c}} + \frac{1}{2} \sum_l F^l_i t_{jl\overline{k}}^{ab\overline{c}} + \frac{1}{8} \sum_{de} W^{ab}_{de} t_{ij\overline{k}}^{de\overline{c}} \\
    &+ \frac{1}{2} \sum_{\overline{d}e} W^{b\overline{c}}_{e\overline{d}} t_{ij\overline{k}}^{ae\overline{d}} + \frac{1}{8} \sum_{lm} W^{lm}_{ij} t_{lm\overline{k}}^{ab\overline{c}} + \frac{1}{2} \sum_{l\overline{m}} W^{l\overline{m}}_{i\overline{k}} t_{lj\overline{m}}^{ab\overline{c}} \\
    &+ \sum_{ld} W^{al}_{id} t_{lj\overline{k}}^{db\overline{c}} + \sum_{\overline{l}\overline{d}} W^{a\overline{l}}_{i\overline{d}} t_{j\overline{l}\overline{k}}^{b\overline{d}\overline{c}} - \frac{1}{2} \sum_{l\overline{d}} \overline{W}^{l\overline{c}}_{i\overline{d}} t_{lj\overline{k}}^{ab\overline{d}} \\
    &- \frac{1}{2} \sum_{\overline{l}d} \overline{W}^{a\overline{l}}_{d\overline{k}} t_{ij\overline{l}}^{db\overline{c}} + \frac{1}{4} \sum_{ld} W^{l\overline{c}}_{d\overline{k}} t_{ijl}^{abd} + \frac{1}{4} \sum_{\overline{ld}} W^{\overline{c}\overline{l}}_{\overline{k}\overline{d}} t_{ij\overline{l}}^{ab\overline{d}} \bigg)\\
\end{aligned}
\end{equation}
The additional mixed-spin intermediates required for $r_{ij\overline{k}}^{ab\overline{c}}$ are
\begin{subequations}
\begin{align}
    &W^{b\overline{c}}_{e\overline{d}} = \tilde{v}^{b\overline{c}}_{e\overline{d}} + \sum_{l\overline{m}} \tilde{v}^{l\overline{m}}_{e\overline{d}} t_{l\overline{m}}^{b\overline{c}}, \\
    &\overline{W}^{l\overline{c}}_{i\overline{d}} = W^{l\overline{c}}_{i\overline{d}} - \frac{1}{2} \sum_{\overline{m} e} \tilde{v}^{l\overline{m}}_{e\overline{d}} t_{i\overline{m}}^{e\overline{c}}, \\
    &\overline{W}^{a\overline{l}}_{d\overline{k}} = W^{a\overline{l}}_{d\overline{k}} - \frac{1}{2} \sum_{m\overline{e}} \tilde{v}^{m\overline{l}}_{d\overline{e}} t_{m\overline{k}}^{a\overline{e}},
\end{align}
\vspace{-5mm}
\begin{equation}
\begin{aligned}
    &W^{b\overline{c}}_{d\overline{k}} = \tilde{v}^{b\overline{c}}_{d\overline{k}} + \sum_{el} \tilde{v}^{lb}_{ed} t_{l\overline{k}}^{e\overline{c}} + \sum_{\overline{e}\overline{l}} \tilde{v}^{b\overline{l}}_{d\overline{e}} t_{\overline{l}\overline{k}}^{\overline{e}\overline{c}} - \sum_{\overline{e}l} \tilde{v}^{l \overline{c}}_{d \overline{e}} t_{l\overline{k}}^{b\overline{e}} \\
    &+ \sum_{l\overline{m}} \tilde{v}^{l \overline{m}}_{d \overline{k}} t_{l\overline{m}}^{b\overline{c}}  - \frac{1}{2} \sum_{lme} \tilde{v}^{lm}_{de} t_{lm\overline{k}}^{be\overline{c}} -  \sum_{l\overline{m}\overline{e}} \tilde{v}^{l\overline{m}}_{d\overline{e}} t_{l\overline{m}\overline{k}}^{b\overline{e}\overline{c}}, \\
\end{aligned}
\end{equation}
\begin{equation}
\begin{aligned}
    &W^{l\overline{c}}_{j\overline{k}} = \tilde{v}^{l\overline{c}}_{j\overline{k}} + \sum_{d} \tilde{f}^l_d t_{j\overline{k}}^{d\overline{c}} +  \sum_{dm} \tilde{v}^{ml}_{dj} t_{m\overline{k}}^{d\overline{c}} + \sum_{\overline{d}\overline{m}} \tilde{v}^{l\overline{m}}_{j\overline{d}} t_{\overline{m}\overline{k}}^{\overline{d}\overline{c}} \\
    &- \sum_{d\overline{m}} \tilde{v}^{l\overline{m}}_{d\overline{k}} t_{j\overline{m}}^{d\overline{c}} + \sum_{d\overline{e}} \tilde{v}^{l\overline{c}}_{d \overline{e}} t_{j\overline{k}}^{d\overline{e}} + \frac{1}{2} \sum_{mde} \tilde{v}^{lm}_{de} t_{jm\overline{k}}^{de\overline{c}} + \sum_{\overline{m}d\overline{e}} \tilde{v}^{l\overline{m}}_{d\overline{e}} t_{j\overline{m}\overline{k}}^{d\overline{e}\overline{c}}, \\
\end{aligned}
\end{equation}
\begin{equation}
\begin{aligned}
    &W^{b\overline{c}}_{j\overline{d}} = \tilde{v}^{b\overline{c}}_{j\overline{d}} - \sum_{e\overline{l}} \tilde{v}^{b\overline{l}}_{e\overline{d}} t_{j\overline{l}}^{e\overline{c}} + \sum_{el} \tilde{v}^{l\overline{c}}_{e\overline{d}} t_{jl}^{be} + \sum_{\overline{el}} \tilde{v}^{\overline{l}\overline{c}}_{\overline{e}\overline{d}} t_{j\overline{l}}^{b\overline{e}} \\
    &+ \sum_{\overline{l}m} \tilde{v}^{m\overline{l}}_{j\overline{d}} t_{m\overline{l}}^{b\overline{c}} - \sum_{\overline{l}me} \tilde{v}^{m\overline{l}}_{e\overline{d}} t_{jm\overline{l}}^{be\overline{c}} - \frac{1}{2} \sum_{\overline{l}\overline{m}\overline{e}} \tilde{v}^{\overline{l}\overline{m}}_{\overline{d}\overline{e}} t_{j\overline{m}\overline{l}}^{b\overline{e}\overline{c}},\\
\end{aligned}
\end{equation}
\begin{equation}
\begin{aligned}
    &W^{a\overline{l}}_{j\overline{k}} = \tilde{v}^{a\overline{l}}_{j\overline{k}} + \sum_{\overline{d}} \tilde{f}^{\overline{l}}_{\overline{d}} t_{j\overline{k}}^{a\overline{d}} - \sum_{\overline{d}m} \tilde{v}^{m\overline{l}}_{j \overline{d}} t_{m\overline{k}}^{a\overline{d}} + \sum_{dm} \tilde{v}^{m\overline{l}}_{d\overline{k}} t_{jm}^{ad} \\
    &+ \sum_{\overline{d}\overline{m}} \tilde{v}^{\overline{m}\overline{l}}_{\overline{d}\overline{k}} t_{j\overline{m}}^{a\overline{d}} +  \sum_{d\overline{e}} \tilde{v}^{a\overline{l}}_{d\overline{e}} t_{j\overline{k}}^{d\overline{e}} + \sum_{m\overline{d}e} \tilde{v}^{m\overline{l}}_{e\overline{d}} t_{jm\overline{k}}^{ae\overline{d}} + \frac{1}{2} \sum_{\overline{m}\overline{d}\overline{e}} \tilde{v}^{\overline{l}\overline{m}}_{\overline{d}\overline{e}} t_{j\overline{m}\overline{k}}^{a\overline{e}\overline{d}} \\
\end{aligned}
\end{equation}
\end{subequations}

The expression for the $\alpha\beta\beta$ triples residual $r_{i\overline{j}\overline{k}}^{a\overline{b}\overline{c}}$ is obtained from the $\alpha\alpha\beta$ residual $r_{ij\overline{k}}^{ab\overline{c}}$ by interchanging barred and unbarred indices.

The leading $O(N^8)$ operation counts for the UCCSDT residual contractions are as follows. For the full $T_3$ implementation:
\begin{subequations}
\begin{align}
    r_{ijk}^{abc}: &N_{\mathrm{o}}^3 N_{\mathrm{v}}^5 + N_{\mathrm{o}}^4 N_{\mathrm{v}}^4 + N_{\mathrm{o}}^3 N_{\mathrm{v}}^3 N_{\bar{\mathrm{o}}} N_{\bar{\mathrm{v}}} + N_{\mathrm{o}}^5 N_{\mathrm{v}}^3 \nonumber
\end{align}
\begin{equation}
\begin{aligned}
    r_{ij\overline{k}}^{ab\overline{c}}: & N_{\mathrm{o}}^2 N_{\mathrm{v}}^4 N_{\bar{\mathrm{o}}} N_{\bar{\mathrm{v}}} + N_{\mathrm{o}}^2 N_{\mathrm{v}}^3 N_{\bar{\mathrm{o}}} N_{\bar{\mathrm{v}}}^2 + 2 N_{\mathrm{o}}^3 N_{\mathrm{v}}^3 N_{\bar{\mathrm{o}}} N_{\bar{\mathrm{v}}} \\
    & + 2 N_{\mathrm{o}}^2 N_{\mathrm{v}}^2 N_{\bar{\mathrm{o}}}^2 N_{\bar{\mathrm{v}}}^2 + N_{\mathrm{o}}^3 N_{\mathrm{v}}^2 N_{\bar{\mathrm{o}}} N_{\bar{\mathrm{v}}}^2 + N_{\mathrm{o}}^2 N_{\mathrm{v}}^3 N_{\bar{\mathrm{o}}}^2 N_{\bar{\mathrm{v}}} \\
    &+ N_{\mathrm{o}}^4 N_{\mathrm{v}}^2 N_{\bar{\mathrm{o}}} N_{\bar{\mathrm{v}}} + N_{\mathrm{o}}^3 N_{\mathrm{v}}^2 N_{\bar{\mathrm{o}}}^2 N_{\bar{\mathrm{v}}} \nonumber
\end{aligned}
\end{equation}
\end{subequations}
For the compact triangular $T_3$ implementation considered here, using the large-system approximations $N_{\mathrm{o}}(N_{\mathrm{o}}-1)(N_{\mathrm{o}}-2)/N_{\mathrm{o}}^3 \approx 1$, $N_{\mathrm{o}}(N_{\mathrm{o}}-1)/N_{\mathrm{o}}^2 \approx 1$ (and similarly for $N_{\mathrm{v}}$), the leading $O(N^8)$ operation counts become
\begin{subequations}
\begin{align}
    r_{ijk}^{abc}: &\frac{1}{12} N_{\mathrm{o}}^3 N_{\mathrm{v}}^5 + \frac{1}{4} N_{\mathrm{o}}^4 N_{\mathrm{v}}^4 + \frac{1}{4} N_{\mathrm{o}}^3 N_{\mathrm{v}}^3 N_{\bar{\mathrm{o}}} N_{\bar{\mathrm{v}}} + \frac{1}{12} N_{\mathrm{o}}^5 N_{\mathrm{v}}^3 \nonumber,
\end{align}
\begin{equation}
\begin{aligned}
    r_{ij\overline{k}}^{ab\overline{c}}: &\frac{1}{4} N_{\mathrm{o}}^2 N_{\mathrm{v}}^4 N_{\bar{\mathrm{o}}} N_{\bar{\mathrm{v}}} + \frac{1}{2} N_{\mathrm{o}}^2 N_{\mathrm{v}}^3 N_{\bar{\mathrm{o}}} N_{\bar{\mathrm{v}}}^2 + \frac{5}{4} N_{\mathrm{o}}^3 N_{\mathrm{v}}^3 N_{\bar{\mathrm{o}}} N_{\bar{\mathrm{v}}} \\
    &+ \frac{5}{4} N_{\mathrm{o}}^2 N_{\mathrm{v}}^2 N_{\bar{\mathrm{o}}}^2 N_{\bar{\mathrm{v}}}^2 + \frac{1}{2} N_{\mathrm{o}}^3 N_{\mathrm{v}}^2 N_{\bar{\mathrm{o}}} N_{\bar{\mathrm{v}}}^2 + \frac{1}{2} N_{\mathrm{o}}^2 N_{\mathrm{v}}^3 N_{\bar{\mathrm{o}}}^2 N_{\bar{\mathrm{v}}} \\
    &+ \frac{1}{4} N_{\mathrm{o}}^4 N_{\mathrm{v}}^2 N_{\bar{\mathrm{o}}} N_{\bar{\mathrm{v}}} + \frac{1}{2} N_{\mathrm{o}}^3 N_{\mathrm{v}}^2 N_{\bar{\mathrm{o}}}^2 N_{\bar{\mathrm{v}}} \nonumber.
\end{aligned}
\end{equation}
\end{subequations}
See Table~\ref{tab:uccsdt_operation_count} for details. For a spin-restricted closed-shell system, only the $r_{ij\overline{k}}^{ab\overline{c}}$ sector is needed, giving a total compact-$T_3$ operation count of $\frac{3}{4} N_{\mathrm{o}}^3 N_{\mathrm{v}}^5 + \frac{7}{2} N_{\mathrm{o}}^4 N_{\mathrm{v}}^4 + \frac{3}{4} N_{\mathrm{o}}^5 N_{\mathrm{v}}^3$. This is more than $\frac{3}{2}$ times the RCCSDT compact-$T_3$ cost of $\frac{1}{2} N_{\mathrm{o}}^3N_{\mathrm{v}}^5 + 2N_{\mathrm{o}}^4N_{\mathrm{v}}^4 + \frac{1}{2} N_{\mathrm{o}}^5N_{\mathrm{v}}^3$.

For a closed-shell system, the $T_3$ amplitudes in the spin-adapted RCCSDT can be reconstructed from the spin-unrestricted UCCSDT amplitudes as
\begin{equation}
    \check{t}_{ijk}^{abc} = - \frac{1}{6} t_{ijk}^{abc} + \frac{1}{3} \left(t_{ij\overline{k}}^{ab\overline{c}} + t_{ik\overline{j}}^{ac\overline{b}} + t_{jk\overline{i}}^{bc\overline{a}} \right).
\end{equation}

\section{Permutation symmetry and redundant components in spin-free excitation operators
\label{appendix:nosa}}

The spin-free excitation operators defined in Eq.~\eqref{eq:spinfree_e} generate spin-free excited configurations by summing over spin cases. The resulting configurations are generally nonorthogonal, and starting at triples, different virtual-index permutations become linearly dependent: some nonzero linear combinations have zero norm with respect to the spin-free overlap metric and therefore do not represent independent spin-free excitations. After the residuals are computed, these redundant components must be removed before the amplitudes are updated, so that the amplitudes remain in the linearly independent spin-free subspace.

To describe this redundancy explicitly for a given set of occupied and virtual labels, we consider the $n!$ spin-free operators obtained by permuting only the virtual labels.
\begin{equation}
    \mathcal{E}_n = \left\{ \hat{E}_{i_1\cdots i_n}^{a_{\pi(1)} \cdots a_{\pi(n)}}: \pi \in S_n \right\},
\end{equation}
where $S_n$ is the set of all permutations of $n$ objects. Their overlap matrix is~\cite{wang2018simple}
\begin{equation}
    M_{\pi,\tau} = \Big\langle \Phi \Big| \hat{E}_{a_{\pi(1)}\cdots a_{\pi(n)}}^{i_1\cdots i_n} \hat{E}_{i_1\cdots i_n}^{a_{\tau(1)}\cdots a_{\tau(n)}} \Big| \Phi \Big\rangle = m(\pi^{-1}\tau),
\end{equation}
where $|\Phi\rangle$ is the reference determinant. Thus, the overlap between two permuted operators depends only on their relative permutation $\rho=\pi^{-1}\tau$. For the spin-free excitation operator in Eq.~\eqref{eq:spinfree_e},
\begin{equation}
m(\rho)=\mathrm{sgn}(\rho) 2^{c(\rho)} .
\label{eq:spinfree_overlap_class_function}
\end{equation}
Here, $c(\rho)$ is the number of cycles in $\rho$, including one-cycles, and $\mathrm{sgn}(\rho)$ is the sign of the permutation.

The overlap matrix can be separated into symmetry sectors of the permutation group $S_n$. We label these sectors by Young diagrams $\lambda$. If $d_\lambda$ is the dimension of the sector $\lambda$, the corresponding eigenvalue of the overlap matrix is
\begin{equation}
\label{eq:character_eigenvalue_formula}
\mu_{\lambda} =
\frac{1}{d_{\lambda}}
\sum_C |C| m(C) \chi_{\lambda}(C),
\end{equation}
where $C$ runs over the conjugacy classes of $S_n$, $|C|$ is the size of the class, and $\chi_\lambda(C)$ is the character for the sector $\lambda$. Sectors with zero eigenvalue form the redundant subspace, while sectors with nonzero eigenvalue form the nonredundant spin-free subspace.

For triples, we use the following character table for $S_3$ and the corresponding class data. Here and in the rest of this appendix, permutations are written in cycle notation, with one-cycles omitted. For example, in $S_3$, $(12)$ denotes the transposition $\pi$ with $\pi(1)=2$, $\pi(2)=1$, and $\pi(3)=3$, while $e$ denotes the identity permutation.
\begin{equation}
\begin{array}{c|ccc}
    & e & (12) & (123) \\
    \hline
    {[3]}     & 1 & 1  & 1  \\
    {[2,1]}   & 2 & 0  & -1 \\
    {[1^3]}   & 1 & -1 & 1  \\
    \hline
    |C|     & 1 & 3  & 2  \\
    m(C)    & 8 & -4 & 2
\end{array}
\end{equation}
Using Eq.~\eqref{eq:character_eigenvalue_formula}, the eigenvalues of the triple overlap matrix are
\begin{equation}
\mu_{[3]} = 0, \quad
\mu_{[2,1]} = 6, \quad
\mu_{[1^3]} = 24 .
\end{equation}
Thus, the fully symmetric sector $[3]$ is redundant. The non-null sectors are $[2,1]$ and $[1^3]$, giving $\mathcal{V}_3^{\mathrm{phys}} = 2[2,1]\oplus [1^3]$ with dimension $2^2+1^2=5$. Therefore, only five of the six virtual permutations are independent. The triple residual must satisfy
\begin{equation}
    \sum_{\pi \in S_3} \check{r}^{\pi(abc)}_{ijk} \equiv \sum_{\pi \in S_3} \check{r}^{a_{\pi(1)}a_{\pi(2)}a_{\pi(3)}}_{i_1i_2i_3} = 0
\label{eq:r3_null_constraint}
\end{equation}
or, equivalently, it is projected as
\begin{equation}
    R_3 \leftarrow (1 - \hat{\Pi}_{[3]})R_3, \quad \hat{\Pi}_{[3]} = \frac{1}{6}\sum_{\pi\in S_3}\hat{P}_{\pi}.
\end{equation}
Here, $\hat{P}_{\pi}$ permutes the virtual indices according to $\pi$.

For quadruples, the character table for $S_4$ and the required class data are
\begin{equation}
\begin{array}{c|ccccc}
    & e & (12) & (12)(34) & (123) & (1234) \\
    \hline
    {[4]}       & 1 & 1  & 1  & 1  & 1  \\
    {[3,1]}     & 3 & 1  & -1 & 0  & -1 \\
    {[2^2]}     & 2 & 0  & 2  & -1 & 0  \\
    {[2,1^2]}   & 3 & -1 & -1 & 0  & 1  \\
    {[1^4]}     & 1 & -1 & 1  & 1  & -1 \\
    \hline
    |C|        & 1  & 6  & 3  & 8  & 6  \\
    m(C)       & 16 & -8 & 4  & 4  & -2
\end{array}
\end{equation}
The eigenvalues of the quadruple overlap matrix are
\begin{equation}
\begin{aligned}
    &\mu_{[4]} = 0, \qquad
    \mu_{[3,1]} = 0, \\
    &\mu_{[2^2]} = 12, \quad
    \mu_{[2,1^2]} = 24, \quad
    \mu_{[1^4]} = 120 .
\end{aligned}
\end{equation}
Thus, the redundant part is $[4]\oplus 3[3,1]$, with dimension $1^2+3^2=10$. The non-null sectors are $[2^2]$, $[2,1^2]$, and $[1^4]$, giving $\mathcal{V}_4^{\mathrm{phys}} = 2[2^2]\oplus 3[2,1^2]\oplus [1^4]$ with dimension $2^2+3^2+1^2=14$. Therefore, only 14 of the 24 virtual permutations are independent. The raw quadruple residual is projected onto this nonredundant subspace as
\begin{equation}
    R_4 \leftarrow \Big(1 - \hat{\Pi}_{[4]} - \hat{\Pi}_{[3,1]} \Big) R_4,
\label{eq:T4ortho}
\end{equation}
where
\begin{equation}
    \hat{\Pi}_{[4]} = \frac{1}{24}\sum_{\pi\in S_4}\hat{P}_{\pi}, \quad \hat{\Pi}_{[3,1]} = \frac{3}{24} \sum_{\pi\in S_4} \chi_{[3,1]}(\pi) \hat{P}_{\pi}.
\end{equation}

In the implementation, Eq.~\eqref{eq:T4ortho} is applied as a polynomial in the sum of all virtual-index transpositions,
$\hat{\Omega} = \hat{P}_{(12)} + \hat{P}_{(13)} + \hat{P}_{(14)} + \hat{P}_{(23)} + \hat{P}_{(24)} + \hat{P}_{(34)}$:
\begin{equation}
    1 - \hat{\Pi}_{[4]} - \hat{\Pi}_{[3,1]} = \frac{(\hat{\Omega} - 6)(\hat{\Omega} - 2)(2\hat{\Omega}^2 + 19\hat{\Omega} + 48)}{576}.
\end{equation}

The non-null sectors identified above are consistent with the general spin-free selection rule. For a spin-free Hamiltonian, the spin and spatial parts transform according to conjugate Young diagrams.~\cite{pauncz1995symmetric} Since each electron has only two spin states, the spin Young diagrams can have at most two rows. Therefore, the corresponding spatial Young diagrams can have at most two columns. In terms of the spatial Young diagram $\lambda$, this condition is $\lambda_1\leq 2$, where $\lambda_1$ is the length of the first row. Sectors with $\lambda_1>2$ have zero norm in the spin-free overlap metric and therefore form the null space of the spin-free overlap matrix.

More generally, the number of independent spin-free virtual-permutation directions at excitation order $n$ is
\begin{equation}
    \dim\left( \mathcal{V}_n^{\mathrm{phys}} \right) = \sum_{\lambda :\ \lambda_1\leq 2} d_\lambda^2,
\end{equation}
where $\mathcal{V}_n^{\mathrm{phys}}$ denotes the nonredundant spin-free subspace, and $d_\lambda$ is the dimension of the $S_n$ symmetry sector labeled by $\lambda$. For $n=3$, $4$, and $5$, this formula gives $5$, $14$, and $42$ independent directions out of $6$, $24$, and $120$, respectively.

\begin{table*}[ht]
\caption{
Dominant contractions and leading operation counts for the most expensive $O(N^8)$ terms in the RCCSDT triples residual equations. The full representation stores $T_3$ as a rank-six tensor over all occupied triples $(i,j,k)$, whereas the compact representation stores only ordered triples $i \leq j \leq k$. In the compact case, the displayed expressions are representative contractions for a fixed ordered triple, and permutation-related contributions are generated using $\mathcal{P}$ while accumulating only the compact residual sector. Here, $\mathcal{P}_{(ia)(jb)(kc)}$ denotes simultaneous permutations of the paired columns $(i,a)$, $(j,b)$, and $(k,c)$. Additional subscripts impose ordering constraints after the paired columns have been permuted; for example, $\mathcal{P}_{(ia)(jb)(kc),2<3}$ retains, from the six possible paired-column permutations, only the three terms in which the paired column in the second position is ordered before that in the third position. Operation counts are reported in the large-$N_{\mathrm{o}}$ limit and in units of multiply-add pairs, including only the leading tensor-contraction cost. Permutation/indexing overhead, lower-order operations, and memory movement are omitted.}
\label{tab:rccsdt_operation_count}
\centering
\renewcommand{\arraystretch}{1.25}
\begin{tabular}{cc|cc}
\hline \hline
\multicolumn{2}{c|}{Full $T_3$ representation} & \multicolumn{2}{c}{Compact $T_3$ representation} \\
Contraction & Leading operation count & Representative contraction & Leading operation count \\
\hline 
$\displaystyle\frac{1}{4} \sum_{ld} \overline{W}^{la}_{di} \check{t}_{\check{l}jk}^{\check{d}bc}$ & $\displaystyle N_{\mathrm{o}}^4 N_{\mathrm{v}}^4$ & $\displaystyle\frac{1}{2} \mathcal{P}_{(ia)(jb)(kc),2<3} \bigg[\sum_{ld} \overline{W}^{la}_{di} \check{t}_{\check{l}jk}^{\check{d}bc} \bigg]$ & $\displaystyle\frac{1}{2}  N_{\mathrm{o}}^4 N_{\mathrm{v}}^4$ \\
$\displaystyle- \frac{1}{2} \sum_{ld} \overline{W}^{la}_{id} \check{t}_{jlk}^{dbc}$ & $\displaystyle N_{\mathrm{o}}^4 N_{\mathrm{v}}^4$ & $\displaystyle- \frac{1}{2} \mathcal{P}_{(ia)(jb)(kc),2<3} \bigg[ \sum_{ld} \overline{W}^{la}_{id} \left(\check{t}_{jlk}^{dbc} + \check{t}_{klj}^{dcb} \right)\bigg] $ & $\displaystyle\frac{1}{2} N_{\mathrm{o}}^4 N_{\mathrm{v}}^4$ \\
$\displaystyle- \sum_{ld} \overline{W}^{lb}_{id} \check{t}_{jlk}^{dac}$ & $\displaystyle N_{\mathrm{o}}^4 N_{\mathrm{v}}^4$ & $\displaystyle- \mathcal{P}_{(ia)(jb)(kc)} \bigg[\sum_{ld} \overline{W}^{lb}_{id} \check{t}_{jlk}^{dac} \bigg]$ & $\displaystyle N_{\mathrm{o}}^4 N_{\mathrm{v}}^4$ \\
$\displaystyle\frac{1}{2} \sum_{lm} W_{ij}^{lm} \check{t}_{lmk}^{abc}$ & $\displaystyle N_{\mathrm{o}}^5 N_{\mathrm{v}}^3$ & $\displaystyle \mathcal{P}_{(ia)(jb)(kc),1<2} \bigg[ \sum_{lm} W^{lm}_{ij} \check{t}_{lmk}^{abc} \bigg]$ & $\displaystyle\frac{1}{2} N_{\mathrm{o}}^5 N_{\mathrm{v}}^3$ \\
$\displaystyle\frac{1}{2} \sum_{de} W_{de}^{ab} \check{t}_{ijk}^{dec}$ & $\displaystyle N_{\mathrm{o}}^3 N_{\mathrm{v}}^5$ & $\displaystyle \mathcal{P}_{(ia)(jb)(kc),1<2} \bigg[ \sum_{de}  W^{ab}_{de} \check{t}_{ijk}^{dec} \bigg]$ & $\displaystyle\frac{1}{2} N_{\mathrm{o}}^3 N_{\mathrm{v}}^5$ \\
\hline \hline
\end{tabular}
\end{table*}

\begin{table*}[ht]
\caption{Dominant contractions and leading operation counts for the most expensive $O(N^{10})$ terms in the RCCSDTQ quadruples residual equations, evaluated in the large-$N_{\mathrm{o}}$ limit. Notation and operation-count conventions follow Table~\ref{tab:rccsdt_operation_count}.}
\label{tab:rccsdtq_operation_count}
\centering
\renewcommand{\arraystretch}{1.25}
\begin{tabular}{cc|cc}
\hline \hline
\multicolumn{2}{c|}{Full $T_4$ representation} & \multicolumn{2}{c}{Compact $T_4$ representation} \\
Contraction & Leading operation count & Representative contraction & Leading operation count \\
\hline 
$\displaystyle\frac{1}{12} \sum_{em} \overline{W}_{ei}^{ma} \check{t}_{\check{m}jkl}^{\check{e}bcd}$ & $\displaystyle N_{\mathrm{o}}^5 N_{\mathrm{v}}^5$ &
$\displaystyle \frac{1}{2} \mathcal{P}_{(ia)(jb)(kc)(ld), 2<3<4} \bigg[\sum_{em} \overline{W}_{ei}^{ma} \check{t}_{\check{m}jkl}^{\check{e}bcd}\bigg]$
& $\displaystyle\frac{1}{6} N_{\mathrm{o}}^5 N_{\mathrm{v}}^5$ \\
$\displaystyle- \frac{1}{4} \sum_{em} \overline{W}_{ie}^{ma} \check{t}_{jmkl}^{ebcd}$ & $\displaystyle N_{\mathrm{o}}^5 N_{\mathrm{v}}^5$ &
$\displaystyle- \frac{1}{2} \mathcal{P}_{(ia)(jb)(kc)(ld), 2<3<4} \bigg[\sum_{em} \overline{W}_{ie}^{ma} \left(\check{t}_{jmkl}^{ebcd} + \check{t}_{kmjl}^{ecbd} + \check{t}_{lmjk}^{edbc} \right)\bigg]$
& $\displaystyle\frac{1}{6} N_{\mathrm{o}}^5 N_{\mathrm{v}}^5$ \\
$\displaystyle- \frac{1}{2} \sum_{em} \overline{W}_{ie}^{mb} \check{t}_{jmkl}^{eacd}$ & $\displaystyle N_{\mathrm{o}}^5 N_{\mathrm{v}}^5$ & $\displaystyle
\mathcal{P}_{(ia)(jb)(kc)(ld), 3<4} \bigg[\sum_{em} \overline{W}_{ie}^{mb} \check{t}_{jmkl}^{eacd}\bigg] $
& $\displaystyle\frac{1}{2} N_{\mathrm{o}}^5 N_{\mathrm{v}}^5$ \\
$\displaystyle \frac{1}{4} \sum_{mn} W_{ij}^{mn} \check{t}_{mnkl}^{abcd}$ & $\displaystyle  N_{\mathrm{o}}^6 N_{\mathrm{v}}^4$ &
$\displaystyle \mathcal{P}_{(ia)(jb)(kc)(ld), 1<2,3<4} \bigg[ \sum_{mn} W_{ij}^{mn} \check{t}_{mnkl}^{abcd} \bigg] $
& $\displaystyle \frac{1}{4} N_{\mathrm{o}}^6 N_{\mathrm{v}}^4$ \\
$\displaystyle \frac{1}{4} \sum_{ef} W_{ef}^{ab} \check{t}_{ijkl}^{efcd}$ & $\displaystyle N_{\mathrm{o}}^4 N_{\mathrm{v}}^6$ &
$\displaystyle \mathcal{P}_{(ia)(jb)(kc)(ld), 1<2, 3<4} \bigg[ \sum_{ef} W_{ef}^{ab}\check{t}_{ijkl}^{efcd} \bigg]$
& $\displaystyle \frac{1}{4} N_{\mathrm{o}}^4 N_{\mathrm{v}}^6$ \\
$\displaystyle \frac{1}{8} \sum_{em} W_{eij}^{mab} \check{t}_{\check{m}kl}^{\check{e}cd}$ & $\displaystyle  N_{\mathrm{o}}^5 N_{\mathrm{v}}^5$ & 
$\displaystyle \frac{1}{2} \mathcal{P}_{(ia)(jb)(kc)(ld), 1<2, 3<4} \bigg[\sum_{em} W_{eij}^{mab} \check{t}_{\check{m}kl}^{\check{e}cd}\bigg]$
& $\displaystyle \frac{1}{4} N_{\mathrm{o}}^5 N_{\mathrm{v}}^5$ \\
$\displaystyle - \frac{1}{2} \sum_{em} W_{iej}^{mab} \check{t}_{kml}^{ecd}$ & $\displaystyle N_{\mathrm{o}}^5 N_{\mathrm{v}}^5$ & 
$\displaystyle - \frac{1}{2} \mathcal{P}_{(ia)(jb)(kc)(ld), 1<2, 3<4} \bigg[ \sum_{em} \left(W_{iej}^{mab} + W_{jei}^{mba} \right) \left(\check{t}_{kml}^{ecd} + \check{t}_{lmk}^{edc} \right)\bigg]$ & $\displaystyle \frac{1}{4} N_{\mathrm{o}}^5 N_{\mathrm{v}}^5$ \\
$\displaystyle - \sum_{em} W_{iej}^{mcb} \check{t}_{kml}^{ead}$ & $\displaystyle N_{\mathrm{o}}^5 N_{\mathrm{v}}^5$ & $\displaystyle - \mathcal{P}_{(ia)(jb)(kc)(ld)} \bigg[\sum_{em} W_{iej}^{mcb} \check{t}_{kml}^{ead}\bigg]$ & $\displaystyle N_{\mathrm{o}}^5 N_{\mathrm{v}}^5$ \\
$\displaystyle \frac{1}{2} \sum_{mn} W_{ijk}^{amn} \check{t}_{mnl}^{bcd}$ & $\displaystyle N_{\mathrm{o}}^6 N_{\mathrm{v}}^4$ & $\displaystyle \mathcal{P}_{(ia)(jb)(kc)(ld), 2<3} \bigg[\sum_{mn} W_{ijk}^{amn} \check{t}_{mnl}^{bcd}\bigg]$ & $\displaystyle \frac{1}{2} N_{\mathrm{o}}^6 N_{\mathrm{v}}^4$ \\
\hline \hline
\end{tabular}
\end{table*}

\begin{table*}[ht]
\caption{Dominant contractions and leading operation counts for the most expensive $O(N^{8})$ terms in the UCCSDT triples residual equations, evaluated in the large-$N_{\mathrm{o}}$ and large-$N_{\mathrm{v}}$ limits. The notation and operation-count conventions follow Table~\ref{tab:rccsdt_operation_count}. Here, $\mathcal{A}_{ijk}$ and $\mathcal{A}^{abc}$ denote antisymmetrizers over occupied and virtual indices, respectively. Optional subscripts, such as $1<2$ and $2<3$, impose compact-ordering constraints on the generated permutations after antisymmetrization. The four rows above the horizontal separator correspond to the $\alpha\alpha\alpha$ residual, whereas the ten rows below correspond to the $\alpha\alpha\beta$ residual.}
\label{tab:uccsdt_operation_count}
\centering
\renewcommand{\arraystretch}{1.25}
\setlength{\extrarowheight}{2pt}
\begin{tabular}{cc|cc}
\hline \hline
\multicolumn{2}{c|}{Full $T_3$ representation} & \multicolumn{2}{c}{Compact $T_3$ representation} \\
Contraction & Leading operation count & Representative contraction & Leading operation count \\
\hline 
$\displaystyle \frac{1}{24} \sum_{de} W^{ab}_{de} t_{ijk}^{dec}$ & $\displaystyle N_{\mathrm{o}}^3 N_{\mathrm{v}}^5$ & $\displaystyle \frac{1}{2} \mathcal{A}^{abc,1<2} \bigg[ \sum_{de} W^{ab}_{de} t_{ijk}^{dec} \bigg] $ & $\displaystyle \frac{1}{12} N_{\mathrm{o}}^3 N_{\mathrm{v}}^5$ \\
$\displaystyle \frac{1}{24} \sum_{lm} W^{lm}_{ij} t_{lmk}^{abc}$ & $\displaystyle N_{\mathrm{o}}^5 N_{\mathrm{v}}^3$ & $\displaystyle \frac{1}{2} \mathcal{A}_{ijk,1<2} \bigg[\sum_{lm} W^{lm}_{ij} t_{lmk}^{abc}\bigg]$ & $\displaystyle \frac{1}{12} N_{\mathrm{o}}^5 N_{\mathrm{v}}^3$ \\
$\displaystyle \frac{1}{4} \sum_{ld} \overline{W}^{al}_{id} t_{ljk}^{dbc}$ & $\displaystyle N_{\mathrm{o}}^4 N_{\mathrm{v}}^4$ & $\displaystyle \mathcal{A}^{abc,2<3} \mathcal{A}_{ijk,2<3} \bigg[\sum_{ld} \overline{W}^{al}_{id} t_{ljk}^{dbc}\bigg]$ & $\displaystyle \frac{1}{4} N_{\mathrm{o}}^4 N_{\mathrm{v}}^4$ \\
$\displaystyle \frac{1}{4} \sum_{\overline{ld}} \overline{W}^{a\overline{l}}_{i\overline{d}} t_{jk\overline{l}}^{bc\overline{d}}$ & $\displaystyle N_{\mathrm{o}}^3 N_{\mathrm{v}}^3 N_{\overline{\mathrm{o}}} N_{\overline{\mathrm{v}}}$  & $\displaystyle \mathcal{A}^{abc,2<3} \mathcal{A}_{ijk,2<3}  \bigg[ \sum_{\overline{ld}} \overline{W}^{a\overline{l}}_{i\overline{d}} t_{jk\overline{l}}^{bc\overline{d}}\bigg]$ & $\displaystyle \frac{1}{4} N_{\mathrm{o}}^3 N_{\mathrm{v}}^3 N_{\overline{\mathrm{o}}} N_{\overline{\mathrm{v}}}$  \\
\hline \rule[-1ex]{0pt}{5ex}
$\displaystyle \frac{1}{8} \sum_{de} W^{ab}_{de} t_{ij\overline{k}}^{de\overline{c}}$ & $\displaystyle N_{\mathrm{o}}^2 N_{\mathrm{v}}^4 N_{\overline{\mathrm{o}}} N_{\overline{\mathrm{v}}}$ & $\displaystyle \frac{1}{2} \sum_{de} W^{ab}_{de} t_{ij\overline{k}}^{de\overline{c}}$ & $\displaystyle \frac{1}{4} N_{\mathrm{o}}^2 N_{\mathrm{v}}^4 N_{\overline{\mathrm{o}}} N_{\overline{\mathrm{v}}}$ \\
$\displaystyle \frac{1}{2} \sum_{\overline{d}e} W^{b\overline{c}}_{e\overline{d}} t_{ij\overline{k}}^{ae\overline{d}}$ & $\displaystyle N_{\mathrm{o}}^2 N_{\mathrm{v}}^3 N_{\overline{\mathrm{o}}} N_{\overline{\mathrm{v}}}^2$ & $\displaystyle \mathcal{A}^{ab} \bigg[\sum_{\overline{d}e} W^{b\overline{c}}_{e\overline{d}} t_{ij\overline{k}}^{ae\overline{d}} \bigg]$ & $\displaystyle \frac{1}{2} N_{\mathrm{o}}^2 N_{\mathrm{v}}^3 N_{\overline{\mathrm{o}}} N_{\overline{\mathrm{v}}}^2$ \\
$\displaystyle \frac{1}{8} \sum_{lm} W^{lm}_{ij} t_{lm\overline{k}}^{ab\overline{c}}$ & $\displaystyle N_{\mathrm{o}}^4 N_{\mathrm{v}}^2 N_{\overline{\mathrm{o}}} N_{\overline{\mathrm{v}}}$ & $\displaystyle \frac{1}{2} \sum_{lm} W^{lm}_{ij} t_{lm\overline{k}}^{ab\overline{c}}$ & $\displaystyle \frac{1}{4} N_{\mathrm{o}}^4 N_{\mathrm{v}}^2 N_{\overline{\mathrm{o}}} N_{\overline{\mathrm{v}}}$ \\
$\displaystyle \frac{1}{2} \sum_{l\overline{m}} W^{l\overline{m}}_{i\overline{k}} t_{lj\overline{m}}^{ab\overline{c}}$ & $\displaystyle N_{\mathrm{o}}^3 N_{\mathrm{v}}^2 N_{\overline{\mathrm{o}}}^2 N_{\overline{\mathrm{v}}}$ & $\displaystyle \mathcal{A}_{ij} \bigg[ \sum_{l\overline{m}} W^{l\overline{m}}_{i\overline{k}} t_{lj\overline{m}}^{ab\overline{c}} \bigg]$ & $\displaystyle \frac{1}{2} N_{\mathrm{o}}^3 N_{\mathrm{v}}^2 N_{\overline{\mathrm{o}}}^2 N_{\overline{\mathrm{v}}}$ \\
$\displaystyle \sum_{ld} W^{al}_{id} t_{lj\overline{k}}^{db\overline{c}}$ & $\displaystyle N_{\mathrm{o}}^3 N_{\mathrm{v}}^3 N_{\overline{\mathrm{o}}} N_{\overline{\mathrm{v}}}$ & $\displaystyle \mathcal{A}^{ab} \mathcal{A}_{ij} \bigg[\sum_{ld} W^{al}_{id} t_{lj\overline{k}}^{db\overline{c}}\bigg]$ & $\displaystyle N_{\mathrm{o}}^3 N_{\mathrm{v}}^3 N_{\overline{\mathrm{o}}} N_{\overline{\mathrm{v}}}$ \\
$\displaystyle \sum_{\overline{l}\overline{d}} W^{a\overline{l}}_{i\overline{d}} t_{j\overline{l}\overline{k}}^{b\overline{d}\overline{c}}$ & $\displaystyle N_{\mathrm{o}}^2 N_{\mathrm{v}}^2 N_{\overline{\mathrm{o}}}^2 N_{\overline{\mathrm{v}}}^2$ & $\displaystyle \mathcal{A}^{ab} \mathcal{A}_{ij} \bigg[\sum_{\overline{l}\overline{d}} W^{a\overline{l}}_{i\overline{d}} t_{j\overline{l}\overline{k}}^{b\overline{d}\overline{c}}\bigg]$ & $\displaystyle N_{\mathrm{o}}^2 N_{\mathrm{v}}^2 N_{\overline{\mathrm{o}}}^2 N_{\overline{\mathrm{v}}}^2$ \\
$\displaystyle - \frac{1}{2} \sum_{l\overline{d}} \overline{W}^{l\overline{c}}_{i\overline{d}} t_{lj\overline{k}}^{ab\overline{d}}$ & $\displaystyle N_{\mathrm{o}}^3 N_{\mathrm{v}}^2 N_{\overline{\mathrm{o}}} N_{\overline{\mathrm{v}}}^2$  & $\displaystyle - \mathcal{A}_{ij} \bigg[ \sum_{l\overline{d}} \overline{W}^{l\overline{c}}_{i\overline{d}} t_{lj\overline{k}}^{ab\overline{d}} \bigg]$ & $\displaystyle \frac{1}{2} N_{\mathrm{o}}^3 N_{\mathrm{v}}^2 N_{\overline{\mathrm{o}}} N_{\overline{\mathrm{v}}}^2$ \\
$\displaystyle - \frac{1}{2} \sum_{\overline{l}d} \overline{W}^{a\overline{l}}_{d\overline{k}} t_{ij\overline{l}}^{db\overline{c}}$ & $\displaystyle N_{\mathrm{o}}^2 N_{\mathrm{v}}^3 N_{\overline{\mathrm{o}}}^2 N_{\overline{\mathrm{v}}}$ & $\displaystyle - \mathcal{A}^{ab} \bigg[\sum_{\overline{l}d} \overline{W}^{a\overline{l}}_{d\overline{k}} t_{ij\overline{l}}^{db\overline{c}}\bigg]$ & $\displaystyle \frac{1}{2} N_{\mathrm{o}}^2 N_{\mathrm{v}}^3 N_{\overline{\mathrm{o}}}^2 N_{\overline{\mathrm{v}}}$ \\
$\displaystyle \frac{1}{4} \sum_{ld} W^{\overline{c}l}_{\overline{k}d} t_{ijl}^{abd}$ & $\displaystyle N_{\mathrm{o}}^3 N_{\mathrm{v}}^3 N_{\overline{\mathrm{o}}} N_{\overline{\mathrm{v}}}$ & $\displaystyle \sum_{ld} W^{\overline{c}l}_{\overline{k}d} t_{ijl}^{abd}$ & $\displaystyle \frac{1}{4} N_{\mathrm{o}}^3 N_{\mathrm{v}}^3 N_{\overline{\mathrm{o}}} N_{\overline{\mathrm{v}}}$ \\
$\displaystyle \frac{1}{4} \sum_{\overline{ld}} W^{\overline{c}\overline{l}}_{\overline{k}\overline{d}} t_{ij\overline{l}}^{ab\overline{d}}$ & $\displaystyle N_{\mathrm{o}}^2 N_{\mathrm{v}}^2 N_{\overline{\mathrm{o}}}^2 N_{\overline{\mathrm{v}}}^2$ & $\displaystyle \sum_{\overline{ld}} W^{\overline{c}\overline{l}}_{\overline{k}\overline{d}} t_{ij\overline{l}}^{ab\overline{d}}$ & $\displaystyle \frac{1}{4} N_{\mathrm{o}}^2 N_{\mathrm{v}}^2 N_{\overline{\mathrm{o}}}^2 N_{\overline{\mathrm{v}}}^2$ \\
\hline \hline
\end{tabular}
\end{table*}

\section{Tabulated absolute energies}

The converged energies (in Hartree) for the systems studied in this work are listed in Table~\ref{tab:raw_data}.

\begin{table*}[ht]
\caption{Converged energies (in Hartree) for the systems studied in this work. HF values are total energies, while post-HF values are correlation energies. Numbers of occupied ($N_{\mathrm{o}}$) and virtual ($N_{\mathrm{v}}$) orbitals used in the post-HF calculations are also given. Counterpoise-corrected calculations were performed for the benzene, naphthalene, and polyene monomers, leading to larger numbers of virtual orbitals than in the corresponding isolated monomer calculations. Energies are reported to six decimal places in Hartree; energy differences in the main text were computed from higher-precision raw data, so minor discrepancies of approximately $0.01~\mathrm{kJ/mol}$ may arise from rounding.}
\label{tab:raw_data}
\centering
\renewcommand{\arraystretch}{1.1}
\begin{tabular}{llrrrrrrrrr}
\hline \hline
Systems & Basis set & HF & $N_{\mathrm{o}}$ & $N_{\mathrm{v}}$ & MP2 & CCSD & CCSD(T) & CCSDT & CCSDT(Q) & CCSDTQ \\
\hline
Benzene dimer & cc-pVDZ & $-461.440246$ & 30 & 186 & $-1.570905$ & $-1.646810$ & $-1.718779$ & $-1.719352$ & $-1.724845$ & -- \\
Naphthalene dimer & cc-pVDZ & $-766.761078$ & 48 & 292 & $-2.602530$ & $-2.694935$ & $-2.822051$ & $-2.821394$ & $-2.831976$ & -- \\
Benzene monomer & cc-pVDZ & $-230.723107$ & 15 & 207 &  $-0.781636$ & $-0.820928$ & $-0.856479$ & $-0.856862$ & $-0.859550$ & -- \\
Naphthalene monomer & cc-pVDZ & $-383.386000$ & 24 & 326 & $-1.292847$ & $-1.342356$ & $-1.404941$ & $-1.404867$ & $-1.410008$ & -- \\
\hline 
(C$_2$H$_4$)$_2$ & cc-pVDZ & $-156.079775$ & 12 & 80 & $-0.548636$ & $-0.608691$ & $-0.628032$ & $-0.629273$ & $-0.630500$ & -- \\
(C$_4$H$_6$)$_2$ & cc-pVDZ & $-309.867797$ & 22 & 142 & $-1.068523$ & $-1.160624$ & $-1.203762$ & $-1.205702$ & $-1.208783$ & -- \\
(C$_6$H$_8$)$_2$ & cc-pVDZ & $-463.656935$ & 32 & 204 & $-1.591555$ & $-1.713544$ & $-1.781207$ & $-1.783752$ & $-1.788867$ & -- \\
(C$_8$H$_{10}$)$_2$ & cc-pVDZ & $-617.446263$ & 42 & 266 & $-2.115896$ & $-2.266870$ & $-2.359326$ & $-2.362435$ & $-2.369673$ & -- \\
(C$_{10}$H$_{12}$)$_2$ & cc-pVDZ & $-771.235886$ & 52 & 328 & $-2.640483$ & $-2.820151$ & $-2.937480$ & $-2.941134$ & $-2.950534$ & -- \\
C$_2$H$_4$ & cc-pVDZ & $-78.040341$ & 6 & 88 & $-0.273952$ & $-0.304043$ & $-0.313678$ & $-0.314302$ & $-0.314911$ & -- \\
C$_4$H$_6$ & cc-pVDZ & $-154.935461$ & 11 & 157 & $-0.532652$ & $-0.579106$ & $-0.600502$ & $-0.601496$ & $-0.603007$ & -- \\
C$_6$H$_8$ & cc-pVDZ & $-231.831065$ & 16 & 226 & $-0.792751$ & $-0.854608$ & $-0.888100$ & $-0.889430$ & $-0.891925$ & -- \\
C$_8$H$_{10}$ & cc-pVDZ & $-308.726851$ & 21 & 295 & $-1.053333$ & $-1.130224$ & $-1.175924$ & $-1.177576$ & $-1.181093$ & -- \\
C$_{10}$H$_{12}$ & cc-pVDZ & $-385.622671$ & 26 & 364 & $-1.314092$ & $-1.405872$ & $-1.463825$ & $-1.465792$ & $-1.470350$ & -- \\
\hline
Cr(CO)$_6$ & def2-TZVPP & $-1719.905841$ & 37 & 382 & $-3.079567$ & $-2.871712$ & $-3.060640$ & $-3.047737$ & $-3.073617$ & -- \\
Cr(CO)$_5$ & def2-TZVPP & $-1607.091168$ & 32 & 327 & $-2.655989$ & $-2.484185$ & $-2.647211$ & $-2.637153$ & $-2.661585$ & -- \\
CO & def2-TZVPP & $-112.784617$ & 5 & 55 & $-0.354160$ & $-0.357527$ & $-0.374439$ & $-0.374641$ & $-0.375797$ & -- \\
\hline
Semibullvalene reactant & cc-pVDZ(d,s) & $-307.508206$ & 20 & 100 & $-1.015488$ & $-1.066692$ & $-1.111381$ & $-1.111603$ & $-1.115270$ & $-1.114831$ \\
Semibullvalene TS& cc-pVDZ(d,s) & $-307.480944$ & 20 & 100 & $-1.037741$ & $-1.075146$ & $-1.125937$ & $-1.125664$ & $-1.130170$ & $-1.129567$ \\
\hline \hline
\end{tabular}
\end{table*}

\bibliography{main}
\end{document}